\documentclass[twocolumn,showpacs,preprintnumbers,eqsecnum,amsmath,amssymb]{revtex4}
\usepackage[dvips]{color}
\usepackage{amsmath}
\usepackage{graphicx}
\usepackage{bm}
\begin{document}
\title{\large \bf Junctions of three quantum wires for spin 1/2 electrons}

\author{Chang-Yu Hou}
\author{Claudio Chamon}
\affiliation{Department of Physics, Boston University, Boston,
MA 02215}

\date{\today}

\begin{abstract}
We study the effects of electron-electron interactions on the
transport properties of a junction of three quantum wires enclosing a
magnetic flux. The wires are modeled as single channel spin-1/2
Tomonaga-Luttinger liquids. The system exhibits a rich phase diagram as a function of the electronic interaction strength, which includes a chiral fixed point with an asymmetric current flow highly sensitive to the sign of the flux, and another fixed point where pair tunneling dominates, similarly to the case of spinless electrons. While in the case of spinless electrons the perturbations that correspond to unequal couplings between the three wires are always irrelevant, we find that, when the electron spin is included, there are small regions in the phase diagram where a current flows only between two of the wires and the third wire is decoupled.
\end{abstract}

\maketitle

\section{Introduction}
\label{sec:intro}

The transport properties of quantum wire systems have been the subject
of intensive investigation in the past one and a half decade, both
because they can have practical applications in nano-electronic circuits
and because they provide an experimentally realizable way for
understanding exotic properties of one-dimensional interacting
electron systems. As opposed to two- and three-dimensional electronic
systems, one-dimensional (1D) systems cannot be described by Fermi
Liquid theory; instead, they are described as Tomonaga-Luttinger
liquids (TLL)~\cite{Tomonaga,Luttinger,Mattis&Lieb,Haldane}, and
experiments on carbon nanotubes are consistent with this
description~\cite{Bockrath,Yao}.

Recently, there has been a number of studies of junctions of multiple
TLL wires~\cite{COA,Nayak,Rao,Rao-SenPRB2004,Egger,Egger2,Lederer,Safi,Moore,Yi,KVF2004,Akira2004,Das,Sodano2005,Meden1,Meden2,Meden3,Benoit2005,white}. Such
studies are relevant because junctions of three or more quantum wires would inevitably appear in any quantum circuit. New tools and methods for investigating junctions
of three quantum wires with {\it spinless} electrons were proposed in
Ref.~\cite{COA}. These new methods allowed for the identification of a
low energy chiral fixed point with an asymmetric current flow that is
highly sensitive to the sign of the magnetic flux enclosed at the
junction. There are, however, important outstanding issues in the
three-wire junctions problem, for instance, the behavior of the more
realistic model in which the electron spin is taken into account.

Even in the case of tunneling between two wires, the inclusion of spin
degrees of freedom already brings about a rich phase diagram in the
charge and spin interaction parameter space~\cite{Kane, Furusaki,WongAffleck}. For example,
one can find a situation where the charge conductance vanishes while
the spin conductance does not, or vice versa. In the case of junctions
of three quantum wires (Y-junction) for electrons with spin, the phase
diagram becomes much richer.  In addition to a phase similar to the
two-wire case in which electron pair tunneling dominates while the
spin conductance vanishes, or vice versa, the chiral fixed point
revealed previously in the spinless Y-junction case persists when the
spin degrees of freedom are taken into account. Moreover, while the
strong asymmetric limit in which one of the wires is totally decoupled
was proved always unstable for the spinless case~\cite{COA}, we find a small region in the coupling parameter space in which this asymmetric fixed point is stable.


The paper is organized as follows.
In Sec.~\ref{sec:results} we summarize the results and present the phase diagram in coupling constant space at zero temperature. We also discuss the conductance tensor corresponding to each stable fixed point.
In Sec.~\ref{sec:model} we present our effective model for the
junctions of three quantum wires for spin-1/2 electrons.
In Sec.~\ref{sec:DEBC} we review the Delayed Evaluation of Boundary
Condition (DEBC) method introduced previously for identifying the
stable fixed points for the Y-junction system with spinless
electrons. Taking advantage of charge/spin separation in one
dimension, we generalize this approach to the system with spin-1/2
electrons and illustrate the method by describing the phase diagram of a junction of two quantum wires.
In Sec.~\ref{sec:DEBC-three-junction} we apply the DEBC method described in Sec.~\ref{sec:DEBC} to the Y-junction with spin 1/2 electrons and determine the stability of each fixed point.
In Sec.~\ref{sec:BCFT-Review} we briefly review the method of Boundary Conformal Field Theory (BCFT) and apply it, as a warm up, to study a junction of two quantum wires. 
We then apply BCFT methods to the Y-junction problem in 
Sec.~\ref{sec:BCFT-three-junction}, and obtain results consistent with those found with the DEBC method. 
%
We briefly conclude in Sec.~\ref{sec:Conclusion}.

\section{Summary of Results}
\label{sec:results}
\begin{figure}
\includegraphics[width=0.60\linewidth]{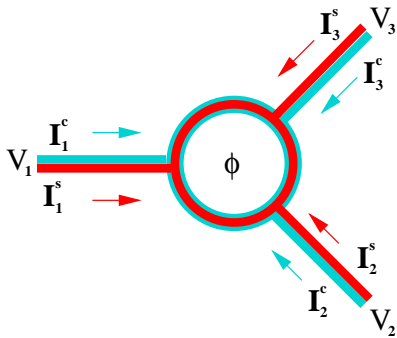}
\caption{[Color online] A junction of three quantum wires with a magnetic flux
threading through the ring. The $I_{1,2,3}^{c(s)}$ are the charge and spin currents
arriving at the junction from each of the three wires.  The lighter
line (Blue online) indicates the charge sector, and the darker line
(Red online) presents the spin sector.}
\label{fig:deviceA}
\end{figure}

In this section, we summarize our study of the stability of the fixed
points associated with the different boundary conditions of Y-junctions
for spin-1/2 electrons and their corresponding conductance
tensors. Based on the stability of each phase, we propose a
zero-temperature phase diagram in terms of the TLL interaction
parameters, $g_{c}$ and $g_s$. The geometry of the device is depicted in
Fig.~\ref{fig:deviceA}, where three identical quantum wires are
attached to a ring, with equal couplings and thus having a $Z_3$
symmetry. The ring can be threaded by a magnetic flux and the quantum
wires are modelled as TLLs characterized by the interaction parameters
$g_c$ and $g_s$. We will only consider spin conserving transport.

Due to the charge and spin separation in the bulk
of TLL quantum wires, independent boundary conditions can be imposed on the charge and spin degrees of freedom. Hence, we introduce
the notation $B_cB_s$ in which $B_c$ and $B_s$ represent the
corresponding boundary condition in the charge and spin sector
respectively. For instance, DN represents Dirichlet boundary condition
(B.C.) in the charge sector and Neumann B.C. in the spin
sector. Moreover, since each combination of the boundary conditions
corresponds to a fixed point in the framework of the renormalization group
(RG), the fixed point corresponding to $B_cB_s$ BC will be referred to as {\it $B_cB_s$ fixed point}.  The physical meaning of
each boundary condition will be discussed in this section and the
detailed analysis and assumptions of the model are given in later
sections.

The charge and spin conductance tensors associated with each fixed point
are important physical response functions characterizing the
Y-junction for spin 1/2 electrons. Within the linear response theory, the
total current $I_j$ flowing into the junction from wire $j$ is related to
the voltage applied at the wire $k$ through
\begin{equation}
I_j^{c} = \sum_k G_{jk}^{c} V_k,
\label{eq:Def-of-conduc-tensor}
\end{equation}
where $j,k = 1,2,3$ and $G_{jk}$ is the $3 \times 3$ conductance
tensor. One can similarly define,
\begin{equation}
I_j^{s} = \sum_k G_{jk}^{s} {(\Delta_{\mu})}_k/e,
\label{eq:Def-of-spin-conduc-tensor}
\end{equation}
where $ {\Delta_{\mu}}_k=\mu^\uparrow_k-\mu^\downarrow_k$ is the chemical potential difference between up and down spins in lead $k$, and $e$ is the electron charge (with this definition, spin and charge conductances are both measured in units of $e^2/h$).

Note that current conservation implies that:
\begin{equation}
\sum_j I_j^{c(s)}=0.
\end{equation}
Furthermore, a common voltage applied to all three wires results
in zero current. Thus:
\begin{equation}
 \sum_j G_{jk}^{c(s)}= \sum_k G_{jk}^{c(s)}=0 \;.
\label{eq:conduc-aspen}
\end{equation}

For $Z_3$ symmetry junctions, the conductance tensor takes the
form~\cite{COA}
\begin{equation}
 G_{jk}^{c(s)} =
\frac{G_S^{c(s)}}{2} \;
(3\delta_{jk}-1)+\frac{G_A^{c(s)}}{2}\; \epsilon_{jk} \; ,
\label{eq:Z3tensor}
\end{equation}
where we separate the symmetric and anti-symmetric components of the
tensor, and $G_S$ and $G_A$ are scalar conductances. (The
$\epsilon_{ij}$ are defined as follows:
$\epsilon_{12}=\epsilon_{23}=\epsilon_{31}=1$, $\epsilon_{21}
=\epsilon_{32}=\epsilon_{13}=-1$ and
$\epsilon_{11}=\epsilon_{22}=\epsilon_{33}=0$.) The anti-symmetric
component $G_A^{c(s)}$ will only appear when time reversal invariance
is broken. Even in the presence of the magnetic flux which would normally break time reversal symmetry (TRS), $G_A$ may vanish at some low
energy fixed points, in which case time reversal symmetry is
restored. However, in the absence of the $Z_3$ symmetry (in the
asymmetric fixed point), the condition of Eq.~(\ref{eq:Z3tensor})
becomes unnecessary. Observe that $G_s=G_{jj}$ represents the
conductance of each wire when zero voltage is applied to the other two
wires ({\it i.e.}, a potential difference is applied between one of
the wires and the other two, which are held at the same potential).

We are mostly interested in the stable RG fixed points which describe
the physics in the low energy limit, i.e. low voltage bias and low
temperature. As in the case of junctions of two quantum wires, the
interaction parameters control the stability of the fixed points and
determine the phase diagram. Below we list our results for conductance
tensors and for the basins of attraction around each fixed point (with
their corresponding boundary conditions) as function of the interaction
parameters, $g_c$ and $g_s$.


\subsection{Neumann BC in both charge and spin sector}

\begin{figure}
\includegraphics[width=0.70\linewidth]{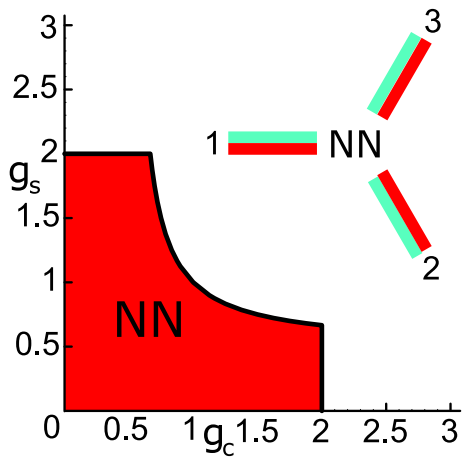}
\caption{[Color Online] The painted area (Red) shows the attractive basin of NN fixed point. The border shows the marginal line of scaling dimension $\Delta = 1$ for all leading order perturbations. The inset depicts the physical consequence of the fixed point related to the NN BC, a disconnection for both charge and spin sectors.}
\label{fig:phase-NN}
\end{figure}

The NN boundary condition corresponds to a fixed point in which the hopping amplitudes between the wires are zero and the three quantum wires are totally decoupled from each other, as illustrated in the inset of Fig.~\ref{fig:phase-NN}. When the interactions in the quantum wire are repulsive, the hopping amplitudes decrease along the RG flow both in junctions of two quantum wires~\cite{Kane,Furusaki, WongAffleck} and in junctions of three quantum wires for spinless electrons~\cite{COA}. Generically, the window of stability for a fixed point with decoupled wires should be independent of the number of wires in the junction. Notice that the effective flux $\phi$ has no effect in the decoupled limit.

The three most relevant perturbations at the NN fixed point are:

(1) Single electron hopping between two wires, with scaling dimension $\Delta_{NN}=\frac{1}{2}(\frac{1}{g_c}+\frac{1}{g_s})$.

(2) Electron pair singlet hopping between two wires, with scaling dimension $\Delta_{NN}=\frac{2}{g_c}$.

(3) Exchange of electrons with opposite spins between two of the wires (thus carrying a spin current), with scaling dimension $\Delta_{NN}=\frac{2}{g_s}$.

By requiring that all scaling dimensions corresponding to these leading order perturbations $\Delta_{NN}>1$, we can identify the attractive basin in the interaction parameter space, as shown in Fig.~\ref{fig:phase-NN}. The conductance tensor for the NN fixed point is zero for both the charge and spin sector.
\begin{equation}
G_{jk}^{c(s)}|_{NN}=0\; 
\end{equation}


\subsection{Dirichlet BC in both charge and spin sector}

\begin{figure}
\includegraphics[width=0.70\linewidth]{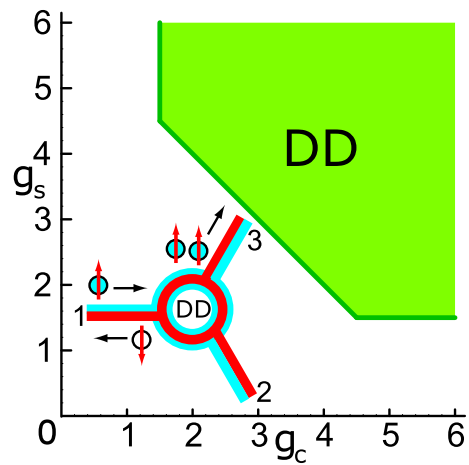}
\caption{[Color Online]  The painted area (green) shows the attractive basin for the DD fixed point. The border shows the marginal line when all leading order perturbations have $\Delta = 1$. The inset shows one of the tunneling processes associated with the DD fixed point. The conductances in both charge and spin sectors are enhanced by the pair hopping (or Andreev-like) processes in both spin and charge sectors.}
\label{fig:phase-DD}
\end{figure}

The DD BC corresponds to a fixed point where the paired-electron hopping and Andreev reflecting processes, depicted in the inset of the Fig.~\ref{fig:phase-DD}, dominate the tunneling processes in both charge and spin degrees of freedom. Hence, the charge and spin conductance will be enhanced. Moreover, the conductance tensor takes the symmetric form in which the scalar conductance $G^{c(s)}_S=2\times (4g_{c(s)}/3)(e^2/h)$ and the antisymmetric component $G_A=0$, and is given by
\begin{equation}
G_{jk}^{c(s)}|_{DD}= 2 \times \frac{2 g_{c(s)}}{3} \; (3\delta_{jk}-1).
\label{eq:DD-Conduc}
\end{equation}
Here, the factor of $2$ comes from the doubling of the degrees of freedom due to the spin. Since $G_A=0$, the presence of the effective flux $\phi$ which breaks TRS has no physical consequence and the TRS is restored in DD fixed point. 

We shall study the leading order perturbations to determine the stability of the DD fixed point. Since several boundary operators possess the same scaling dimensions, we only list the dimensions of the leading order perturbations without specifying the corresponding operators,
\begin{equation}
\Delta_{DD}^1=\frac{1}{6}(g_c+g_s), \;\; \Delta_{DD}^2= \frac{2 g_c}{3}, \;\; \Delta_{DD}^3=\frac{2 g_s}{3}.
\end{equation}
The attractive basin of the DD fixed point, shown in Fig.~\ref{fig:phase-DD}, can be obtained by requiring the dimensions of these leading order perturbations to be larger than one. 


\subsection{ND and DN BC}

\begin{figure}
\includegraphics[width=0.85\linewidth]{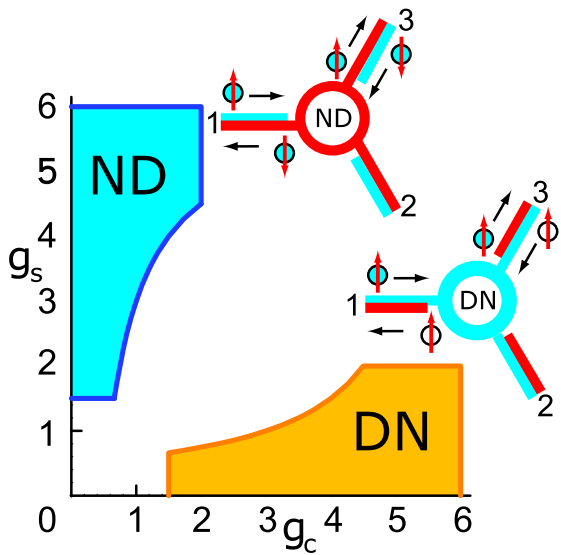}
\caption{[Color Online] The painted area shows the stable region for the ND
and DN fixed points. The borders (Blue for ND BC and Yellow for DN BC) show the marginal line of scaling dimensions $\Delta = 1$ for all leading order perturbations. The inset shows a pictorial representation of the ND and DN fixed points. The fundamental process associated with the ND fixed point leads to an enhancement of the spin conductance due to a spin exchange between two wires, while the fundamental process associated with the DN fixed point leads to an enhancement of the charge conductance due to a electron-hole exchange. Notice that the processes described here occur between  any two arbitrary wires.}
\label{fig:phase-ND-DN}
\end{figure}

We have seen that the charge and spin current terminate or flow at the same direction for both NN and DD fixed point. However, phases with decoupled charge and spin degrees of freedom are also possible. Indeed, the ND and DN BC correspond to fixed points where the charge or spin degrees of freedom are disentangled at the boundary. The ND fixed point possesses a pure spin current; likewise, the DN fixed point possesses a pure charge current. The dominant processes corresponding to these two fixed points are illustrated in the insets of Fig.~\ref{fig:phase-ND-DN}. Notice that similar phases also exist in the system of junction of two quantum wires for spin-1/2 electrons.

Observe that the dominating process of the ND fixed points carries no net charge current while that of DN fixed point carries no net spin current, hence the charge and spin conductance vanish at the ND and DN fixed point, respectively. The conductance tensors, taking the symmetric form with $Z_3$ symmetry, are given by
\begin{eqnarray}
G_{jk}^{s}|_{ND}&=& 2 \frac{2 g^{s}}{3} \; (3\delta_{jk}-1),\qquad G_{jk}^{c}|_{ND}=0,
\\
G_{jk}^{c}|_{DN}&=& 2 \frac{2 g^{c}}{3} \; (3\delta_{jk}-1),\qquad  G_{jk}^{s}|_{DN}=0\; .
\end{eqnarray}

The leading order perturbations of the ND BC have the dimensions, discussed in detail in Sec.~\ref{sec:DEBC-three-junction} and \ref{sec:BCFT-three-junction},
\begin{equation}
\Delta_{ND}^1=\frac{1}{2g_c}+\frac{g_s}{6}, \;\;
\Delta_{ND}^2=\frac{2 g_s}{3}, \;\; \Delta_{ND}^3=\frac{2}{g_c}.
\label{eq:ND-Scal-Dim}
\end{equation}
The basin of attraction relative to the ND fixed point is obtained by requiring all $\Delta_{ND}^j \geq 1$ and is depicted in Fig.~\ref{fig:phase-ND-DN}. By exchanging $g_c$ and $g_s$ in Eq.(\ref{eq:ND-Scal-Dim}), we can obtain the dimensions of the leading order perturbations for the DN fixed points. The basins of attraction corresponding to the ND and DN fixed points thus show a mirror symmetry with respect to the line $g_c=g_s$ in the interaction parameter space. 

Unlike the case of the junction of two quantum wires, where basins of attraction corresponding to the NN, DD, ND, and DN fixed points patch the whole space of the interaction parameters with some overlap, see Sec.~\ref{sec:DEBC}, a section of the parameter space remains uncovered. This implies the existence of nontrivial fixed points. Notice that both ND and DN fixed points restore TRS even in the presence of the magnetic field.


\subsection{$\chi^{\ }_+\chi^{\ }_+$ and $\chi^{\ }_-\chi^{\ }_-$ B.C}

\begin{figure}
\includegraphics[width=0.85\linewidth]{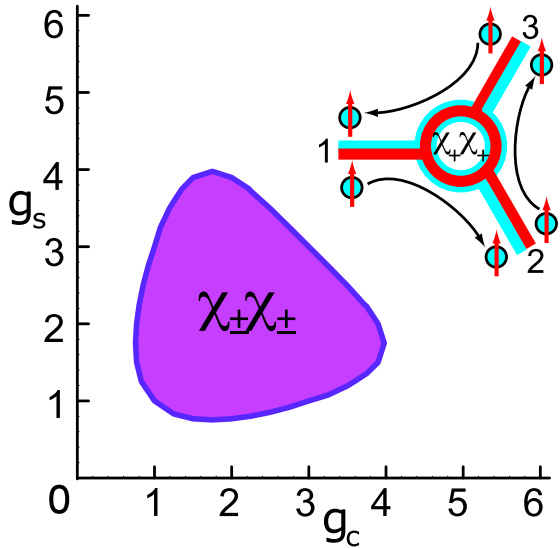}
\caption{ [Color Online] The painted area is the basin of attraction of both $\chi^{\ }_\pm \chi^{\ }_\pm$ fixed points. This area sits between and osculates the stable regions of the NN, DD, ND, and DN fixed points. The border shows the marginal line for scaling dimension $\Delta = 1$ of Eq.(\ref{eq:Scal-Dim-chiral}). The pictorial representation of the tunneling processes associated with $\chi^{\ }_+\chi^{\ }_+$ fixed point is shown in the inset. An incoming electron from wire $i$ will always divert to wire $i+1$. The currents associated with the $\chi^{\ }_-\chi^{\ }_-$ fixed point flow in the inverse direction.} 
\label{fig:phase-chiral}
\end{figure}

The most striking consequences of $\chi^{\ }_+$ and $\chi^{\ }_-$ fixed point are that TRS is broken explicitly and the stability of the fixed points is determined both by the effective flux, $\phi$, and the interaction parameters. For spinless electrons, the $\chi^{\ }_+$ fixed point can be stable when $0<\phi<\pi$ while the $\chi^{\ }_-$ can be stable when $-\pi<\phi<0$~\cite{COA}. As a result, the charge and spin current always flow together for spin 1/2 electrons, i.e. mixed chiral fixed points, $\chi^{\ }_{\pm}\chi^{\ }_{\mp}$, are never stable. Moreover, among all combinations of BCs corresponding to chiral BC, only $\chi^{\ }_\pm \chi^{\ }_\pm$ fixed points are stable. One can conjecture that the criterion of stability of the $\chi_\pm$ BC due to the flux for spin 1/2 electrons is the same as for spinless electrons. The boundary operators of leading order perturbations, such as backscattering processes, have scaling dimension in terms of interaction parameters
\begin{equation}
\Delta=\frac{2 g_c}{3+g_c^2}+\frac{2 g_s}{3+g_s^2}, 
\label{eq:Scal-Dim-chiral}
\end{equation}
and provide the only constraint for determining the stability. The basin of attraction is thus common for both $\chi_\pm \chi_\pm$ fixed points and is depicted in the Fig.~\ref{fig:phase-chiral}. However, which phase is preferred in the low energy limit is completely determined by the effective flux.

The conductance tensors at the $\chi^{\ }_\pm \chi^{\ }_\pm$ fixed points takes the $Z_3$ symmetric form in the Eq.(\ref{eq:Z3tensor})
\begin{equation}
  G_{jk}^{c(s)}|_{\chi_\pm} = \frac{4g_{c(s)}}{3+ g_{c(s)}^2} \frac{e^2}{h} \;
  \left[(3\delta_{jk}-1)\pm g_{c(s)}\,\epsilon_{jk}\right]
\label{eq:chi-Cond}
\end{equation}
with the scalar conductance $G_S^{c(s)}=\frac{8g_{c(s)}}{3+ g_{c(s)}^2} \frac{e^2}{h}$, and the antisymmetric component $G_A^{c(s)}=g_{c(s)}G_S^{c(s)}$. The dominating hopping processes are schematically shown in the inset of the Fig.~\ref{fig:phase-chiral}. Notice that the conductance at the $\chi^{\ }_\pm \chi^{\ }_\pm$ fixed points becomes perfect transmission of charge and spin from wire $j$ to wire $j \pm 1$ when $g_{c(s)}\to 1$. 

\subsection{Asymmetric Boundary Conditions, $D_A D_A$ BC}

\begin{figure}
\includegraphics[width=0.85\linewidth]{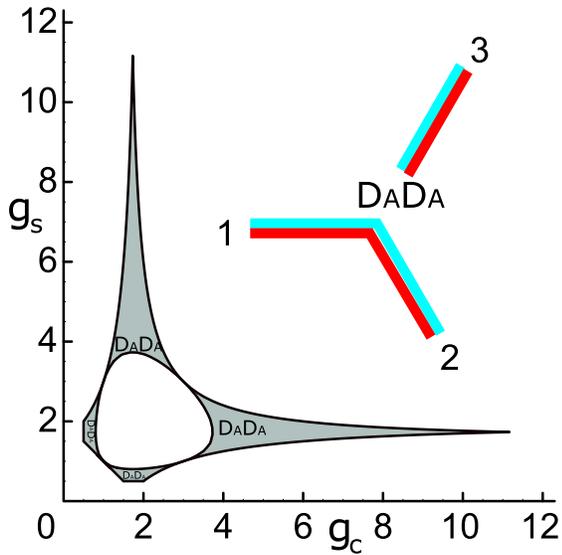}
\caption{[Color Online] The painted areas are the basins of attraction for the $D_AD_A$ fixed point. The overlap of the attractive basin for the $D_A D_A$ fixed point and others is very complicated and will be illustrated in Fig.~\ref{fig:phase-All}. A pictorial representation of the $D_AD_A$ fixed point with the decoupled third wire is shown in the inset. Notice that there is a three fold degeneracy of this fixed point, {\it i.e.}, the decoupled wire can be arbitrary, resulting from breaking the ${\mathcal Z}_3$ symmetry.}
\label{fig:phase-AA}
\end{figure}

We shall consider the simplest $Z_3$ asymmetric BC, $D_A$, corresponding to a situation where as shown in the inset of Fig~\ref{fig:phase-AA}, two of the wires are strongly coupled while the third one is decoupled from the rest of system. Note that the $D_A$ fixed point has been proven unstable in the Y-junction system for spinless electrons~\cite{COA}. However, for spin-1/2 electrons, there is a window in which the $D_AD_A$ fixed point is stable. Of course, it depends on the detailed structure of the hopping amplitudes to determine which wire will be decoupled. 

As shown in Fig.~\ref{fig:phase-AA}, the stable area of $D_A D_A$ fixed point emerges from the borders of the attractive basins of other fixed points. Several boundary operators contribute to the leading order perturbations and need to be considered for determining the stability. Here, we will only present the basins of attraction corresponding to $D_A D_A$ BC and postpone the discussion of scaling dimensions.

The conductance at the asymmetric fixed points will not take the form in Eq.(\ref{eq:Z3tensor}) due to the broken $\mathcal{Z}_3$ symmetry. The conductance tensor at the $D_AD_A$ fixed point is given in the matrix representation
\begin{equation}
G^{c(s)}|_{D_AD_A}=2 g_{c(s)} \frac{e^2}{h}\left(%
\begin{array}{ccc}
  1 & -1 & 0 \\
  -1 & 1 & 0 \\
  0 & 0 & 0 \\
\end{array}%
\right)\;.
\end{equation}
Despite an extra component representing the decoupled third wire, the conductance tensor is exactly the same as that at $DD$ fixed point for junction of two quantum wires.



\begin{figure}
\includegraphics[width=0.9\linewidth]{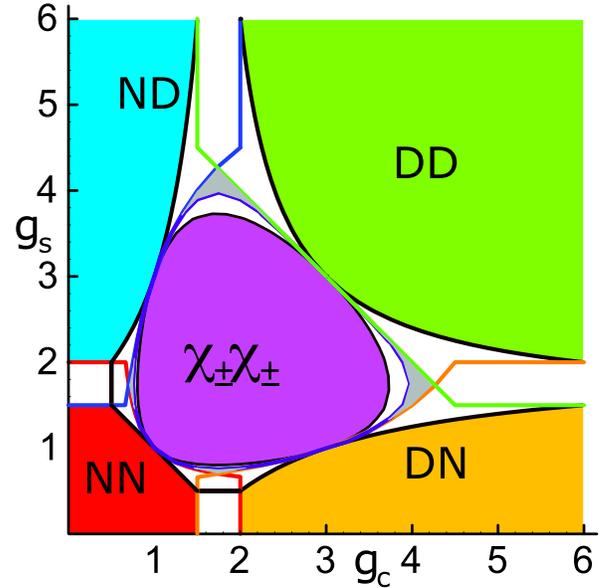}
\caption{[Color Online] The proposed phase diagram - The colored lines indicate the marginal boundary of $\Delta = 1$ for each fixed point upon the identification of the leading order perturbations. The painted areas shows the regions with only one stable fixed point while the unpainted regions represent those where two or three fixed points have an overlapping region of stability. Notice that the $D_AD_A$ fixed point is the sole stable one in four tiny gray areas surrounded by the unpainted regions. }
\label{fig:phase-All}
\end{figure}

We shall conclude this section by proposing a phase diagram based on the results discussed above. The stability of phases at some area in the parameter space is uniquely defined because there exists only one stable fixed point. As a result, the painted areas in the Fig.~\ref{fig:phase-All} with red, green, blue, orange, purple and grey colors unambiguously correspond to the $NN$, $DD$, $ND$, $DN$, $\chi_{\pm} \chi_{\pm}$, and $D_A D_A$ fixed points, respectively. When there are more than one possible candidates for stable fixed points, determining the stability of phases becomes tricky and generically non-universal. Hence, the phases in the low energy limit are defined not only by the interaction parameters, $g_{c(s)}$ but also by the details of the device. For instance, the different hopping amplitudes of single electron tunneling between wires may determine the final destination of the RG flow and the phase at the overlaps of $DD$ and $D_A D_A$ fixed points. However, determining the stable fixed points in these cases is beyond the scope of our methods and can only be conjectured. 

Let us consider the system along the $SU_s(2)$ invariant line, $g_s=1$: the NN fixed point is stable when $g_c<1$, the chiral fixed points are stable when $1<g_c<3$, and the DN fixed point is stable when $g_c>3$. This result is expected due to the similarity with a system of spinless electrons where N fixed point is stable when $g_c<1$, the chiral fixed points are stable when $1<g_c<3$, and D fixed point is stable when $g_c>3$. Notice that four common tangential points of marginal lines when $(g_c,g_s)=(1,1),(1,3),(3,1),(3,3)$ remain marginal for all adjacent phases.


\section{Model and Bosonization}
\label{sec:model}

In this section, we define the simplest model including the TLL effects and spin degree of freedom. As depicted in Fig.~\ref{fig:deviceA}, we study three identical single channel quantum wires with spin 1/2 electrons joined with a ring enclosing a magnetic flux. We will ignore phonons and impurities, and assume that electron-electron interaction is short-ranged in the wires. The Euclidean action of a semi-infinite interacting single channel TTL with spin-1/2 electrons is described in terms of the independent charge and spin boson fields, 
\begin{eqnarray}
\nonumber S= \sum_{j=1}^3 && \int d\tau dx \{ \frac{v_c g_c}{4\pi} [(\partial_x \varphi_{j,c})^2+ \frac{1}{v_c^2}  (\partial_\tau \varphi_{j,c})^2]
\\
&+&
\label{eq:BulkAction-model-1} \frac{v_s g_s}{4 \pi} [(\partial_x \varphi_{j,s})^2+ \frac{1}{v_s^2}  (\partial_\tau \varphi_{j,s})^2] \},
\end{eqnarray}
where the subscript $j=1,2,3$ represent each wire and $v_{c(s)}$ is the sound velocity for the charge and spin degree of freedom respectively, or in terms of dual field $\theta$
\begin{eqnarray}
\nonumber S= \sum_{j=1}^3 && \int d\tau dx \{ \frac{v_c}{4\pi g_c} [(\partial_x \theta_{j,c})^2+ \frac{1}{v_c^2}  (\partial_\tau \theta_{j,c})^2]
\\
&+&
\label{eq:BulkAction-model-2} \frac{v_s}{4 \pi g_s} [(\partial_x \theta_{j,s})^2+ \frac{1}{v_s^2}  (\partial_\tau \theta_{j,s})^2] \} .
\end{eqnarray}
Here, $\varphi_{c(s)}$ and $\theta_{c(s)}$ are phase fields which follow the canonical commutation relation, $[\varphi_{c(s)}(x),\theta_{c(s)}(x')]=-i\Theta(x-x')$, where $\Theta(x)$ is the Heaviside step function. Notice that $g_c=g_s=1$ corresponds to the non-interacting point, and in the absence of a magnetic field and any spin-dependent interactions, we must take $g_s=1$ in order to respect the underlying $SU(2)$ symmetry~\cite{Kane}.

The fields with spin up and down degrees of freedom can be represented as linear combinations of these charge and spin boson fields,
\begin{equation}
\label{eq:up-down-theta-model} \varphi_{\sigma,i} =\frac{\varphi_{c,i}+
\sigma \varphi_{s,i}}{\sqrt{2}}, \; {\rm and} \qquad
\theta_{\sigma,i}=\frac{\theta_{c,i}+
\sigma \theta_{si,}}{\sqrt{2}},
\end{equation}
where the commutation relation between $\varphi$ and $\theta$ fields is still followed. It is convenient to introduce left and right mover representations
\begin{equation}
\label{eq:RL-field-model} \phi^{L}_{\sigma,i}= \frac{\varphi_{\sigma,i}+\theta_{\sigma,i}}{2}, \; {\rm and} \qquad \phi^R_{\sigma,i}= \frac{\varphi_{\sigma,i}-\theta_{\sigma,i}}{2} ,
\end{equation}
and to identify the electron fermion field in terms of the boson field 
\begin{equation}
\label{eq:BosonRepre-model}\Psi^{L(R)}_{j,\sigma}= \;
\eta_{\sigma,j}^{L(R)} \; e^{i \sqrt{2} \phi_{\sigma,j}^{L(R)}} ,
\end{equation}
with Klein factors $\eta_{\mu}$ that satisfy the anti-commutation relation, $\{ \eta_\mu,\eta_\nu \}=2 \delta_{\mu,\nu}$ and commute with the boson fields.

The effect of interactions at the boundary in the action Eq.(\ref{eq:BulkAction-model-1}) and Eq.(\ref{eq:BulkAction-model-2}) will be enter in the form of tunneling operators between wires. These boundary operators must conserve both charge and spin and therefore they are constrained to respect the corresponding U(1) symmetries.


\section{DEBC}
\label{sec:DEBC}

The Delayed Evaluation of Boundary Condition (DEBC) method was introduced recently for determining the stability of the boundary conditions (fixed points) of the Y-junction for spinless electrons~\cite{COA}. In this section we will generalize this method to the case of spin-1/2 electrons and use this method to determine the phase diagram of a junction of two quantum wires.

Generically, fermionic operators can be represented in terms of the boson fields, $\varphi_{c(s)}(x,t)$, and their conjugate fields, $\theta_{c(s)}(x,t)$ up to a Klein factor in an infinite wire. For a semi-infinite wire, the relation between left and right moving fields leads to an analytical continuation $\phi^{R}(-x,t)=\phi^L(x,t)$. This is the familiar unfolded picture where the right mover and the left mover are related by a particular choice of the boundary condition, namely the N BC, at $x=0$. The DEBC method is an extension of the unfolded picture to different boundary conditions.

Within the DEBC framework, an arbitrary boundary operator should be first represented in terms of the independent bulk boson fields $\phi_{j}^{L(R)}$ without specifying the boundary condition. Then the boundary conditions that must be imposed on $\phi_{j}^{L(R)}(t,x=0)$ ($\phi_i$ and $\theta_i$) are determined {\it a posteriori}. Because $\phi_{j}^{L(R)}$ are functions of $x\pm t $, imposing a particular boundary condition relates $\varphi$ and $\theta$ fields in the bulk, and thus eliminates the redundancy of working with both $\varphi$ and $\theta$ fields in the semi-infinite wire.

The scaling dimension $\Delta$ of a boundary operator $\mathcal{O}_{B}$, with two-point correlation
\begin{equation}
\langle \mathcal{O}^{\ }_B(t) \mathcal{O}_B^\dag(t') \rangle \sim |t-t'|^{-2 \Delta},
\end{equation}
depends on the boundary conditions and can be used to determine whether a perturbation consisting of the boundary operators is relevant or not. When $\Delta>1$ the perturbation is irrelevant and when $\Delta=1$ the perturbation is marginal. Thus, a boundary condition is stable when all boundary operators either have scaling dimension, $\Delta=0$, or are irrelevant $\Delta>1$.

Now, let us introduce the generic representation of boundary operators and discuss how to obtain their scaling dimensions given a boundary condition.

\subsection{Boundary Operators and Scaling dimensions}

For applying the DEBC scheme, the left and right moving fields appearing in the boundary operators initially are treated as independent regardless of the boundary condition. The boson representation of fermions is given in Eq.(\ref{eq:BosonRepre-model})
\begin{equation}
\Psi^{L(R)}_{j,\sigma}(t,x=0) \propto \; \eta_{\sigma,j}^{L(R)} \; e^{i
\sqrt{2} \phi_{j,\sigma}^{L(R)}(t,x=0)}.
\end{equation}
All boundary operators are constructed by combining creation and
annihilation operators of fermions in each wire. The Klein factors and extra phases produced by commuting the boson fields will not affect the scaling dimension of a boundary operator
\begin{equation}
\mathcal{O}_B \sim \exp \left[i \sqrt{2} \sum_{i,\sigma,a} n_{i,\sigma}^a \; \phi^a_{i,\sigma}\right] ,
\end{equation}
near stable fixed points. Here, $n_{i,\sigma}^a$ uniquely defines the tunneling processes at the boundary, where $i$, $\sigma$, and $a$ represent the wire, the spin, and the chirality of the fermions, respectively. We shall refer hereafter to $n_{i,\sigma}^a$ as the particle number vector. Since the boundary operators are constructed from the full electron, the charge and spin degrees of freedom are coupled at the boundary. The total charge and spin conservation imply
\begin{equation}
\sum_{i,\sigma,a} n_{i,\sigma}^a=0,\; {\rm and} \qquad  \sum_{i,\sigma,a} \sigma n_{i,\sigma}^a=0\; ,
\end{equation}
respectively. For instance, a tunneling process where a up-spin right-mover at wire 1 scatters into a up-spin left-mover at wire 2 leads to $n_{1,\uparrow}^R=1=- n_{2,\uparrow}^L$ with the sign convention followed from the bosonic representation of the fermions.

It is convenient to introduce scaled bosonic fields
\begin{equation}
\tilde{\varphi}_{c(s)}=\sqrt{g_{c(s)}}\varphi_{c(s)} \; {\rm and} \qquad  \tilde{\theta}_{c(s)}=\frac{\theta_{c(s)}}{\sqrt{g_{c(s)}}}\; ,
\end{equation}
such that the commutation relation between $\tilde{\varphi}_{c(s)}$ and $\tilde{\theta}_{c(s)}$ is still followed. The action of the rescaled fields becomes independent of the interaction parameters, and their correlation functions are given by
\begin{equation}
\langle \tilde \theta_{c(s)}(z,\bar{z}) \tilde \theta_{c(s)}(0)\rangle= - \frac{1}{2} \ln |z|^2 .  
\end{equation}
In terms of the rescaled boson fields, the original left and right moving fields become
\begin{eqnarray}
\sqrt{2} \nonumber \phi_{i,\sigma}^a &=& ( \cosh \alpha_c \;
\tilde{\phi}^a_{i,c}+ \sigma \cosh \alpha_s \;
\tilde{\phi}^a_{i,s})
\\
\label{eq:LR-mixing}&&+ (\sinh \alpha_c \;
\tilde{\phi}^{\bar{a}}_{i,c}+ \sigma \sinh \alpha_s \;
\tilde{\phi}^{\bar{a}}_{i,s}) ,
\end{eqnarray}
where
\begin{equation}
\tilde \phi^{L}_{\sigma,i}= \frac{\tilde \varphi_{\sigma,i}+\tilde \theta_{\sigma,i}}{2}, \; {\rm and} \qquad \tilde \phi^R_{\sigma,i}= \frac{\tilde \varphi_{\sigma,i}-\tilde \theta_{\sigma,i}}{2} ,
\end{equation}
and $\cosh \alpha=(\frac{1}{\sqrt{g}}+\sqrt{g})/2$ and $\sinh \alpha=(\frac{1}{\sqrt{g}}-\sqrt{g})/2$. Further, $\bar{a}$ is defined such that $\bar{a}=L$ when $a=R$ and vice versa. Also, shorthanded notations, $\bar{R}=L$ and $\bar{L}=R$, are used. 

In terms of the particle number vector and non-interacting bosonic fields, the bosonic argument of the boundary operators becomes 
\begin{widetext}
\begin{equation}
\label{eq:Boundary-boson}
\sqrt{2} \sum_{i,\sigma,a} n_{i,\sigma}^a \; \phi^a_{i,\sigma} =\sum_{i,\sigma,a} \{ n_{i,\sigma}^a \;(\cosh \alpha_c \;
\tilde \phi^a_{i,c} + \sinh \alpha_c \; \tilde \phi^{\bar{a}}_{i,c} )
+ n_{i,\sigma}^a \; \sigma \; (\cosh \alpha_s\; 
\tilde \phi^a_{i,s} + \sinh \alpha_s; \tilde \phi^{\bar{a}}_{i,s}) \}.
\end{equation}
\end{widetext}
The non-trivial scaling behaviors of the boundary operators are attributed to the mixing structure of the left and right movers in Eq.(\ref{eq:LR-mixing}).

Let us introduce vectors $\vec{\phi}^{R(L)}_{c(s)}$ whose $i^{th}$ components are the fields $\tilde{\phi}^{R(L)}_{i,c(s)}$. Eq.(\ref{eq:Boundary-boson}) becomes
\begin{equation}
\label{eq:GenBounOpe-DEBC}(\vec{v}_c^R \cdot
\vec{\phi}^R_c+\vec{v}_s^R \cdot \vec{\phi}^R_s)+ (\vec{v}_c^L
\cdot \vec{\phi}^L_c+\vec{v}_s^L \cdot \vec{\phi}^L_s) \; ,
\end{equation}
where the vectors $\vec{v}_{c(s)}^{R(L)}$ are defined as
\begin{subequations}
\label{eq:electron-number-vector}
\begin{eqnarray}
\label{eq:electron-number-vector-1}
(v_c^R)_i &=& \sum_{\sigma} (n_{i,\sigma}^R \; \cosh \alpha_c
+n_{i,\sigma}^L \; \sinh \alpha_c)
\\
\label{eq:electron-number-vector-2}
(v_s^R)_i &=& \sum_{\sigma} \sigma (n_{i,\sigma}^R \; \cosh
\alpha_s +n_{i,\sigma}^L  \; \sinh \alpha_s)
\\
\label{eq:electron-number-vector-3}
(v_c^L)_i &=& \sum_{\sigma} (n_{i,\sigma}^L \; \cosh \alpha_c
+n_{i,\sigma}^R \; \sinh \alpha_c)
\\
\label{eq:electron-number-vector-4}
(v_s^L)_i &=& \sum_{\sigma}  \sigma (n_{i,\sigma}^L  \; \cosh
\alpha_s +n_{i,\sigma}^R \;  \sinh \alpha_s) .
\end{eqnarray}
\end{subequations}
Here $\vec{\phi}^{\bar{L}}=\vec{\phi}^R$ and $\vec{\phi}^{\bar{R}}=\vec{\phi}^L$ were used explicitly. Observe that the charge and spin degrees of freedom differ only by an extra $\sigma$ term.

The boundary conditions can be generally identified as
\begin{equation}
\vec{\phi}^R=\mathcal{R}^{-1} \vec{\phi}^L+\vec{C} |_{x=0},
\end{equation}
where $\vec{C}$ is a constant vector and $\mathcal{R}$ is a rotation matrix. The total charge and spin conservation imposes the N BC
\begin{equation}
\label{eq:N-BC-0-mode-DEBC}\Phi_{0,c(s)}^R=\Phi^L_{0,c(s)},
\end{equation}
on the center of mass mode, $\Phi_{0,c(s)}=\frac{1}{\sqrt{N}} \sum_{i=1}^N \phi_{i,c(s)}$, which in turn constraints the rotation matrix $\mathcal{R}$. Moreover, $\mathcal{R}$ has to be an orthogonal transformation to preserve the total length of the fields~\cite{Bellazzini}. The scaling dimension $\Delta$ of the boundary operators for a boundary condition  reads
\begin{equation}
\label{eq:ScaDim-DEBC} \Delta_{\mathcal{R}}(n_{i,\sigma}^a)=\frac{1}{4} \big|
\mathcal{R}_c \vec{v}_c^R +\vec{v}^L_c \big|^2 + \frac{1}{4} \big|
\mathcal{R}_s \vec{v}_s^R +\vec{v}^L_s \big|^2 .
\end{equation}

The physical processes responsible for pinning a boundary condition determine the rotation $\mathcal{R}$, for which $\Delta_{\mathcal{R}}=0$ for that specific particle number vector. The corresponding operators thus act as the identity operator at this fixed point. Once the boundary condition is picked, the stability of the fixed point can be analyzed by evaluating the scaling dimensions of all other boundary operators. In particular, the fixed point will be stable if all these dimensions, for all other particle number vectors, satisfy $\Delta_{\mathcal{R}}(n_{i,\sigma}^a)>1$.

To illustrate the method, here we apply the DEBC scheme to the case of a junction of two quantum wires for spin 1/2 electrons and find results in agreement with Ref.~\cite{Kane,Furusaki,WongAffleck}. 

\subsection{Boundary conditions of junction of two quantum wires for spin 1/2 electrons}

When requiring the spin and charge conservation and orthogonality, only two boundary conditions, Neumann and Dirichlet, are possible in the case of a junction of two quantum wires for spin 1/2 electrons. The $\mathcal{R}$ matrix takes a particularly simple form in both cases. The N BC corresponds to the total reflection fixed point and has $\mathcal{R}_N=\mathbf{1}$; The D BC corresponds to the perfect transport fixed point and has
\begin{equation}
\label{eq:DRotamatr-2-DEBC}\mathcal{R}_D= \left(
\begin{array}{cc}
  0 & 1 \\
  1 & 0 \\
\end{array}
\right).
\end{equation}
Since the N and D BC can be imposed independently in the charge and spin sector, we can now discuss the contributions of the charge and spin degree of freedom to the scaling dimensions separately.

\subsubsection{N BC in charge sector}

Focusing on the charge sector of Eq.(\ref{eq:ScaDim-DEBC}) with $\mathcal{R}_c=\mathbf{1}$, the scaling dimension of the charge sector reads
\begin{eqnarray}
\label{eq:scaling-NBC-Charge}&& \frac{1}{4} \big| \mathcal{R}_c \vec{v}_c^R +\vec{v}^L_c \big|^2
\\
\nonumber &=&  \frac{1}{4} \big| \sum_{i,\sigma,a} (n_{i,\sigma}^a \; \cosh
\alpha_c +n_{i,\sigma}^a \; \sinh \alpha_c) \hat{e}_{i,c} \big|^2
\\
\nonumber &=&  \frac{1}{4} \big| \sum_{\sigma,a} n_{1,\sigma}^a (\frac{1}{\sqrt{g_c}})
\hat{e}_{1,c} + \sum_{\sigma,a} n_{2,\sigma}^a (\frac{1}{\sqrt{g_c}})
\hat{e}_{2,c} \big|^2 ,
\end{eqnarray}
where we introduce $\hat{e}_{i,c}$ as the basis vectors of $\vec{v}_c$ and use $\cosh \alpha_{c}=(\frac{1}{\sqrt{g_c}}+\sqrt{g_c})/2$ and $\sinh \alpha_{c}=(\frac{1}{\sqrt{g_c}}-\sqrt{g_c})/2$ explicitly in the second equality. Due to the charge conservation, we can introduce
\begin{equation}
\label{eq:Defi-n-DEBC}n := \sum_{\sigma,a}
n_{1,\sigma}^a=-\sum_{\sigma,a} n_{2,\sigma}^a,
\end{equation}
and the contribution to the scaling dimension reads
\begin{eqnarray}
\Delta_{N_c} = \frac{1}{2g_c} n^2 .
\end{eqnarray}

\subsubsection{N BC in spin sector}

The contribution of the spin degree of freedom to the scaling dimension is very similar to the charge sector with a substitution of $\vec{v}_s$ to $\vec{v}_c$ in Eq.(\ref{eq:scaling-NBC-Charge}). Hence, the scaling dimension with N BC in the spin sector is given
\begin{eqnarray}
&& \frac{1}{4} \big| \mathcal{R}_s \vec{v}_s^R +\vec{v}^L_s \big|^2
\\
\nonumber &=&  \frac{1}{4} \big| \sum_{\sigma,a} n_{1,\sigma}^a \; \sigma
(\frac{1}{\sqrt{g_s}}) \hat{e}_{1,s} + \sum_{\sigma,a}
n_{2,\sigma}^a \;  \sigma(\frac{1}{\sqrt{g_s}})\hat{e}_{2,s} \big|^2.
\end{eqnarray}
Due to the spin conservation, we define
\begin{equation}
\label{eq:Defi-s-DEBC} s := \sum_{\sigma,a} n_{1,\sigma}^a
\sigma=-\sum_{\sigma,a} n_{2,\sigma}^a \sigma.
\end{equation}
with the relation $n=s$ (mod2). The scaling dimension for the N BC on the spin degree of freedom reads
\begin{equation}
\Delta_{N_s}= \frac{1}{2g_s} s^2.
\end{equation}

\subsubsection{D BC in charge sector}

The rotation matrix of D BC in Eq.(\ref{eq:DRotamatr-2-DEBC}) implies the mixing of two quantum wires. Inserting the matrix into the charge sector of the Eq.(\ref{eq:ScaDim-DEBC}), we obtain
\begin{widetext}
\begin{eqnarray}
\label{eq:scaling-DBC-charge} \big| \mathcal{R}_c \vec{v}_c^R +\vec{v}^L_c \big| 
=
\big| \sum_{\sigma} [ (n_{2,\sigma}^R + n_{1,\sigma}^L) \; \cosh
\alpha_c +(n_{2,\sigma}^L + n_{1,\sigma}^R) \; \sinh \alpha_c ] \;
\hat{e}_{1,c}\nonumber 
+ \sum_{\sigma} [ (n_{1,\sigma}^R + n_{2,\sigma}^L) \; \cosh
\alpha_c + (n_{1,\sigma}^L + n_{2,\sigma}^R) \; \sinh \alpha_c ]
\; \hat{e}_{2,c} \big| \nonumber 
\end{eqnarray}
By using charge conservation, we introduce a new variable
\begin{eqnarray}
\label{eq:Defi-tilden-DEBC}\tilde{n}=n_+=\sum_{l,\sigma,a}
n_{l,\sigma}^a \; \Theta(+ \;l\;a)=\sum_{\sigma} (n_{1,\sigma}^L
+n_{2,\sigma}^R) =-n_-=-\sum_{l,\sigma,a} n_{l,\sigma}^a \;
\Theta(- \;l\;a)=-\sum_{\sigma} (n_{1,\sigma}^R+n_{2,\sigma}^L)
.
\end{eqnarray}
Here, when applied inside $\Theta$-function, the indices of wires $l=1,2$ become $l=+,-$ respectively, and the left and right mover indices become $a=+,-$ respectively. In terms of the new variable, $\tilde{n}$, the scaling dimension is given by
\begin{eqnarray}
\Delta_{D_c}=\frac{1}{4} \big| \mathcal{R}_c \vec{v}_c^R
+\vec{v}^L_c \big|^2=\frac{1}{4} \big| (\sqrt{g_c} \; \tilde{n})
\; \hat{e}_{1,c}-(\sqrt{g_c} \; \tilde{n}) \; \hat{e}_{2,c}
\big|^2= \frac{1}{2} g_c \tilde{n}^2
\end{eqnarray}
\end{widetext}

\subsubsection{D B.C in Spin Sector}

Observe that the scaling dimensions attributed to the spin sector can be obtained by replacing $\vec{v}_c$ with $\vec{v}_s$ in Eq.(\ref{eq:scaling-DBC-charge}). Since we have spin conservation rather than charge conservation, we can define a new variable $\tilde{s}$
\begin{widetext}
\begin{eqnarray}
\label{eq:Defi-tildes-DEBC}\tilde{s}=s_+=\sum_{l,\sigma,a} \sigma 
n_{l,\sigma}^a \; \Theta(+ \;l\;a)=\sum_{\sigma}
(\sigma n_{1,\sigma}^L + \sigma n_{2,\sigma}^R )
=-s_-=-\sum_{l,\sigma,a}  \sigma n_{l,\sigma}^a \; \Theta(-
\;l\;a)=-\sum_{\sigma} (\sigma n_{1,\sigma}^R  + \sigma n_{2,\sigma}^L).
\end{eqnarray}
Also one can show the relation $\tilde{n}=\tilde{s}$ (mod2). The scaling dimension corresponding to the spin degree of freedom with the D BC is given by 
\begin{equation}
\Delta_{D_s}=\frac{1}{4} \big| \mathcal{R}_s \vec{v}_s^R +\vec{v}^L_s
\big|^2=\frac{1}{4} \big| (\sqrt{g_s} \; \tilde{s}) \;
\hat{e}_{1,s}-(\sqrt{g_s} \; \tilde{s}) \; \hat{e}_{2,s} \big|^2=
\frac{1}{2} g_s \tilde{s}^2
\end{equation}
\end{widetext}

\subsection{Stability of the fixed points for junction of two quantum wires}

Since both the total charge and spin are conserved separately, the boundary conditions can be imposed independently in the charge and spin sector. We follow the categorization in Sec.~\ref{sec:results} by relating a fixed point to a $B_cB_s$ BC, and explore its instability. The scaling dimensions of boundary operators given a BC are obtained by adding the contributions from both the charge and spin degrees of freedom. In light of the discussion in this section, the possible scaling dimensions of boundary operators are given by
\begin{subequations}
\label{eq:scaling-dimensions-2-DEBC} 
\begin{eqnarray}
\label{eq:NN-2-DEBC} \Delta_{NN}&=&\frac{1}{2}\left[\frac{1}{g_c}
n^2 + \frac{1}{g_s} s^2 \right]\; ,
\\
\label{eq:DD-2-DEBC} \Delta_{DD}&=&\frac{1}{2}\left[{g_c} {\tilde
n}^2 + {g_s} {\tilde s}^2 \right]\; ,
\\
\label{eq:ND-2-DEBC} \Delta_{ND}&=&\frac{1}{2}\left[\frac{1}{g_c}
n^2 + {g_s} {\tilde s}^2 \right]\; ,
\\
\label{eq:DN-2-DEBC} \Delta_{DN}&=&\frac{1}{2}\left[{g_c} {\tilde
n}^2 + \frac{1}{g_s} s^2 \right] \; ,
\end{eqnarray}
\end{subequations}
with the constraint $n= s\;({\rm mod}\; 2)$ and $\tilde n= \tilde
s\;({\rm mod}\; 2)$. Notice that there is no constraint between  $n,\tilde s$ and between $\tilde n,s$.

Let us compute, as an example, the scaling dimension of the following boundary operators given NN BC
\begin{subequations}
\begin{eqnarray}
&& \rm{{\bf 1}} . \;\;\;
T_B={\Psi^{R}_{1,\uparrow}}^\dagger\Psi^{L}_{1,\uparrow} \;
\propto  e^{i \sqrt{2} [-\phi_{1,\uparrow}^R+
\phi_{1,\uparrow}^L]}
\\
&& \rm{{\bf 2}}. \;\;\; T_F={\Psi^{R}_{2,\uparrow}}^\dagger
\Psi^{L}_{1,\uparrow} \;  \propto  e^{i \sqrt{2}
[-\phi_{2,\uparrow}^R+ \phi_{1,\uparrow}^L]}
\\
&& \rm{{\bf 3}}. \;\;\;
T_{\ }={\Psi^{R}_{2,\uparrow}}^\dagger{\Psi^{R}_{2,\downarrow}}^\dagger
\Psi^{L}_{1,\uparrow}\Psi^{L}_{1,\downarrow}
\\
&& \label{eq:PairExc-DEBC}\rm{{\bf 4}}. \;\;\;
T_{\ }={\Psi^{R}_{2,\uparrow}}^\dagger{\Psi^{R}_{1,\downarrow}}^\dagger
\Psi^{L}_{1,\uparrow}\Psi^{L}_{2,\downarrow} . 
\end{eqnarray}
\end{subequations}
The case 1 corresponds to the backscattering process with $n_{1,\uparrow}^L=1$ and $n_{1,\uparrow}^R=-1$, hence $n=0=s$ from Eq.(\ref{eq:Defi-n-DEBC}) and Eq.(\ref{eq:Defi-s-DEBC}). Notice that $\Delta^1_{NN}=0$, and thus backscattering is the physical process that fixes this boundary condition. The case 2 corresponds to the forward scattering with $n_{1,\uparrow}^L=1$ and $n_{2,\uparrow}^R=-1$ such that $n=1=s$ and $\Delta^2_{NN}=\frac{1}{2} (1/g_c+1/g_s)$. The case 3 corresponds to pair tunneling process with $n_{1,\uparrow}^L=1=n_{1,\downarrow}^L$ and $n_{2,\uparrow}^R=-1=n_{2,\downarrow}^R$. One can show that $n=2$ and $s=0$ and obtains $\Delta^3_{NN}=2/g_c$. The last case corresponds to pair exchange tunneling process with $n_{1,\uparrow}^L=1$, $n_{1,\downarrow}^R=-1$, $n_{2,\downarrow}^L=1$ and
$n_{2,\uparrow}^L=-1$ such that $n=0$ and $s=2$ and $\Delta_2^{NN}=2/g_s$. The last three boundary operators describe the leading order perturbations for the NN fixed point. The basin of attraction of the NN fixed point is determined when all $\Delta_{2,3,4}>1$.

\begin{figure}
\includegraphics[width=8cm]{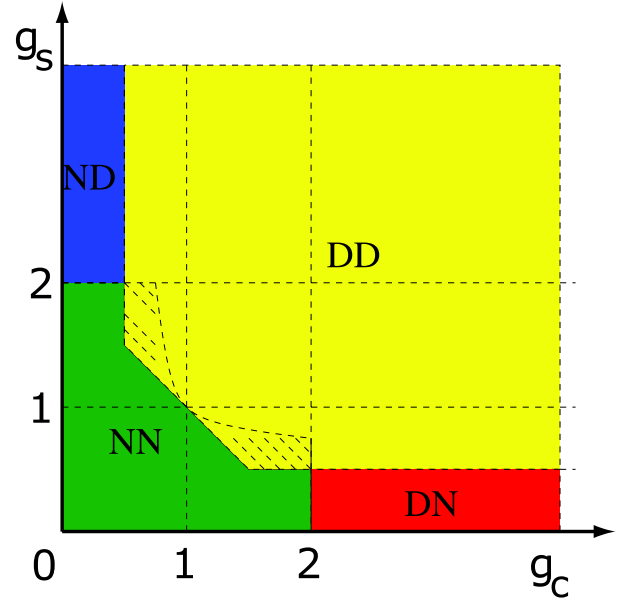}
\caption{Phase diagram of a junction of two quantum wires- The shaded area is common to the NN and DD stable regions. There should be an unstable intermediary fixed point in this range of $g_{c(s)}$.}
\label{fig:N2phase-diagram}
\end{figure}

In principle, we need to know scaling dimensions of all boundary operators given a boundary condition in order to determine the basin of attraction for each fixed point. However, Eq.(\ref{eq:scaling-dimensions-2-DEBC}) provides a compact way to determine the stability of the fixed points. Since the basin of attraction is defined when the scaling dimensions of the perturbations have $\Delta>1$, we can obtain the possible scaling dimensions in an increasing order by inserting the smallest integers for the quantities $n, s$, $\tilde n$, and $\tilde s$ with their respective parity constraints. Consequently, the stability of the fixed points can be completely determined from this minimum construction. The phase diagram determined by DEBC for junction of two quantum wires for spin-1/2 electrons is depicted in Fig.~\ref{fig:N2phase-diagram} and agrees with the results in \cite{Kane, Furusaki,WongAffleck}.

We conclude this section by reviewing the DEBC approach. First, the boundary operators should be expressed in terms of redundant fields $\varphi$ and $\theta$ or $\phi^{R(L)}$. Second, one has to identify the rotation matrix $\mathcal{R}$ corresponding to a given boundary condition. Finally, the scaling dimensions of boundary operators given a boundary condition are evaluated using the rotation matrix $\mathcal{R}$, and the basin of attraction of the fixed point can be determined by requiring all dimensions to be larger than one.


\section{Junction of three quantum wires for spin-1/2 electrons-DEBC}
\label{sec:DEBC-three-junction}

Following the DEBC scheme developed in the last section, we have to first identify the rotation matrix representations of boundary conditions for junctions of three quantum wires for spin 1/2 electrons. The conservation of the total charge and spin at the boundary can be implemented by imposing the N BC on the zero mode (total charge or spin mode) as shown in Eq.(\ref{eq:N-BC-0-mode-DEBC}). Also, the $\mathcal{R}$ has to be orthogonal to preserve the length of the field vector. Observe that any rotation around the unit vector, $\vec{u}_0=\frac{1}{\sqrt{3}} (1,1,1)$, will leave the total charge and spin mode invariant, hence naively $\mathcal{R}$ can be an arbitrary rotation around $\vec{u}_0$. However, there are some boundary conditions corresponding to discrete transformations, where $\mathcal{R}$ cannot be totally classified by a rotation. Moreover, as discussed in the Appendix~\ref{sec:conductance}, the conductance of fixed points can be directly related to the rotation matrices, Eq.(\ref{eq:conductance})s
\begin{equation}
\label{eq:conductance-master}
G_{ij,c(s)}=2 g_{c(s)}\frac{e^2}{h} (\delta_{ij}-\mathcal{R}_{ji}).
\end{equation}
We will compute the conductance by the construction of the rotation matrices associated with different boundary conditions.


\subsection{Boundary Conditions}

We will construct in this subsection the transformations $\mathcal{R}$ corresponding to different boundary conditions. Generically, any rotation matrix satisfying the constraint of conservation and orthogonality can be a candidate for a physical boundary condition. However, only some of them correspond to stable fixed points. Below, we will calculate the $\mathcal{R}$ matrices for physically motivated boundary conditions.

\subsubsection{Neumann boundary condition}

The Neumann boundary condition, 
\begin{equation}
\frac{\partial}{\partial x} \vec{\varphi}(x)|_{x=0}=0\Longleftrightarrow \vec{\theta}|_{x=0}=0 \; ,
\end{equation}
corresponds to a fixed point where the three wires are decoupled from each other. The relation $\vec{\theta}|_{x=0}=0$ can be represented in terms of the left and right moving fields, $\vec{\phi}^R=\vec{\phi}^L|_{x=0}$. The backscattering processes thus yield zero scaling dimension and dominate the low energy physics. The rotation matrix of the N BC is given $\mathcal{R}_N=\mathbf{1}$.

By using Eq.(\ref{eq:conductance-master}) and the rotation matrix of N BC, the conductance is 
\begin{equation}
G^{c(s)}_{ij,N}=2 g_{c(s)}\frac{e^2}{h} (\delta_{ij}-\delta_{ji})=0
\end{equation}

\subsubsection{Dirichlect B.C-Andreev reflection fixed point}

Since the D BC, $\varphi(x)|_{x=0}=C$, cannot be imposed on all boson fields due to the N BC imposed on the center of mass mode, it can be only imposed on two independent fields
\begin{subequations}
\begin{eqnarray}
\Phi^1(x)&=&\frac{1}{\sqrt{2}}(\varphi_1-\varphi_2)(x)|_{x=0}=0,
\\
\Phi^2(x)&=&\frac{1}{\sqrt{6}}(\varphi_1+\varphi_2-2\varphi_3)(x)|_{x=0}=0,
\end{eqnarray}
\end{subequations}
where $\Phi^1$ and $\Phi^2$ are orthogonal to the center of mass mode. Hence, the D BC is given by the condition
\begin{eqnarray}
\Phi^{R,1(2)}_{c(s)}(x)&=&-\Phi^{L,1(2)}_{c(s)}(x).
\end{eqnarray}
This transformation can be described by a rotation around $\vec{u}_0$ by $\pi$, and is given by
\begin{subequations}
\begin{equation}
\mathcal{R}_D=\left(%
\begin{array}{ccc}
  -\frac{1}{3} & \frac{2}{3} & \frac{2}{3} \\
  \frac{2}{3} & -\frac{1}{3} & \frac{2}{3} \\
  \frac{2}{3} & \frac{2}{3} & -\frac{1}{3} \\
\end{array}%
\right)
\end{equation}
or 
\begin{equation}
\mathcal{R}_{ij}=\frac{1}{3}(2-3\delta_{i,j}).
\end{equation}
\end{subequations}
Then the conductance of the D BC is
\begin{subequations}
\begin{eqnarray}
G^{c(s)}_{ij,D}&=&2 g_{c(s)} \frac{e^2}{h}
(\delta_{i,j}-\frac{1}{3}(2-3\delta_{i,j}))
\\
&=&2 \frac{ 2g_{c(s)}}{3} \frac{e^2}{h}(3\delta_{i,j}-1),
\end{eqnarray}
\end{subequations}
with $G_S^{c(s)}= \frac{ 8g_{c(s)}}{3}$. 

\subsubsection{Chiral boundary condition}

It is trickier to obtain the transformation corresponding to the chiral fixed points since the rotation matrix depends on the coupling constants. However, the pinning processes with $\Delta=0$ for the chiral fixed points provide the conditions for constructing the rotation matrix.

Let us consider the charge sector of the chiral-like tunneling processes: A left mover at wire $i$ scatters to a right mover at wire $i+p$, i.e. $n_{i,\sigma}^L=1$ and $n_{i+p,\sigma}^R=-1$ for all $i$. The scaling dimension of these boundary operators for arbitrary rotation matrix can be written using Eq.(\ref{eq:ScaDim-DEBC})
\begin{eqnarray}
\nonumber \Delta&=&|{\mathcal R} \left(\sinh\alpha_c \;{\hat e}_i-\cosh\alpha_c\;{\hat e}_{i+p}\right)
\\
\label{eq:ScalDimChiral-fixedpoints} &&+\left(\cosh\alpha_c\;{\hat e}_i-\sinh\alpha_c\;{\hat
e}_{i+p}\right) |^2 .
\end{eqnarray}
By using the fact that $\Delta_{\chi_+}=0$ for all chiral boundary operators with $p=1$ and $\Delta_{\chi_-}=0$ for those with $p=2$ and after some algebra, one obtains the rotation matrices
\begin{subequations}
\label{eq:RotMatchi}
\begin{equation}
\label{eq:RotMatchiPlus-fixedpoints}\mathcal{R}_{\chi_+}=\left(%
\begin{array}{ccc}
  \frac{-1+g_c^2}{3+g_c^2} & \frac{2(1+g_c)}{3+g_c^2} & \frac{2(1-g_c)}{3+g_c^2} \\
  \frac{2(1-g_c)}{3+g_c^2} & \frac{-1+g_c^2}{3+g_c^2} & \frac{2(1+g_c)}{3+g_c^2} \\
  \frac{2(1+g_c)}{3+g_c^2} & \frac{2(1-g_c)}{3+g_c^2} & \frac{-1+g_c^2}{3+g_c^2} \\
\end{array}%
\right) ,
\end{equation}
and
\begin{equation}
\label{eq:RotMatchiMinus-fixedpoints}\mathcal{R}_{\chi_-}=\left(%
\begin{array}{ccc}
  \frac{-1+g_c^2}{3+g_c^2} & \frac{2(1-g_c)}{3+g_c^2} & \frac{2(1+g_c)}{3+g_c^2} \\
  \frac{2(1+g_c)}{3+g_c^2} & \frac{-1+g_c^2}{3+g_c^2} & \frac{2(1-g_c)}{3+g_c^2} \\
  \frac{2(1-g_c)}{3+g_c^2} & \frac{2(1+g_c)}{3+g_c^2} & \frac{-1+g_c^2}{3+g_c^2} \\
\end{array}%
\right),
\end{equation}
\end{subequations}
or in the tensor form
\begin{equation}
\mathcal{R}_{ij}^{\chi^\pm}=\frac{1}{3+g^2}[(-3+g^2)\delta_{i,j}+2(1
\pm g \epsilon_{ij})],
\end{equation}
for $\chi_\pm$ BC respectively. Similarly, the $\mathcal{R}$ matrix for the spin sector will be in the same form as Eq.(\ref{eq:RotMatchi}) with a simple substitution, $g_c \to g_s$. For the case of more than three wires, there may exist more than two ``chiral'' boundary conditions. In this paper, we will restrict ourself in the case of three quantum wires.

Using Eq.(\ref{eq:conductance-master}), the conductance is given by
\begin{eqnarray}
\nonumber G^{c(s)}_{ij,\chi^\pm}&=& 2 g_{c(s)} \frac{e^2}{h}
[\delta_{i,j}-\frac{(g_{c(s)}^2-3)\delta_{i,j}+2(1 \pm g_{c(s)}
\epsilon_{j,i})}{3+g_{c(s)}^2}]
\\
&=& 2 g_{c(s)} \frac{e^2}{h}\frac{2}{3+g_{c(s)}^2}[(3 \delta_{i,j}-1) \pm g_{c(s)}
\epsilon_{ij}].
\end{eqnarray}

\subsubsection{Asymmetric fixed points, $D_A$}

There are boundary conditions that cannot be represented by a simple rotation around the $\vec{u}_0$; instead, they correspond to improper rotations. The only important fixed point that falls into this category corresponds to the asymmetric boundary condition ($D_A$ BC), in which one of the wires is totally decoupled from the system. The matrix representation of this boundary condition can be constructed by imposing the N BC on the third wire and the D BC on the first and second wires. ( Here, one can choose to decouple any of the three wires. Hence, there are three distinct $D_A^{i}$ BC, where $i=1,2,3$ represents the decoupled wire. However, they are identical up to a $Z_3$ transformation.) The matrix representation of $D_A^3$ BC is given by
\begin{equation}
\label{eq:RotationMatrix-fixedpoints}
\mathcal{R}_{D_A^3}=
\left(%
\begin{array}{ccc}
  0 & 1 & 0 \\
  1 & 0 & 0 \\
  0 & 0 & 1 \\
\end{array}%
\right)
\end{equation}
According to the Eq.(\ref{eq:conductance-master}), the conductance tensor of the $D_A^3$ BC becomes
\begin{equation}
G_{D_A^3}^{c(s)}= 2 g_{c(s)} \frac{e^2}{h}
\left(%
\begin{array}{ccc}
  1 & -1 & 0 \\
  -1 & 1 & 0 \\
  0 & 0 & 0 \\
\end{array}%
\right).
\end{equation}

Now, we can impose these boundary conditions separately on charge and spin sectors and determine the stability of each fixed point by computing the scaling dimensions of the boundary operators.

\subsection{The boundary operators}

Now, we have to write the boundary operators in terms of redundant bosonic degrees of freedom. In principle, we need to consider all tunneling operators between three wires at $x=0$. However, since more particle processes tend to be less relevant, we will merely consider single and two particle tunneling processes.  The scaling dimensions of these operators given the BC will be evaluated later and provide measures of relevant or irrelevant perturbations. A more systematic method to compute all scaling dimensions of boundary operators, which relies on conformal symmetry of the system, will be introduced in the next section. From the full spectrum of the dimensions, one will conclude that the leading order perturbations indeed come from the few particle tunneling processes. 

The single electron tunneling processes can be classified into four groups and their Hermitian conjugates:

$\bullet$ $+$ cycle:
\begin{subequations}
\label{eq:+cycle}
\begin{eqnarray}
\label{eq:+cycle-1}S^{+}_{1,\sigma}&=&{\Psi^{R}_{2,\sigma}}^\dagger\Psi^{L}_{1,\sigma}\big|_{0}\propto e^{-i\sqrt{2} \phi_{2,\sigma}^R } e^{i\sqrt{2} \phi_{1,\sigma}^L },
\\
\label{eq:+cycle-2}S^{+}_{2,\sigma}&=&{\Psi^{R}_{3,\sigma}}^\dagger\Psi^{L}_{2,\sigma}\big|_{0}\propto e^{-i\sqrt{2} \phi_{3,\sigma}^R } e^{i\sqrt{2} \phi_{2,\sigma}^L },
\\
\label{eq:+cycle-3}S^{+}_{3,\sigma}&=&{\Psi^{R}_{1,\sigma}}^\dagger\Psi^{L}_{3,\sigma}\big|_{0} \propto e^{-i\sqrt{2} \phi_{1,\sigma}^R } e^{i\sqrt{2} \phi_{3,\sigma}^L }.
\end{eqnarray}
\end{subequations}
These boundary operators can be categorized by the non-zero elements of particle number vectors $n^a_{j,\sigma}$ for $S^+_{i,\sigma}$: $n_{i,\sigma}^L=1$ and $n_{i+1,\sigma}^R=-1$.

$\bullet$ $-$ cycle:
\begin{subequations}
\label{eq:-cycle}
\begin{eqnarray}
\label{eq:-cycle-1}S^{-}_{1,\sigma}&=&{\Psi^{R}_{3,\sigma}}^\dagger\Psi^{L}_{1,\sigma}\big|_{0} \propto e^{-i\sqrt{2} \phi_{3,\sigma}^R } e^{i\sqrt{2} \phi_{1,\sigma}^L },
\\
\label{eq:-cycle-2}S^{-}_{2,\sigma}&=&{\Psi^{R}_{1,\sigma}}^\dagger\Psi^{L}_{2,\sigma}\big|_{0}\propto e^{-i\sqrt{2} \phi_{1,\sigma}^R } e^{i\sqrt{2} \phi_{2,\sigma}^L },
\\
\label{eq:-cycle-3} S^{-}_{3,\sigma}&=&{\Psi^{R}_{2,\sigma}}^\dagger\Psi^{L}_{3,\sigma}\big|_{0}\propto e^{-i\sqrt{2} \phi_{2,\sigma}^R } e^{i\sqrt{2} \phi_{3,\sigma}^L },
\end{eqnarray}
\end{subequations}
and also classified by the non-zero elements of particle number vector for $S^-_{i,\sigma}$: $n_{i,\sigma}^L=1$ and $n_{i-1,\sigma}^R=-1$.

$\bullet$ Backscattering:
\begin{subequations}
\label{eq:Backscattering}
\begin{eqnarray}
\label{eq:Backscattering-1} S^{B}_{1,\sigma}&=&{\Psi^{R}_{1,\sigma}}^\dagger\Psi^{L}_{1,\sigma}\big|_{0} \propto e^{-i\sqrt{2} \phi_{1,\sigma}^R } e^{i\sqrt{2} \phi_{1,\sigma}^L },
\\
\label{eq:Backscattering-2}S^{B}_{2,\sigma}&=&{\Psi^{R}_{2,\sigma}}^\dagger\Psi^{L}_{2,\sigma}\big|_{0}\propto e^{-i\sqrt{2} \phi_{2,\sigma}^R } e^{i\sqrt{2} \phi_{2,\sigma}^L },
\\
\label{eq:Backscattering-3}S^{B}_{3,\sigma}&=&{\Psi^{R}_{3,\sigma}}^\dagger\Psi^{L}_{3,\sigma}\big|_{0}\propto e^{-i\sqrt{2} \phi_{3,\sigma}^R } e^{i\sqrt{2} \phi_{3,\sigma}^L }.
\end{eqnarray}
\end{subequations}
Again, we should identify the representation of the particle number vector for $S^B_{i,\sigma}$: $n_{i,\sigma}^L=1$ and $n_{i,\sigma}^R=-1$.

$\bullet$ $LL-RR$ combinations:
\begin{subequations}
\label{eq:LL-RR}
\begin{eqnarray}
S^{L}_{1,\sigma}&=&{\Psi^{L}_{2,\sigma}}^\dagger\Psi^{L}_{1,\sigma}\big|_{0}\propto e^{-i\sqrt{2} \phi_{2,\sigma}^L } e^{i\sqrt{2} \phi_{1,\sigma}^L }
\\
S^{L}_{2,\sigma}&=&{\Psi^{L}_{3,\sigma}}^\dagger\Psi^{L}_{2,\sigma}\big|_{0}\propto e^{-i\sqrt{2} \phi_{3,\sigma}^L } e^{i\sqrt{2} \phi_{2,\sigma}^L }
\\
S^{L}_{3,\sigma}&=&{\Psi^{L}_{1,\sigma}}^\dagger\Psi^{L}_{3,\sigma}\big|_{0}\propto e^{-i\sqrt{2} \phi_{1,\sigma}^L } e^{i\sqrt{2} \phi_{3,\sigma}^L }
\\
S^{R}_{1,\sigma}&=&{\Psi^{R}_{2,\sigma}}^\dagger\Psi^{R}_{1,\sigma}\big|_{0}\propto e^{-i\sqrt{2} \phi_{2,\sigma}^R } e^{i\sqrt{2} \phi_{1,\sigma}^R }
\\
S^{R}_{2,\sigma}&=&{\Psi^{R}_{3,\sigma}}^\dagger\Psi^{R}_{2,\sigma}\big|_{0}\propto e^{-i\sqrt{2} \phi_{3,\sigma}^R } e^{i\sqrt{2} \phi_{2,\sigma}^R }
\\
S^{R}_{3,\sigma}&=&{\Psi^{R}_{1,\sigma}}^\dagger\Psi^{R}_{3,\sigma}\big|_{0}\propto e^{-i\sqrt{2} \phi_{1,\sigma}^R } e^{i\sqrt{2} \phi_{3,\sigma}^R },
\end{eqnarray}
\end{subequations}
Then, we can identify the representation of the particle number vector for $S^{L(R)}_{i,\sigma}$: $n_{i,\sigma}^{L(R)}=1$ and $n_{i+1,\sigma}^{L(R)}=-1$.

Here, $S$ indicates ``single'' particle processes. Notice that there can be no spin flips in single particle processes. However, the scaling dimensions of multi-particle tunneling processes depend on the spin degree of freedom. The multi-particle operators can be constructed from the combinations of single particle boundary operators and their hermitian conjugates. Here, we list some of the two particle processes that will be useful for identifying the leading order perturbations.

$\bullet$ Pair Tunneling in $+$ cycle:
\begin{subequations}
\label{eq:+Pair-tunneling}
\begin{eqnarray}
\label{eq:+Pair-tunneling-1}PT^+_{1}&=&S^{+}_{1,\uparrow}S^{+}_{1,\downarrow},
\\
\label{eq:+Pair-tunneling-2}PT^+_{2}&=&S^{+}_{2,\uparrow}S^{+}_{2,\downarrow},
\\
\label{eq:+Pair-tunneling-3}PT^+_{3}&=&S^{+}_{3,\uparrow}S^{+}_{3,\downarrow}.
\end{eqnarray}
\end{subequations}
Then, we can identify the representation of the particle number vector for $PT^{+}_{i}$: $n_{i,\uparrow}^{L}=n_{i,\downarrow}^{L}=1$ and $n_{i+1,\uparrow}^{R}=n_{i+1,\downarrow}^{R}=-1$.

$\bullet$ Pair Tunneling in $-$ cycle:
\begin{subequations}
\label{eq:-Pair-tunneling}
\begin{eqnarray}
\label{eq:-Pair-tunneling-1}PT^-_{1}&=&S^{-}_{1,\uparrow}S^{-}_{1,\downarrow}
\\
\label{eq:-Pair-tunneling-2}PT^-_{2}&=&S^{-}_{2,\uparrow}S^{-}_{2,\downarrow}
\\
\label{eq:-Pair-tunneling-3}PT^-_{3}&=&S^{-}_{3,\uparrow}S^{-}_{3,\downarrow}
\end{eqnarray}
\end{subequations}
In terms of the particle number vector, we obtain for $PT^{-}_{i}$: $n_{i,\uparrow}^{L}=n_{i,\downarrow}^{L}=1$ and $n_{i-1,\uparrow}^{R}=n_{i-1,\downarrow}^{R}=-1$.

$\bullet$ Pair Tunneling in LL-RR combinations with net spin:
\begin{subequations}
\label{eq:Pair-tunneling-LR-net-spin}
\begin{eqnarray}
\label{eq:Pair-tunneling-LR-net-spin-1}PTS^{LR}_{1,\sigma}&=&S^{L}_{1,\sigma}S^{R}_{1,\sigma}
\\
\label{eq:Pair-tunneling-LR-net-spin-2}PTS^{LR}_{2,\sigma}&=&S^{L}_{2,\sigma}S^{R}_{2,\sigma}
\\
\label{eq:Pair-tunneling-LR-net-spin-3}PTS^{LR}_{3,\sigma}&=&S^{L}_{3,\sigma}S^{R}_{3,\sigma}
\end{eqnarray}
\end{subequations}
In terms of the particle number vector, we obtain for $PTS^{LR}_{i,\sigma}$: $n_{i,\sigma}^{L}=n_{i,\sigma}^{R}=1$ and $n_{i+1, \sigma}^{L}=n_{i+1, \sigma}^{R}=-1$.

$\bullet$ Pair Tunneling in LL-RR combinations without net spin:
\begin{subequations}
\label{eq:Pair-tunneling-LR-no-net-spin}
\begin{eqnarray}
\label{eq:Pair-tunneling-LR-no-net-spin-1}PT^{LR}_{1,\sigma}&=&S^{L}_{1,\sigma}S^{R}_{1,-\sigma},
\\
\label{eq:Pair-tunneling-LR-no-net-spin-2}PT^{LR}_{2,\sigma}&=&S^{L}_{2,\sigma}S^{R}_{2,-\sigma},
\\
\label{eq:Pair-tunneling-LR-no-net-spin-3}PT^{LR}_{3,\sigma}&=&S^{L}_{3,\sigma}S^{R}_{3,-\sigma}.
\end{eqnarray}
\end{subequations}
One can read the the particle number vector off for $PT^{LR}_{i,\sigma}$: $n_{i,\sigma}^{L}=n_{i,-\sigma}^{R}=1$ and $n_{i+1, \sigma}^{L}=n_{i+1, -\sigma}^{R}=-1$.

$\bullet$ Pair Backscattering in the same wire:
\begin{subequations}
\label{eq:Pair-bachscattering-same-wire}
\begin{eqnarray}
\label{eq:Pair-bachscattering-same-wire-1}PB_{1}&=&S^{B}_{1,\uparrow}S^{B}_{1,\downarrow}
\\
\label{eq:Pair-bachscattering-same-wire-2}PB_{2}&=&S^{B}_{2,\uparrow}S^{B}_{2,\downarrow}
\\
\label{eq:Pair-bachscattering-same-wire-3}PB_{3}&=&S^{B}_{3,\uparrow}S^{B}_{3,\downarrow}
\end{eqnarray}
\end{subequations}
The representations of non-zero elements of particle number vector become for $PB_{i}$: $n_{i,\uparrow }^{L}=n_{i,\downarrow}^{L}=1$ and $n_{i, \uparrow}^{R}=n_{i, \downarrow}^{R}=-1$.

$\bullet$ Pair Backscattering in the different wires without net spin:
\begin{subequations}
\label{eq:Pair-bachscattering-diff-wire-no-spin}
\begin{eqnarray}
\label{eq:Pair-bachscattering-diff-wire-no-spin-1}PB_{12,\sigma}&=&S^{B}_{1,\sigma}S^{B}_{2,-\sigma}
\\
\label{eq:Pair-bachscattering-diff-wire-no-spin-2}PB_{23,\sigma}&=&S^{B}_{2,\sigma}S^{B}_{3,-\sigma}
\\
\label{eq:Pair-bachscattering-diff-wire-no-spin-3}PB_{31,\sigma}&=&S^{B}_{3,\sigma}S^{B}_{1,-\sigma}
\end{eqnarray}
\end{subequations}
The non-zero elements of the particle number vector are given for $PB_{ij,\sigma}$: $n_{i,\sigma}^{L}=n_{j,-\sigma}^{L}=1$ and $n_{i,\sigma}^{R}=n_{j,-\sigma}^{R}=-1$.

$\bullet$ Pair Backscattering in the different wires with net spin:
\begin{subequations}
\label{eq:Pair-bachscattering-diff-wire-spin}
\begin{eqnarray}
\label{eq:Pair-bachscattering-diff-wire-spin-1}PBS_{12,\sigma}&=&S^{B}_{1,\sigma}S^{B}_{2,\sigma}
\\
\label{eq:Pair-bachscattering-diff-wire-spin-2}PBS_{23,\sigma}&=&S^{B}_{2,\sigma}S^{B}_{3,\sigma}
\\
\label{eq:Pair-bachscattering-diff-wire-spin-3}PBS_{31,\sigma}&=&S^{B}_{3,\sigma}S^{B}_{1,\sigma}
\end{eqnarray}
\end{subequations}
The non-zero elements of the particle number vector are given for $PBS_{ij,\sigma}$: $n_{i,\sigma}^{L}=n_{j,\sigma}^{L}=1$ and $n_{i,\sigma}^{R}=n_{j,\sigma}^{R}=-1$.

$\bullet$ Pair Exchange processes:
\begin{subequations}
\label{eq:Pair-exchange}
\begin{eqnarray}
\label{eq:Pair-exchange-1}PE_{1,\sigma}&=&S^{+}_{1,\sigma}S^{-}_{2,-\sigma}
\\
\label{eq:Pair-exchange-2}PE_{2,\sigma}&=&S^{+}_{2,\sigma}S^{-}_{3,-\sigma}
\\
\label{eq:Pair-exchange-3}PE_{3,\sigma}&=&S^{+}_{3,\sigma}S^{-}_{1,-\sigma}
\end{eqnarray}
\end{subequations}
The non-zero elements of the particle number vector are given for $PE_{i,\sigma}$: $n_{i,\sigma}^{L}=n_{i+1,-\sigma}^{L}=1$ and $n_{i+1,\sigma}^{R}=n_{i,-\sigma}^{R}=-1$.

$\bullet$ Particle Hole Pair Tunneling in $+$ cycle:
\begin{subequations}
\label{eq:particle-hole-pair+}
\begin{eqnarray}
\label{eq:particle-hole-pair+1}PH^+_{1}&=&S^{+}_{1,\uparrow}{S^{+}_{1,\downarrow}}^\dagger
\\
\label{eq:particle-hole-pair+2}PH^+_{2}&=&S^{+}_{2,\uparrow}{S^{+}_{2,\downarrow}}^\dagger
\\
\label{eq:particle-hole-pair+3}PH^+_{3}&=&S^{+}_{3,\uparrow}{S^{+}_{3,\downarrow}}^\dagger
\end{eqnarray}
\end{subequations}
The non-zero elements of the particle number vector are given for $PH^+_{i}$: $n_{i,\uparrow}^{L}=n_{i+1,\downarrow}^{R}=1$ and $n_{i,\downarrow}^{L}=n_{i+1,\uparrow}^{R}=-1$.

$\bullet$ Particle Hole Pair Tunneling in $-$ cycle:
\begin{subequations}
\label{eq:particle-hole-pair-}
\begin{eqnarray}
\label{eq:particle-hole-pair-1}PH^{-}_{1}&=&S^{-}_{1,\uparrow}{S^{-}_{1,\downarrow}}^\dagger
\\
\label{eq:particle-hole-pair-2}PH^{-}_{2}&=&S^{-}_{2,\uparrow}{S^{-}_{2,\downarrow}}^\dagger
\\
\label{eq:particle-hole-pair-3}PH^{-}_{3}&=&S^{-}_{3,\uparrow}{S^{-}_{3,\downarrow}}^\dagger
\end{eqnarray}
\end{subequations}
The non-zero elements of the particle number vector are given for $PH^-_{i}$: $n_{i,\uparrow}^{L}=n_{i-1,\downarrow}^{R}=1$ and $n_{i,\downarrow}^{L}=n_{i-1,\uparrow}^{R}=-1$.

$\bullet$ Particle Hole Exchange processes:
\begin{subequations}
\label{eq:particle-hole-exchange}
\begin{eqnarray}
\label{eq:particle-hole-exchange-1}PHE_{1,\sigma}&=&S^{+}_{1,\sigma}{S^{-}_{2,-\sigma}}^\dagger
\\
\label{eq:particle-hole-exchange-2}PHE_{2,\sigma}&=&S^{+}_{2,\sigma}{S^{-}_{3,-\sigma}}^\dagger
\\
\label{eq:particle-hole-exchange-3}PHE_{3,\sigma}&=&S^{+}_{3,\sigma}{S^{-}_{1,-\sigma}}^\dagger
\end{eqnarray}
\end{subequations}
Here, the non-zero elements of the particle number vector are given for $PHE_{i,\sigma}$: $n_{i,\sigma}^{L}=n_{i,-\sigma}^{R}=1$ and $n_{i,\sigma}^{R}=n_{i+1,-\sigma}^{L}=-1$.

Using the particle number vectors of each boundary operator, the dimensions of each operator given the BC can be evaluated using Eq.(\ref{eq:ScaDim-DEBC}). Hence, the stability of each BC can be determined by considering the relevant or irrelevant perturbations induced by the boundary operators. 

\subsection{Stability of fixed points}

This section is organized in terms of the different boundary conditions. The boundary conditions are imposed separately on the charge and spin degrees of freedom. We will evaluate the scaling dimensions of the boundary operators corresponding to the leading order perturbations.

\subsubsection{NN BC}

The NN BC corresponds to the totally reflective fixed point in both charge and spin degrees of freedom. Hence, all backscattering processes, Eqs.(\ref{eq:Backscattering}, \ref{eq:Pair-bachscattering-same-wire}, \ref{eq:Pair-bachscattering-diff-wire-no-spin}, \ref{eq:Pair-bachscattering-diff-wire-spin}) and their hermitian conjugates will have zero scaling dimensions. Now we identify the boundary operators associated with the leading order perturbations. 

All other single particle processes contribute to the leading order perturbations with scaling dimension
\begin{equation}
\Delta^{S}_{NN}=\frac{1}{2g_c}+\frac{1}{2g_s}.
\end{equation}
The pair tunneling operators in the $+$ and $-$ cycle, Eqs.(\ref{eq:+Pair-tunneling}) and (\ref{eq:-Pair-tunneling}), the pair tunneling in LL-RR operators without net spin degree of freedom in Eq.(\ref{eq:Pair-tunneling-LR-no-net-spin}) and the particle-hole exchange operators in Eq.(\ref{eq:particle-hole-exchange}) are also leading order perturbations with the scaling dimension,
\begin{equation}
\Delta_{NN}^{PT^\pm}=\Delta_{NN}^{PT^{LR}} = \Delta_{NN}^{PHE}=\frac{2}{g_c} .
\end{equation}
Finally, the pair exchange operators in Eq.(\ref{eq:Pair-exchange}) have scaling dimension
\begin{equation}
\Delta_{NN}^{PE}=\frac{2}{g_s},
\end{equation}
and are therefore leading order perturbations.

Now, we can identify the stable region when all leading order perturbations are irrelevant, $\Delta>1$, and obtain the basin of attraction shown in Fig.~\ref{fig:phase-NN}.

\subsubsection{DD BC}

The DD BC corresponds to the Andreev reflection fixed point in which all pair tunneling LL-RR operators with net spin, Eq.(\ref{eq:Pair-tunneling-LR-net-spin}), have zero scaling dimension. Note that the dominant processes are multi-particle operators.

First, the single particle processes in the $+$ and $-$ cycles in Eqs.(\ref{eq:+cycle}) and (\ref{eq:-cycle}) are leading order operators and have the scaling dimension
\begin{equation}
\Delta_{DD}^{S_\pm} = \frac{1}{6}(g_c+g_s).
\end{equation}
In addition, the pair tunneling operators in $\pm$ cycles in Eqs.(\ref{eq:+Pair-tunneling}, \ref{eq:-Pair-tunneling}), and the pair exchange operators in Eq.(\ref{eq:Pair-exchange}) are also leading order perturbations with scaling dimension
\begin{equation}
\Delta_{DD}^{PT_\pm} = \frac{2}{3}g_c\; .
\end{equation}
Moreover, the particle-hole pair tunneling operators in Eqs.(\ref{eq:particle-hole-pair+}, \ref{eq:particle-hole-pair-}) and the particle hole exchange operators in Eq.(\ref{eq:particle-hole-exchange}) have scaling dimension
\begin{equation}
\Delta_{DD}^{PHE_i} =\Delta_{DD}^{PH^\pm_i}= \frac{2}{3}g_s\; .
\end{equation}

Thus, the basin of the attraction corresponding to the DD fixed point can be obtained by requiring $\Delta>1$ and is shown in Fig.~\ref{fig:phase-DD}. Unlike the case in junction of two quantum wires, the stable region of NN BC and DD BC do not overlap.

\subsubsection{ND BC}

The ND BC corresponds to a fixed point which describes a charge insulator but spin conductor. The pair exchange operators in Eq.(\ref{eq:Pair-exchange}) fix the boundary condition with $\Delta_{ND}^{PE_{i,\sigma}}=0$. Intuitively, the pair exchange processes induce a pure spin current depicted in the inset of Fig.~\ref{fig:phase-ND-DN}. Furthermore, the processes of pair backscattering in the same wire Eq.(\ref{eq:Pair-bachscattering-same-wire}) are pinned as well and indicate that there is no charge current.

The leading order perturbations are attributed to the operators for the single particle tunneling in the $\pm$ cycles, Eqs.(\ref{eq:+cycle}, \ref{eq:-cycle}), with the scaling dimension
\begin{equation}
\Delta^{S^\pm}_{ND}= \frac{1}{2 g_c}+\frac{g_s}{6}\; .
\end{equation}
The single particle backscattering processes in Eq.(\ref{eq:Backscattering}), the pair backscattering processes in Eq.(\ref{eq:Pair-bachscattering-diff-wire-spin}) and the particle hole pair tunneling in Eqs.(\ref{eq:particle-hole-pair+}, \ref{eq:particle-hole-pair-}) are also leading order perturbations with
\begin{equation}
\Delta^{S^B}_{ND}=\Delta^{PBS}_{ND}= \frac{2g_s}{3}\; .
\end{equation}
Moreover, the pair tunneling operators in Eqs.(\ref{eq:+Pair-tunneling}, \ref{eq:-Pair-tunneling}, \ref{eq:Pair-tunneling-LR-net-spin}) are also leading order perturbations with scaling dimension
\begin{equation}
\Delta_{ND}^{PT^\pm}=\Delta_{ND}^{PTS^{LR}}= \frac{2}{g_c}
\end{equation}

Again, we can identify the basin of attraction for the ND fixed point by requiring all $\Delta>1$ as shown in Fig.~\ref{fig:phase-ND-DN}.

\subsubsection{DN BC}

The DN BC is the counterpart of the ND BC and corresponds to a fixed point where the system becomes a charge conductor but spin insulator. Here, the particle hole exchange operators in Eq.(\ref{eq:particle-hole-exchange}) fix the boundary condition with $\Delta_{DN}^{PHE_{i,\sigma}}=0$. As illustrated in the inset of Fig.~\ref{fig:phase-ND-DN}, there is no spin current in these tunneling processes. Moreover, the pair tunneling operators in LL-RR channel without net spin Eq.(\ref{eq:Pair-tunneling-LR-no-net-spin}) are also pinned at DN fixed point since they also represent the processes with only net charge current.

The leading order perturbations are attributed to the single particle processes in the $\pm$ cycles in Eqs.(\ref{eq:+cycle}, \ref{eq:-cycle}) with scaling dimension
\begin{equation}
\Delta^{S^\pm}_{DN}= \frac{g_c}{6}+\frac{1}{2g_s} .
\end{equation}
The backscattering in Eq.(\ref{eq:Backscattering}), the pair tunneling operators both in $\pm$ cycles in Eqs.(\ref{eq:+Pair-tunneling}, \ref{eq:-Pair-tunneling}) and the pair backscattering processes involving different wires in Eqs.(\ref{eq:Pair-bachscattering-diff-wire-no-spin}, \ref{eq:Pair-bachscattering-diff-wire-spin}) are leading order perturbations with
\begin{equation}
\Delta^{S^B}_{DN}=\Delta^{PT^\pm}_{DN}=\Delta^{PB_{ij}}_{DN}= \Delta^{PBS_{ij}}_{DN} = \frac{2g_c}{3} .
\end{equation}
Finally, the pair tunneling operators in LL-RR channel with the net spin in Eq.(\ref{eq:Pair-tunneling-LR-net-spin}) and the particle-hole pair tunneling in both $\pm$ cycles in Eqs.(\ref{eq:particle-hole-pair+}, \ref{eq:particle-hole-pair-}) represent the other set of leading order perturbations with scaling dimension
\begin{equation}
\Delta^{PTS^{LR}}_{DN}=\Delta^{PH^{\pm}}_{DN}= \frac{2}{g_s}
\end{equation}

Again, the basin of attraction can be found for the DN BC by requiring all $\Delta>1$ and is shown in Fig.~\ref{fig:phase-ND-DN}.

\subsubsection{$\chi^{ }_+\chi^{ }_+$ and $\chi^{ }_-\chi^{ }_-$ BC}

The $\chi^{ }_+\chi^{ }_+$ BC describes a fixed point in which both charge and spin current have a preferred flow direction $1 \to 2$, $2 \to 3$ and $3 \to 1$. Thus, at the $\chi^{ }_+\chi^{ }_+$ fixed point, the processes involving current flows in these particular directions will fix the boundary condition. For instance, the pure charge current processes in the $+$ cycle, Eq.(\ref{eq:+cycle}) and Eq.(\ref{eq:+Pair-tunneling}), and the pure spin current processes in the $+$ cycle, Eq.(\ref{eq:particle-hole-pair+}), have zero scaling dimension. 

Notice that the single particle tunneling operators in the $+$ cycle have zero scaling dimension regardless of the spin degree of freedom. However, boundary  operators representing pure spin current, constructed from the particle hole pair tunneling, have zero scaling dimension only when the spin current is in the $+$ cycle, i.e. $\Delta^{PH^-}\neq 0$. Similarly, operators corresponding to the pure charge current, constructed from pair tunneling processes, have zero scaling dimension only when the charge current flows in the $+$ cycle, i.e. $\Delta^{PT^-}\neq 0$.

One can confirm that all single particle processes except the $+$ cycle and some multi-particle processes provide leading order perturbations with scaling dimension
\begin{equation}
\Delta_{\chi^{ }_+\chi^{ }_+}=\frac{2 g_c}{3+ g_c^2}+\frac{2 g_s}{3+ g_s^2} .
\end{equation}

The scenario of $\chi^{ }_-\chi^{ }_-$ BC is very similar to the case of $\chi^{ }_+\chi^{ }_+$ with relative changes from the $+$ cycle to the $-$ cycle. The leading order perturbations have the same scaling dimension as that of the $\chi_+\chi_+$ BC. Thus, both fixed points have exactly the same basin of attraction in the coupling constant space. The stability of the two fixed points is determined by the direction of the magnetic flux threaded through the ring. We then plot the stable region of the fixed points in Fig.~\ref{fig:phase-chiral}.

\subsubsection{$D_A D_A$ B.C}

Without lose of generality, we will only consider in this section the boundary condition where the third wire is decoupled, $D_A^3$, both in charge and spin degrees of freedom. The situations with the first or second wire decoupled are similar.

In the presence of the $Z_3$ symmetry, the operators in the same group listed in the previous section have the same scaling dimension. However, the $D_AD_A$ fixed point breaks the $Z_3$ symmetry, hence the operators in the same group may have different dimensions. For instance, the operators in Eq.(\ref{eq:+cycle-1}, \ref{eq:-cycle-1}), have zero scaling dimension and fix the BC, while the rest of the operators in the $\pm$ cycle are leading order perturbations with the scaling dimension
\begin{equation}
\Delta=\frac{3+g_c^2}{8g_c}+\frac{3+g_s^2}{8g_s}.
\end{equation}
In addition, the single particle backscattering in the third wire in
Eq.(\ref{eq:Backscattering-3}) has zero scaling dimension at the $D_A^3D_A^3$ fixed point while the backscattering at the first or second wire becomes another leading order perturbation
\begin{equation}
\Delta=\frac{1}{2}(g_c+g_s).
\end{equation}
Some of the operators associated with the leading order perturbations are listed below:
\begin{equation}
\begin{tabular}{|c|c|}
  \hline
  Operators & Dimensions \\
  \hline
  $PTS^{LR}_{2(3),\sigma}$ & $\frac{3}{2}(g_c^{-1}+g_s^{-1})$ \\
  \hline
  $PT^{\pm}_{2(3)}$, $PT^{LR}_{2(3),\sigma}$, $PHE_{2(3),\sigma}$ & $\frac{3}{2g_c}+\frac{g_s}{2}$ \\
  \hline
  $PT^{LR}_{1,\sigma}$, $PB_{12,\sigma}$ & $2 g_s$ \\
  \hline
  $PB_{1(2)}$ & $2 g_c$ \\
  \hline
  $PE_{2(3),\sigma}$ & $\frac{g_c}{2}+\frac{3}{2g_s}$ \\
  \hline
\end{tabular}
\end{equation}
Thus, we identify the basin of attraction for $D_AD_A$ BC and plot it in Fig.~\ref{fig:phase-AA}.

\subsubsection{$ND_A$ and $D_A N$ BC- A demonstration of the unstable fixed points}

In principle one can arbitrarily combine different BC in charge and spin sector to obtain the new boundary conditions. However, the rest of them are unstable against perturbations. We discuss here two unstable fixed points, $ND_A$ and $D_A N$ BC, and show explicitly that there are always leading order perturbations with scaling dimensions smaller than one in any region of the interaction parameter space. Considering the case where the third wire is decoupled from the ring, the backscattering in the third wire fix the BC in both $ND_A^3$ and $D_A^3 N$ fixed points. The operators $PH^{+}_1$ and $PH^{-}_{2}$, corresponding to the processes with pure spin current between the first and second wire, have zero scaling dimensions at the $N D_A^3$ BC while the operators $PT^{+}_{1}$ and $PT^{-}_{2}$, corresponding to the processes with pure charge current, have zero dimension at the $D_A^3 N$ BC. Hence, we conclude that $N D_A^3$ BC corresponds to a fixed point with pure spin current between first and second wire, while $D_A^3 N$ BC corresponds to a fixed point with pure charge current between them.

We list below some operators which are crucial for determining the stability of the fixed points with their scaling dimensions,
\begin{eqnarray}
\Delta_{N D_A^3}^{S^{B}_{1(2),\sigma} }&=&\frac{g_s}{2}, \qquad \Delta_{N D_A^3}^{PE_{2(3),\sigma} }=\frac{3}{2 g_s}
\\
\Delta_{D_A^3 N}^{S^{B}_{1(2),\sigma} }&=&\frac{g_c}{2}, \qquad  \Delta_{D_A^3 N}^{ PT^{LR}_{2(3),\sigma} }=\frac{3}{2 g_c}.
\end{eqnarray}
Observe that $\Delta_{N D_A^3}^{S^{B}_{1(2),\sigma} }$ and $\Delta_{N D_A^3}^{PE_{2(3),\sigma} }$ cannot be larger than one at the same point in the parameter space; $\Delta_{D_A^3 N}^{S^{B}_{1(2),\sigma} }$ and $\Delta_{D_A^3 N}^{ PT^{LR}_{2(3),\sigma} }$ cannot either. Hence, we can concludes that $N D_A$ and $D_A N$ fixed points are unstable against the perturbations.

\subsection{Summary of DEBC}

We have demonstrated in this section how to implement the DEBC method for obtaining the scaling dimensions of the boundary operators and determining the stability of different fixed points. We find that the DEBC method provides a simple way to examine the junction systems and to determine the phase diagram. However, the main drawback of this method is that  one has to determine the scaling dimensions of ``all'' boundary operators in each given BC. As mentioned in this section, we have to rely on the conjecture that operators involving more particles are less relevant to simplify the computation. Hence, in the next two section, we shall provide the confirmation of our results through the approach of boundary conformal field theory where the full spectrum of the scaling dimensions of the boundary operators can be identified explicitly.


\section{Review of BCFT}
\label{sec:BCFT-Review}

The application of boundary conformal field theory (BCFT) to the analysis of critical phenomena was developed by Cardy and widely applied to the Kondo problem and one dimensional problems~\cite{Cardy, Cardy-review, Ian-Ludwig}. In the BCFT, comformally invariant boundary conditions are formulated in terms of boundary states. Here, the boundary states belong to the Hilbert space of the theory with the periodic boundary condition in the space direction. Considering the Fermi statistics of the electrons, the structure of the Hilbert space is twisted compared to that of the standard free boson. The twisted structure affects the possible boundary states, the scaling dimensions of the boundary operators and the stability of the fixed points.

In this section, we will first review the mode expansion of bosonic fields $\varphi_{c(s)}$ and $\theta_{c(s)}$ and derive the twisted structure of a TLL quantum wire for spin-1/2 electrons. A simplified version of the BCFT, exclusively for non-resonante tunneling~\cite{Nayak}, will be implemented by projecting out the degree of freedom corresponding to the total charge and spin. As a demonstration, we compute the partition function and obtain the scaling dimensions of the boundary operators corresponding to a boundary state for the case of a junction of two quantum wires.

\subsection{Review of mode expansions of boson fields and  derivation of twisted structure}

The BCFT of free bosons can be applied to the system of interacting electrons via bosonization. However, due to the Fermi statistics of the electrons, there are various differences compared to the standard compactified boson field theory. In this section, the effect of the Fermi statistics will be implemented in terms of a twisted structure in the Hilbert space of the free boson theory.

Since the Matsubara formalism for fermions implies {\it anti-periodic} boundary condition in the imaginary time, the space direction will also have anti-periodic boundary condition after the modular transformation. The boundary condition should be imposed independently on left and right movers,
\begin{equation}
\psi_{R(L), \sigma}(t,x)=-\psi_{R(L), \sigma}(t,x+\beta)\; .
\end{equation}
In the case of free fermions, $g=1$, the boundary condition becomes
\begin{equation}
\label{eq:B-C-fermion-boson} \exp [i \sqrt{2} \phi_{R(L),\sigma}(t,\beta)]= - \exp [i \sqrt{2} \phi_{R(L),\sigma}(t,0)]\; .
\end{equation}
Since $[\phi_{R(L)}, \phi_{R(L)}]\neq0$, it is not clear how to determine the boundary condition on the boson field from Eq.(\ref{eq:B-C-fermion-boson}). We refer to Ref.~\cite{COA} for the proof that the anti-periodic BC on fermions leads to the periodic BC on bosons. Precisely, the right and left moving boson field, $\phi_{R(L)}$ are periodic variables
\begin{equation}
\phi_{R(L),\sigma}(t,x+\beta)\sim \phi_{R(L),\sigma}(t,x)+\sqrt{2} \pi n,
\end{equation}
where $ (n \in \mathbf{Z})$. The mode expansion of the boson fields compactified on a circle of radius $1/\sqrt{2}$ becomes
\begin{widetext}
\begin{subequations}
\begin{eqnarray}
\phi_{R,\sigma}(x_-)&=&\hat{\varphi}_{0,\sigma}^R+ \frac{\sqrt{2}\pi}{\beta} \hat{Q}^R_{\sigma}x_{-}+\frac{1}{\sqrt{2}} \sum_{n=1}^{\infty} \;  \frac{1}{\sqrt{n}}[a_{n,\sigma}^R \; e^{-i\frac{2\pi n}{\beta}  x_{-}}+h.c]
\\
\phi_{L,\sigma}(x_+)&=&\hat{\varphi}_{0,\sigma}^L+ \frac{\sqrt{2}\pi}{\beta} \hat{Q}^L_{\sigma}x_{+} +\frac{1}{\sqrt{2}} \sum_{n=1}^{\infty} \; \frac{1}{\sqrt{n}}[a_{n,\sigma}^L \; e^{-i\frac{2\pi n}{\beta} x_{+}}+h.c] \; ,
\end{eqnarray}
\end{subequations}
where $\sqrt{2} \hat{Q}^{R(L)}_{\sigma}$, the momentum variable conjugating to $\hat{\varphi}_{0,\sigma}^{R(L)}$, have eigenvalues $\sqrt{2}$ times an integer, $a_{n,\sigma}^{R/L}$ are oscillator modes, and $x_{\pm}\equiv t\pm x$.

Following the standard bosonization scheme, the independent charge and spin modes are defined as
\begin{subequations}
\begin{eqnarray}
\phi_{R(L),c}&=&\frac{1}{\sqrt{2}}(\phi_{R(L),\uparrow}+\phi_{R(L),\downarrow})
\\
\phi_{R(L),s}&=&\frac{1}{\sqrt{2}}(\phi_{R(L),\uparrow}-\phi_{R(L),\downarrow})\; .
\end{eqnarray}
\end{subequations}
Hence, the mode expansions of the boson fields corresponding to the charge and spin degrees of freedom become,
\begin{subequations}
\begin{eqnarray}
\label{eq:ModeExpansion1}
\phi_{R,c(s)}&=&\hat{\varphi}^R_{0,c(s)}+\frac{\pi}{\beta}\hat{Q}^R_{c(s)}t_{-}+\frac{1}{\sqrt{2}}\sum_{n=1}^\infty \frac{1}{\sqrt{n}}[a^R_{n,c(s)} e^{-i\frac{2\pi n}{\beta}x_-}+h.c.]
\\
\label{eq:ModeExpansion2}
\phi_{L,c(s)}&=&\hat{\varphi}^L_{0,c(s)}+\frac{\pi}{\beta}\hat{Q}^L_{c
(s)}t_{+}+\frac{1}{\sqrt{2}}\sum_{n=1}^\infty \frac{1}{\sqrt{n}}[a^L_{n,c(s)} e^{-i\frac{2\pi n}{\beta}x_+}+h.c.]\; ,
\end{eqnarray}
with the definitions
\begin{eqnarray}
\hat{\varphi}^{R(L)}_{0,c(s)}=\frac{1}{\sqrt{2}}(\hat{\varphi}^{R(L)}_{0,\uparrow} \pm \hat{\varphi}^{R(L)}_{0,\downarrow}), \qquad a^{R(L)}_{n,c(s)}=\frac{1}{\sqrt{2}}(a^{R(L)}_{n,\uparrow} \pm a^{R(L)}_{n,\downarrow}), \;  { \rm and} \qquad \hat{Q}^{R(L)}_{c(s)}=\hat{Q}^{R(L)}_{\uparrow} \pm \hat{Q}^{R(L)}_{\downarrow}  .
\end{eqnarray}
\end{subequations}
\end{widetext}
Notice that the eigenvalues of the operators $\hat{Q}^{R(L)}_{c(S)}$ follow a special relation
\begin{equation}
\label{eq:GluingCondition1} Q^{R(L)}_c=Q^{R(L)}_s  \;{\rm (mod\; 2)},
\end{equation}
named {\it gluing conditions}. 

Now, let us define the new boson fields in the new basis,
\begin{equation}
\varphi_{c(s)}=\phi_{R,c(s)}+\phi_{L,c(s)}, \;  \theta_{c(s)}=\phi_{L,c(s)}-\phi_{R,c(s)}.
\end{equation}
Following Eq.(\ref{eq:ModeExpansion1}) and Eq.(\ref{eq:ModeExpansion2}), the mode expansion of the new felds is given by
\begin{widetext}
\begin{subequations}
\begin{eqnarray}
\label{eq:Mode-Exp-varphi}\varphi_{c(s)}&=&\hat{\varphi}_{0,c(s)}+\frac{\pi}{\beta}[\hat{Q}_{c(s)} t+\hat{\tilde{Q}}_{c(s)} x]+\frac{1}{\sqrt{2}} \sum_{n=1}^{\infty} \frac{1}{\sqrt{n}}[a^R_{n,c(s)} e^{-i\frac{2 \pi n}{\beta} x_{-}} + a^L_{n,c(s)}e^{-i\frac{2 \pi n}{\beta} x_{+}}+ h.c.]
\\
\label{eq:Mode-Exp-theta}\theta_{c(s)}&=&\hat{\theta}_{0,c(s)}+\frac{\pi}{\beta}[\hat{\tilde{Q}}_{c(s)} t+\hat{Q}_{c(s)} x]+\frac{1}{\sqrt{2}} \sum_{n=1}^{\infty} \frac{1}{\sqrt{n}}[a^L_{n,c(s)} e^{-i\frac{2 \pi n}{\beta} x_{+}} -a^R_{n,c(s)}e^{-i\frac{2 \pi n}{\beta} x_{-}}+ h.c.] \;\; ,
\end{eqnarray}
where
\begin{eqnarray}
\hat{\varphi}_{0,c(s)}=\hat{\varphi}^L_{0,c(s)}+\hat{\varphi}^R_{0,c(s)}, \;\;  \hat{\theta}_{0,c(s)}=\hat{\varphi}^L_{0,c(s)}-\hat{\varphi}^R_{0,c(s)}, \; \;  \hat{Q}_{c(s)}=\hat{Q}^L_{c(s)}+\hat{Q}^R_{c(s)}, \;\; \hat{\tilde{Q}}_{c(s)}=\hat{Q}^L_{c(s)}-\hat{Q}^R_{c(s)} .
\end{eqnarray}
\end{subequations}
\end{widetext}
A new gluing condition can be derived from Eq.(\ref{eq:GluingCondition1})
\begin{subequations}
\label{eq:GluingCondition2}
\begin{eqnarray}
\label{eq:GluingCondition2-1} Q_c&=&Q_s=\tilde{Q}_c=\tilde{Q}_s \;  {\rm (mod 2)},
\\
\label{eq:GluingCondition2-2}Q_c&+&Q_s+\tilde{Q}_c+\tilde{Q}_s=0 \; {\rm(mod 4)}.
\end{eqnarray}
\end{subequations}
Hence, the periodic boundary condition of the fields $\varphi_{c(s)}$ and $\theta_{c(s)}$ are of the form 
\begin{eqnarray}
\label{eq:periodic-identi}
\varphi_{c(s)} \sim \varphi_{c(s)} + \pi \tilde{Q}_{c(s)}, \; 
\theta_{c(s)} \sim \theta_{c(s)} + \pi Q_{c(s)}\; ,
\end{eqnarray}
accompanied by the gluing condition in Eq.(\ref{eq:GluingCondition2}). We refer to this as a ``twisted structure", which reflects the fact that the bosons arise from the bosonized fermions. We have only discussed the mode expansion of the free fermions, $g=1$, so far. However, similar results can be carried over to the case of interacting fermions with arbitrary $g$. The mode expansions of $\varphi_{c(s)}$ and $\theta_{c(s)}$ become
\begin{widetext}
\begin{subequations}
\begin{eqnarray}
\label{eq:Mode-Exp-varphi-int}\varphi_{c(s)}&=&\hat{\varphi}_{0,c(s)}+\frac{\pi}{\beta}[\frac{1}{g_{c(s)}}\hat{Q}_{c(s)} t+\hat{\tilde{Q}}_{c(s)} x]+\frac{1}{\sqrt{2 g_{c(s)}}} \sum_{n=1}^{\infty} \frac{1}{\sqrt{n}}[a^R_{n,c(s)} e^{-i\frac{2 \pi n}{\beta} x_{-}} + a^L_{n,c(s)}e^{-i\frac{2 \pi n}{\beta} x_{+}}+ h.c.]
\\
\label{eq:Mode-Exp-theta-int}\theta_{c(s)}&=&\hat{\theta}_{0,c(s)}+\frac{\pi}{\beta}[\hat{\tilde{Q}}_{c(s)} t+g_{c(s)} \hat{Q}_{c(s)} x]+\sqrt{\frac{g_{c(s)}}{2} } \sum_{n=1}^{\infty} \frac{1}{\sqrt{n}}[a^L_{n,c(s)} e^{-i\frac{2 \pi n}{\beta} x_{+}} -a^R_{n,c(s)}e^{-i\frac{2 \pi n}{\beta} x_{-}}+ h.c.] \;\; ,
\end{eqnarray}
\end{subequations}
\end{widetext}
following the same periodic boundary conditions in Eq.(\ref{eq:periodic-identi}) and gluing conditions in Eq.(\ref{eq:GluingCondition2}).

\subsection{Junction of two quantum wires for spin-1/2 electrons}

In this subsection, we analyze the stability of the junction of two quantum wires for spin-1/2 electrons by computing the partition function and the scaling dimensions of the boundary operators provided that charge and spin are conserved and there is no resonant tunneling. This method has been pursued by Wong and Affleck~\cite{WongAffleck} for a junction of two quantum wires; here, we follow closely their approach, and only differ in that we make use of the conservation laws at the very beginning, thus changing the gluing conditions and simplifying the computation of scaling dimensions obtained from the expansion of the partition function.

\begin{widetext}
\subsubsection{finite-size spectrum and bulk operators}

Let us consider two sets of independent boson fields $\varphi^{i}_{c(s)}$ and $\theta^i_{c(s)}$, where $i=1, 2$ labels the wires 1 and 2. Since total charge and spin are conserved, it is convenient to work in an alternative basis,
\begin{equation}
\label{eq:BosonFiled1}\Phi^0_{c(s)}\equiv \frac{1}{\sqrt{2}}(\varphi^1_{c(s)}+\varphi^2_{c(s)}), \qquad  \Phi_{c(s)} \equiv \frac{1}{\sqrt{2}}(\varphi^1_{c(s)}-\varphi^2_{c(s)})\; ,
\end{equation}
where the total charge and spin modes are explicit. Using the Eq.(\ref{eq:Mode-Exp-varphi-int}) and Eq.(\ref{eq:Mode-Exp-theta-int}), the mode expansions of $\Phi^0_{c(s)}$ and $\Phi_{c(s)}$ become
\begin{subequations}
\begin{eqnarray}
\Phi^0_{c(s)}&=&\hat{\Phi}^0_{0,c(s)}+\frac{\pi}{\sqrt{2}\beta}[\frac{ \hat{Q}^0_{c(s)} }{g_{c(s)}} t+ \hat{\tilde{Q}}^0_{c(s)} x]+\frac{1}{\sqrt{2 g_{c(s)} }} \sum_{n=1}^{\infty} \frac{1}{\sqrt{n}}[a^{R,0}_{n,c(s)} e^{-i\frac{2 \pi n}{\beta} x_{-}} + a^{L,0}_{n,c(s)} e^{-i\frac{2 \pi n}{\beta} x_{+}}+ h.c.]
\\
\Phi_{c(s)}&=&\hat{\Phi}_{0,c(s)}+\frac{\pi}{\sqrt{2}\beta}[ \frac{\hat{Q}_{c(s)}}{ g_{c(s)} } t+ \hat{\tilde{Q}}_{c(s)} x]+\frac{1}{\sqrt{2 g_{c(s)} }} \sum_{n=1}^{\infty}\frac{1}{\sqrt{n}}[a^{R}_{n,c(s)} e^{-i\frac{2 \pi n}{\beta} x_{-}} + a^{L}_{n,c(s)}e^{-i\frac{2 \pi n}{\beta} x_{+}}+ h.c.]\; ,
\end{eqnarray}
where
\begin{eqnarray}
\Phi^0_{0,c(s)}&=&\frac{1}{\sqrt{2}}(\varphi^1_{0,c(s)}+\varphi^2_{0,c(s)}), \qquad
\Phi_{0,c(s)}=\frac{1}{\sqrt{2}}(\varphi^1_{0,c(s)}-\varphi^2_{0,c(s)}),
\\
a^{R/L,0}_{n,c(s)}&=&\frac{1}{\sqrt{2}}(a^{R/L,1}_{n,c(s)}+a^{R/L,2}_{n,c(s)}),\qquad
a^{R/L}_{n,c(s)}=\frac{1}{\sqrt{2}}(a^{R/L,1}_{n,c(s)}-a^{R/L,2}_{n,c(s)}),
\end{eqnarray}
and the winding operators follow 
\begin{eqnarray}
\hat{Q}^0_{c(s)}=\hat{Q}^1_{c(s)}+\hat{Q}^2_{c(s)}, \qquad
\hat{Q}_{c(s)}=\hat{Q}^1_{c(s)}-\hat{Q}^2_{c(s)},\qquad 
\hat{\tilde{Q}}^0_{c(s)}=\hat{\tilde{Q}}^1_{c(s)}+\hat{\tilde{Q}}^2_{c(s)},\qquad
\hat{\tilde{Q}}_{c(s)}=\hat{\tilde{Q}}^1_{c(s)}-\hat{\tilde{Q}}^2_{c(s)}.
\end{eqnarray}
\end{subequations}
The eigenvalues of these winding operators are integer and follow gluing conditions, derived from Eq.(\ref{eq:GluingCondition2}):
\begin{eqnarray}
\label{eq:Gluing-Cond-2W}
\begin{array}{cc}
{\mathbf 1.}\;  Q^0_c=Q^0_s=\tilde{Q}^0_c=\tilde{Q}^0_s=Q_c=Q_s=\tilde{Q}_c=\tilde{Q}_s  \;\;\; \;\; & \qquad  {\rm (mod 2)} \\
{\mathbf 2.} \;  Q^0_c+\tilde{Q}^0_c+Q^0_s+\tilde{Q}^0_s=0 \qquad \qquad \qquad \qquad \qquad  & \qquad {\rm (mod 4) } \\
{\mathbf 3.}\;  Q^0_s+\tilde{Q}^0_c+Q_s+\tilde{Q}_c=0 \qquad \qquad \qquad \qquad\qquad& \qquad {\rm (mod 4)} \\
{\mathbf 4.} \; Q^0_s+Q^0_c+Q_s+Q_c=0 \qquad \qquad  \qquad \qquad \qquad& \qquad {\rm (mod 4)}  \\
{\mathbf 5.}\; Q^0_c+Q^0_s+\tilde{Q}^0_c+\tilde{Q}^0_s+Q_c+Q_s+\tilde{Q}_c+\tilde{Q}_s=0 \; & \qquad {\rm (mod 8)} \\
\end{array}\; .
\end{eqnarray}


Using the Lagrangian density 
\begin{eqnarray}
\nonumber \mathcal{L} =  \sum_{j=1,2} \frac{g_c}{4\pi} (\partial_\mu \varphi_{j,c})^2+ \frac{g_s}{4 \pi} [(\partial_\mu \varphi_{j,s})^2, 
\end{eqnarray}
the Hamiltonian of the two quantum wires can be written in terms of the winding and number operators as follows
\begin{eqnarray}
\nonumber H^P_\beta=\frac{2
        \pi}{\beta}&[&\frac{(\hat{Q}^0_c)^2}{16 g_c}+\frac{g_c
        (\hat{\tilde{Q}}^0_c)^2}{16}+\frac{(\hat{Q}_c)^2}{16
        g_c}+\frac{g_c
        (\hat{\tilde{Q}}_c)^2}{16}+\frac{(\hat{Q}^0_s)^2}{16
        g_s}+\frac{g_s
        (\hat{\tilde{Q}}^0_s)^2}{16}+\frac{(\hat{Q}_s)^2}{16
        g_s}+\frac{g_s (\hat{\tilde{Q}}_s)^2}{16}
        \\
\label{eq:Hamil2wire}&+&\sum_{m=1}^{\infty} m
        (\hat{n}^{L,0}_{m,c}+\hat{n}^{R,0}_{m,c}+\hat{n}^{L,0}_{m,s}+\hat{n}^{R,0}_{m,s})+\sum_{m=1}^{\infty}
        m
        (\hat{n}^{L}_{m,c}+\hat{n}^{R}_{m,c}+\hat{n}^{L}_{m,s}+\hat{n}^{R}_{m,s})]\;.
\end{eqnarray}
The corresponding {\it bulk} primary operators are in the form
\begin{equation}
\exp[ i ( Q^0_c \Phi_c^0 +  \tilde{Q}^0_c \Theta^0_c + Q_c \Phi_c +\tilde{Q}_c \Theta_c +  Q ^0_s \Phi_s^0 + \tilde{Q}^0_s \Theta^0_s + Q_s \Phi_s + \tilde{Q}_s \Theta_s  )/(2 \sqrt{2})],
\end{equation}
with the scaling dimensions 
\begin{equation}
\frac{(Q^0_c)^2}{16 g_c}+\frac{g_c
        (\tilde{Q}^0_c)^2}{16}+\frac{(Q_c)^2}{16
        g_c}+\frac{g_c
        (\tilde{Q}_c)^2}{16}+\frac{(Q^0_s)^2}{16
        g_s}+\frac{g_s
        (\tilde{Q}^0_s)^2}{16}+\frac{(Q_s)^2}{16
        g_s}+\frac{g_s (\tilde{Q}_s)^2}{16}.
\end{equation}
\end{widetext}

\subsubsection{boundary states}

Following Cardy, the boundary conditions can be represented in terms of the corresponding boundary states upon modular transformation. Without going in to the detail, we construct the boundary states for the case of two quantum wires.

Because of the conservation of total charge and spin, it is natural to impose the Neumann(N) boundary condition on the center of mass modes, $\Phi^0_{c(s)}$. Since the N boundary condition on $\Phi^0$ implies Dirichlet boundary condition on the dual field $\Theta^0$, the winding of $\Theta^0$ along the boundary should be zero; hence the corresponding winding number $Q^0_{c(s)}=0$. Because of the gluing conditions, the quantum numbers $\tilde{Q}^0_{c(s)}$ are now restricted to even numbers and follow some extra constraints. The N boundary state for both center of mass modes is
\begin{widetext}
\begin{equation}
\label{eq:B-S-zero-mode} |N^0 \rangle = G^0_N  \exp[\sum_{n=1}^{\infty} {a_{n,c}^{0,L}}^\dag {a_{n,c}^{0,R}}^\dag ] \exp[\sum_{n=1}^{\infty} {a_{n,s}^{0,L}}^\dag {a_{n,s}^{0,R}}^\dag ]  \sum^{\infty\prime}_{\tilde{Q}^0_{c(s)}=-\infty}  \exp[-i\tilde{Q}_c^0 \Theta^0_{0,c}] \exp[-i\tilde{Q}_s^0 \Theta^0_{0,s}] |\tilde{Q}_c^0,0\rangle \otimes |\tilde{Q}_s^0,0\rangle, 
\end{equation}
\end{widetext}
where prime over the summation indicates the gluing conditions, and $G^0_N$ is the ground state degeneracy which will be fixed by Cardy's consistency condition~\cite{WongAffleck, Cardy, gtheorem}. The vacuum states are denoted by the winding number $ |\tilde{Q}_\mu,Q_\mu \rangle$ of each independent boson field. For simplicity, the phases $\Theta^0_{0,c(s)}$ correspond to the applied voltages in the wires and will not affect the boundary physics; hence we will take these phases as zero.

Despite the fixed boundary condition for the center of mass modes, the other degrees of freedom can have different boundary conditions. In the case of a junction of two wires, there are two possible boundary states, Neumann and Dirichlet boundary conditions on the $\varphi$ field (N and D boundary states). Similar to the case of the center of mass mode, the N boundary state for the dynamical fields $\Phi_{c(s)}$ is given by
\begin{widetext}
\begin{equation}
\label{eq:B-S-N} |N_{c(s)} \rangle = G_{N,c(s)}  \exp[\sum_{n=1}^{\infty} {a_{n,c(s)}^{L}}^\dag {a_{n,c(s)}^{R}}^\dag ]  \sum^{\infty\prime}_{\tilde{Q}_{c(s)}=-\infty}    |\tilde{Q}_{c(s)},0\rangle .
\end{equation}
The D boundary state implies that the winding of $\Phi_{c(s)}$ along the boundary is zero. As a result, the quantum number $\tilde{Q}_{c(s)}=0$ and $Q_{c(s)}$ is restricted to even integers with more constraints from the gluing conditions. Hence, the boundary state is given by
\begin{equation}
\label{eq:B-S-D} |D_{c(s)} \rangle = G_{D,c(s)}  \exp[ - \sum_{n=1}^{\infty} {a_{n,c(s)}^{L}}^\dag {a_{n,c(s)}^{R}}^\dag ]  \sum^{\infty\prime}_{Q_{c(s)}=-\infty}    |0, Q_{c(s)}\rangle ,
\end{equation}
where the minus sign inside the exponential function comes from the identification, $\phi^R=-\phi^L+C$. 
\end{widetext}

\subsubsection{projection of center of mass modes}

In principle, we have everything we need to compute the partition function
\begin{equation}
\label{eq:def-partition-function}
\langle B| e^{-l H_\beta^P} |B \rangle,
\end{equation}
with a given boundary state and gluing conditions. Upon modular transformation, one can read off the scaling dimensions of primary boundary operators. However, the gluing conditions complicate the calculation of the partition function. If the charge and spin are conserved, there is a simpler way to find the dimensions of the boundary operators. Since the boundary state of $\Phi^0_{c(s)}$ is Neumann, only the $Q^0_{c(s)}=0$ states occur. Hence the constraints of Eq.(\ref{eq:Gluing-Cond-2W}) reduce to
\begin{widetext}
\begin{eqnarray}
\label{eq:Gluing-Cond-2W-reduced}
\begin{array}{cc}
{\mathbf 1.}\; \tilde{Q}^0_c=\tilde{Q}^0_s=Q_c=Q_s=\tilde{Q}_c=\tilde{Q}_s=0  &   {\rm (mod 2)} \\
{\mathbf 2.}\; \tilde{Q}^0_c+\tilde{Q}^0_s=0 \qquad \qquad \qquad\qquad \qquad \;&  {\rm (mod 4) } \\
{\mathbf 3.}\; \tilde{Q}^0_c+Q_s+\tilde{Q}_c=0 \qquad \qquad\qquad\;\;\;\;\;\;\;& {\rm (mod 4)} \\
{\mathbf 4.}\; Q_s+Q_c=0  \qquad\qquad \qquad\qquad \qquad \;&  {\rm (mod 4)} \\
{\mathbf 5.}\; \tilde{Q}^0_c+\tilde{Q}^0_s+Q_c+Q_s+\tilde{Q}_c+\tilde{Q}_s=0  \;\;&  {\rm (mod 8)} \\
\end{array}.
\end{eqnarray}
From the first constraint, we obtain
\begin{subequations}
\begin{eqnarray}
\label{eq:All-even-constraint}
\tilde{Q}^0_c =2  \tilde{n}^0_c, \qquad \tilde{Q}^0_s=2 \tilde{n}^0_s, \qquad Q_c= 2 n_c, \qquad Q_s=2 n_s, \qquad \tilde{Q}_c= 2 \tilde{n}_c, \qquad \tilde{Q}_s= 2 \tilde{n}_s.
\end{eqnarray}
Here, all $n_\mu$ are integers and follow the constraints below
\begin{eqnarray}
\begin{array}{cc}
{\mathbf 2.}\; \tilde{n}^0_c+\tilde{n}^0_s=0 \qquad \qquad \qquad\qquad \qquad \;&  {\rm (mod 2) } \\
{\mathbf 3.}\; \tilde{n}^0_c+n_s+\tilde{n}_c=0 \qquad \qquad\qquad\;\;\;\;\;\;\;& {\rm (mod 2)} \\
{\mathbf 4.}\; n_s+n_c=0  \qquad\qquad \qquad\qquad \qquad \;&  {\rm (mod 2)} \\
{\mathbf 5.}\; \tilde{n}^0_c+\tilde{n}^0_s+n_c+n_s+\tilde{n}_c+\tilde{n}_s=0  \;\;\;\; &  {\rm (mod 4)} \\
\end{array},
\end{eqnarray}
where the numbering of equations corresponds to that of Eq.(\ref{eq:Gluing-Cond-2W-reduced}). There are four sets of possible parameterizations of $n_\mu$ that satisfy all the constraints
\begin{equation}
\label{eq:RGC-m-sets}
\left\{
\begin{split}
(a)\; &\tilde{n}^0_c= 2\tilde{m}^0_c, \qquad \tilde{n}^0_s= 2\tilde{m}^0_s, \qquad \tilde{n}_c= 2 \tilde{m}_c, \qquad n_c=2 m_c,\qquad \tilde{n}_s= 2 \tilde{m}_s, \qquad n_s=2 m_s
\\
(b) \; &\tilde{n}^0_c= 2\tilde{m}^0_c, \qquad \tilde{n}^0_s= 2\tilde{m}^0_s, \qquad \tilde{n}_c= 2 \tilde{m}_c+1, \; n_c=2 m_c+1,\; \tilde{n}_s= 2 \tilde{m}_s+1, \; n_s=2 m_s+1
\\
(c) \;&\tilde{n}^0_c= 2\tilde{m}^0_c+1, \; \tilde{n}^0_s= 2\tilde{m}^0_s+1, \; \tilde{n}_c= 2 \tilde{m}_c+1, \; n_c=2 m_c,\qquad \tilde{n}_s= 2 \tilde{m}_s+1, \; n_s=2 m_s
\\
(d) \; &\tilde{n}^0_c= 2\tilde{m}^0_c+1, \; \tilde{n}^0_s= 2\tilde{m}^0_s+1, \; \tilde{n}_c= 2 \tilde{m}_c, \qquad n_c=2 m_c+1,\; \tilde{n}_s= 2 \tilde{m}_s, \qquad n_s=2 m_s+1
\end{split} \right. ,
\end{equation}
\end{subequations}
\end{widetext}
with the constraint $\tilde{m}_c^0+\tilde{m}_s^0+\tilde{m}_c+m_c+\tilde{m}_s+m_s=0$ (mod 2) for each set. 

In order to disentangle the Hilbert space of $\Phi^0_{c(s)}$ from the total Hilbert space, we denote a state with quantum numbers $ \tilde{Q}_{c(s)}$ and $Q_{c(s)} $ in the Hilbert space of $\Phi_{c(s)}$ by $|\psi_{\tilde{Q}_{c(s)},Q_{c(s)}} \rangle$.  Note that this state is proportional to $| \tilde{Q}_{c(s)},Q_{c(s)} \rangle$. Although only $|\psi_{\tilde{Q}_{c(s)},0} \rangle$ is non-zero in the N state and only $|\psi_{0,Q_{c(s)}} \rangle$ is non-zero in the D state case, we would like to consider the general case. Using Eq.(\ref{eq:All-even-constraint}) and Eq.(\ref{eq:RGC-m-sets}) and considering that $\tilde{m}^0_{c(s)}= 2 \tilde{l}_{c(s)}$ or $\tilde{m}^0_{c(s)}= 2 \tilde{l}_{c(s)}+1$, the boundary sates corresponding to the center of mass modes can be generally written as
\begin{widetext}
\begin{subequations}
\begin{eqnarray}
|N^0_{0,c(s)}\rangle &=& (4 g_{c(s)})^{1/4}  \exp[\sum_{n=1}^{\infty} {a_{n,c(s)}^{0,L}}^\dag {a_{n,c(s)}^{0,R}}^\dag ]  \sum_{\tilde{l}_{c(s)} \in Z}   |8 \tilde{l}_{c(s)}^0,0\rangle , 
\\
|N^0_{2,c(s)}\rangle &=& (4 g_{c(s)})^{1/4}   \exp[\sum_{n=1}^{\infty} {a_{n,c(s)}^{0,L}}^\dag {a_{n,c(s)}^{0,R}}^\dag ]  \sum_{\tilde{l}_{c(s)} \in Z}   |8 \tilde{l}_{c(s)}^0+4,0\rangle , 
\\
|N^0_{1,c(s)}\rangle &=& (4 g_{c(s)})^{1/4}  \exp[\sum_{n=1}^{\infty} {a_{n,c(s)}^{0,L}}^\dag {a_{n,c(s)}^{0,R}}^\dag ]  \sum_{\tilde{l}_{c(s)} \in Z}   |8 \tilde{l}_{c(s)}^0+2,0\rangle , 
\\
|N^0_{3,c(s)}\rangle &=& (4 g_{c(s)})^{1/4}   \exp[\sum_{n=1}^{\infty} {a_{n,c(s)}^{0,L}}^\dag {a_{n,c(s)}^{0,R}}^\dag ]  \sum_{\tilde{l}_{c(s)} \in Z}   |8 \tilde{l}_{c(s)}^0+6,0\rangle ,  
\end{eqnarray}
\end{subequations}
where combinations $|N^0_{0,c(s)}\rangle$ and $|N^0_{2,c(s)}\rangle$ correspond to the sets (a) and (b) in Eq.(\ref{eq:RGC-m-sets}) and combinations $|N^0_{1,c(s)}\rangle$ and $|N^0_{3,c(s)}\rangle$ correspond to the sets (c) and (d). Hence, the most general boundary state can be written as
\begin{equation}
\label{eq:full-boundary-state}
\begin{split}
|B\rangle & =  (|N^0_{0,c}\rangle \otimes |N^0_{0,s}\rangle +|N^0_{2,c}\rangle  \otimes  |N^0_{2,s}\rangle )  \otimes \sum^{\prime}\left( | \psi_{4 \tilde{m}_c, 4 m_c }  \rangle \otimes | \psi_{4 \tilde{m}_s, 4 m_s }  \rangle \right)
\\
& + (|N^0_{0,c}\rangle \otimes |N^0_{2,s}\rangle +|N^0_{2,c}\rangle  \otimes  |N^0_{0,s}\rangle )  \otimes \sum^{\prime \prime}\left( | \psi_{4 \tilde{m}_c, 4 m_c }  \rangle \otimes | \psi_{4 \tilde{m}_s, 4 m_s }  \rangle \right)
\\
& +(|N^0_{0,c}\rangle \otimes |N^0_{0,s}\rangle +|N^0_{2,c}\rangle  \otimes  |N^0_{2,s}\rangle )  \otimes \sum^{\prime}\left( | \psi_{4 \tilde{m}_c+2, 4 m_c+2 }  \rangle \otimes | \psi_{4 \tilde{m}_s+2, 4 m_s+2 }  \rangle \right)
\\
& + (|N^0_{0,c}\rangle \otimes |N^0_{2,s}\rangle +|N^0_{2,c}\rangle  \otimes  |N^0_{0,s}\rangle )  \otimes \sum^{\prime \prime}\left( | \psi_{4 \tilde{m}_c+2, 4 m_c+2 }  \rangle \otimes | \psi_{4 \tilde{m}_s+2, 4 m_s+2}  \rangle \right)
\\
&+(|N^0_{1,c}\rangle \otimes |N^0_{1,s}\rangle +|N^0_{3,c}\rangle  \otimes  |N^0_{3,s}\rangle )  \otimes \sum^{\prime}\left( | \psi_{4 \tilde{m}_c+2, 4 m_c }  \rangle \otimes | \psi_{4 \tilde{m}_s+2, 4 m_s }  \rangle \right)
\\
&+(|N^0_{1,c}\rangle \otimes |N^0_{3,s}\rangle +|N^0_{3,c}\rangle  \otimes  |N^0_{1,s}\rangle )  \otimes \sum^{\prime \prime }\left( | \psi_{4 \tilde{m}_c+2, 4 m_c }  \rangle \otimes | \psi_{4 \tilde{m}_s+2, 4 m_s }  \rangle \right)
\\
&+(|N^0_{1,c}\rangle \otimes |N^0_{1,s}\rangle +|N^0_{3,c}\rangle  \otimes  |N^0_{3,s}\rangle )  \otimes \sum^{\prime}\left( | \psi_{4 \tilde{m}_c, 4 m_c +2 }  \rangle \otimes | \psi_{4 \tilde{m}_s, 4 m_s +2 }  \rangle \right)
\\
&+(|N^0_{1,c}\rangle \otimes |N^0_{3,s}\rangle +|N^0_{3,c}\rangle  \otimes  |N^0_{1,s}\rangle )  \otimes \sum^{\prime \prime }\left( | \psi_{4 \tilde{m}_c, 4 m_c +2}  \rangle \otimes | \psi_{4 \tilde{m}_s, 4 m_s +2}  \rangle \right)
\end{split}\; ,
\end{equation}
where the prime over the summation indicates the constraint $\tilde{m}_c+m_c+\tilde{m}_s+m_s=0$ (mod 2) for each term and the double prime over the summation indicates the constraint $\tilde{m}_c+m_c+\tilde{m}_s+m_s=1$ (mod 2).

Using Eq.(\ref{eq:def-partition-function}), the diagonal partition function is given by:
\begin{equation}
\begin{split}
Z_{BB} & =  ( Z_{N^0_{0,c}} \; Z_{N^0_{0,s}} + Z_{N^0_{2,c}}\; Z_{N^0_{2,s}} ) \; \sum^{\prime}\left( Z_{ 4 \tilde{m}_c, 4 m_c }  \; Z_{4 \tilde{m}_s, 4 m_s }  \right)
\\
& + ( Z_{N^0_{0,c}} \; Z_{N^0_{2,s}} + Z_{N^0_{2,c}}\; Z_{N^0_{0,s}} ) \; \sum^{\prime\prime}\left( Z_{ 4 \tilde{m}_c, 4 m_c }  \; Z_{4 \tilde{m}_s, 4 m_s }  \right)
\\
& +( Z_{N^0_{0,c}} \; Z_{N^0_{0,s}} + Z_{N^0_{2,c}}\; Z_{N^0_{2,s}} ) \; \sum^{\prime}\left( Z_{ 4 \tilde{m}_c+2, 4 m_c+2 }  \; Z_{4 \tilde{m}_s+2, 4 m_s+2 }  \right)
\\
& +( Z_{N^0_{0,c}} \; Z_{N^0_{2,s}} + Z_{N^0_{2,c}}\; Z_{N^0_{0,s}} ) \; \sum^{\prime \prime }\left( Z_{ 4 \tilde{m}_c+2, 4 m_c+2 }  \; Z_{4 \tilde{m}_s+2, 4 m_s+2 }  \right)
\\
&+( Z_{N^0_{1,c}} \; Z_{N^0_{1,s}} + Z_{N^0_{3,c}}\; Z_{N^0_{3,s}} ) \; \sum^{\prime}\left( Z_{ 4 \tilde{m}_c+2, 4 m_c }  \; Z_{4 \tilde{m}_s+2, 4 m_s }  \right)
\\
&+( Z_{N^0_{1,c}} \; Z_{N^0_{3,s}} + Z_{N^0_{3,c}}\; Z_{N^0_{1,s}} ) \; \sum^{\prime \prime}\left( Z_{ 4 \tilde{m}_c+2, 4 m_c }  \; Z_{4 \tilde{m}_s+2, 4 m_s }  \right)
\\
&+( Z_{N^0_{1,c}} \; Z_{N^0_{1,s}} + Z_{N^0_{3,c}}\; Z_{N^0_{3,s}} ) \; \sum^{\prime}\left( Z_{ 4 \tilde{m}_c, 4 m_c+2 }  \; Z_{4 \tilde{m}_s, 4 m_s+2 }  \right)
\\
&+( Z_{N^0_{1,c}} \; Z_{N^0_{3,s}} + Z_{N^0_{3,c}}\; Z_{N^0_{1,s}} ) \; \sum^{\prime \prime}\left( Z_{ 4 \tilde{m}_c, 4 m_c+2 }  \; Z_{4 \tilde{m}_s, 4 m_s+2 }  \right)
\end{split}\; .
\end{equation}
Here, $Z_{N^0_{i,c(s)}}$ are diagonal partition functions in the $\Phi^0_{c(s)}$ Hilbert space while the $Z_{\tilde{Q}_{c(s)}, Q_{c(s)} }$ are diagonal partition functions in the $\Phi_{c(s)}$ Hilbert space:
\begin{equation}
Z_{\tilde{Q}_{c(s)}, Q_{c(s)}} \equiv \langle \psi_{\tilde{Q}_{c(s)}, Q_{c(s)}} | e^{- l H_\beta^P} | \psi_{\tilde{Q}_{c(s)}, Q_{c(s)}}   \rangle
\end{equation}

The partition functions corresponding to the center of mass mode modes are given by
\begin{subequations}
\begin{eqnarray}
Z_{N^0_{0,c(s)}}&=&\frac{\sqrt{4 g_{c(s)}}}{\eta(\tilde{q})}  \sum_{\tilde{l}_{c(s)}^0} \tilde{q}^{ 2g_{c(s)} (\tilde{l}_{c(s)}^0)^2 }=\frac{1}{\eta(q)} \sum_{Q^0_{c(s)}\in Z} q^{(Q^0_{c(s)})^2 /(8 g_{c(s)})} ,
\\
Z_{N^0_{1,c(s)}}&=&\frac{\sqrt{4 g_{c(s)}}}{\eta(\tilde{q})}  \sum_{\tilde{l}_{c(s)}^0} \tilde{q}^{ 2g_{c(s)} (\tilde{l}_{c(s)}^0+ 1/4)^2 }=\frac{1}{\eta(q)} \sum_{Q^0_{c(s)}\in Z} e^{-i \frac{\pi}{2} Q^0_{c(s)} } q^{(Q^0_{c(s)})^2 /(8 g_{c(s)})},
\\
Z_{N^0_{2,c(s)}}&=&\frac{\sqrt{4 g_{c(s)}}}{\eta(\tilde{q})} \sum_{\tilde{l}_{c(s)}^0} \tilde{q}^{ 2g_{c(s)} (\tilde{l}_{c(s)}^0+ 1/2)^2 }=\frac{1}{\eta(q)} \sum_{Q^0_{c(s)}\in Z} e^{-i \pi Q^0_{c(s)} } q^{(Q^0_{c(s)})^2 /(8 g_{c(s)})}, 
\\
Z_{N^0_{3,c(s)}}&=&\frac{\sqrt{4 g_{c(s)}}}{\eta(\tilde{q})} \sum_{\tilde{l}_{c(s)}^0} \tilde{q}^{ 2g_{c(s)} (\tilde{l}_{c(s)}^0+3/4)^2 }=\frac{1}{\eta(q)} \sum_{Q^0_{c(s)}\in Z} e^{-i \frac{3 \pi}{2} Q^0_{c(s)} } q^{(Q^0_{c(s)})^2 /(8 g_{c(s)})},
\end{eqnarray}
\end{subequations}
\end{widetext}
where the modular transformation has been performed at the second equality and  
\begin{equation}
\tilde{q}\equiv e^{-\frac{4\pi l}{\beta}}, \qquad q \equiv e^{- \frac{\pi \beta}{l}}. 
\end{equation}
Here, we introduce the Dedekind $\eta$-function:
\begin{equation}
\eta(\tilde{q})\equiv \tilde{q}^{1/24} \prod_{n=1}^{\infty}
(1-\tilde{q}^n)\; ,
\end{equation}
which comes from the oscillator modes of the boundary state. One can show that each set of independent oscillator modes contribute a factor of  $1/\eta(\tilde{q})$. Moreover, the modular transformation of the Dedekind $\eta$-function is given by
\begin{equation}
\eta(\tilde{q})=\sqrt{\frac{\beta}{2l}}\eta(q).
\end{equation}
The scaling dimensions in $Z_{N^0_{i,c(s)}}$ correspond to the boundary operators that include $\Phi^0_{c(s)}$ and therefore do not conserve the total charge and spin. For instance, a nontrivial dimension, $(Q^0_{c(s)})^2 /(8 g_{c(s)})$, corresponds to the vertex operator $e^{i Q^0_{c(s)} \Phi^0_{c(s)}/(2 \sqrt{2}) }$. Since we only consider the perturbations which conserve the charge and spin, those vertex operators should not appear. The term $1/\eta(q)$ corresponds to irrelevant or marginal boundary operators which come from the derivatives of $\Phi^{0}_{c(s)}$ and have integer dimensions. 

Thus, we shall concentrate on the boundary operators involving only the dynamical fields $\Phi_{c(s)}$ and replace
\begin{equation}
Z_{N^0_{i,c(s)}}\to 1. 
\end{equation}
The effective partition function becomes 
\begin{equation}
 \begin{split}
Z_{BB} \to &  \;\;\;\; \;  2  \; \sum_{m \in Z } \left( Z_{ 4 \tilde{m}_c, 4 m_c }  \; Z_{4 \tilde{m}_s, 4 m_s }  \right)
\\
& + 2 \; \sum_{m \in Z}\left( Z_{ 4 \tilde{m}_c+2, 4 m_c+2 }  \; Z_{4 \tilde{m}_s+2, 4 m_s+2 }  \right)
\\
& + 2 \; \sum_{m\in Z}\left( Z_{ 4 \tilde{m}_c+2, 4 m_c }  \; Z_{4 \tilde{m}_s+2, 4 m_s }  \right)
\\
& + 2 \; \sum_{m \in Z}\left( Z_{ 4 \tilde{m}_c, 4 m_c+2 }  \; Z_{4 \tilde{m}_s, 4 m_s+2 }  \right)
\end{split}\; .
\end{equation}
Observe that this reduced partition function can be obtained by simply eliminating the charge and spin center of mass modes from the full boundary states. The original gluing conditions can be reduced to a set of constraints on $\tilde{Q}_{c(s)}$ and $Q_{c(s)}$ only, which will be refer to as the ``Reduced Gluing Conditions'' (RGC).

Here we outline the procedures for obtaining the scaling dimensions of boundary operators from the reduced partition function. First, we eliminate the center of mass modes from the original boundary state in Eq.(\ref{eq:full-boundary-state}) to obtain the``Reduced Boundary State". Then, we derive the nontrivial RGC for $\tilde{Q}_{c(s)}$ and $Q_{c(s)}$ from the original gluing conditions. Finally, we compute the reduced partition function upon a modular transformation and obtain complete spectrum of the scaling dimensions of boundary operators for a given boundary state. A similar reduction procedure can be applied to the case of junctions of three quantum wires for spin-1/2 electrons.

\begin{widetext}
\subsection{Partition functions and scaling dimensions of boundary operators for junction of two quantum wires}

In this subsection, we apply the method described above to compute the reduced partition functions and scaling dimensions of boundary operators involving the dynamic field $\Phi_{c(s)}$ for given boundary conditions. 

\subsubsection{{\bf NN Boundary State}}

The reduced NN boundary state can be written as
\begin{equation}
|NN\rangle= g^{\ }_{NN} \exp[\sum_{n=1}^\infty a_{n,c}^{L\dag}a_{n,c}^{R\dag}+\sum_{n=1}^\infty  a_{n,s}^{L\dag}a_{n,s}^{R\dag}] \sum^\prime |(\tilde{Q}_c,0)\rangle \otimes|(\tilde{Q}_s,0)\rangle,
\end{equation}
where $g_{NN}$ is the ground state degeneracy and prime indicates that the RGC should be obeyed. The corresponding partition function for the NN boundary state is given by
\begin{equation}
\label{eq:ReduNNparti1}Z_{NN,NN} = \langle NN|e^{-lH_\beta^P}|NN\rangle =\frac{g_{NN}^2}{\eta(\tilde{q})^2} \sum^\prime \exp[-(\frac{l \pi g_c}{8 \beta}(\tilde{Q}_c)^2+\frac{l \pi g_s}{8 \beta}(\tilde{Q}_s)^2)]\; ,
\end{equation}
\end{widetext}

Now, we shall discuss how to obtain the RGC. Recalling the gluing conditions in Eq.(\ref{eq:Gluing-Cond-2W}) and using $Q^0_{c(s)}=Q_{c(s)}=0$, one concludes that $\tilde{Q}_c$ and $\tilde{Q}_s$ are even. Hence, $\tilde{Q}_{c(s)}=2\tilde{n}_{c(s)}$. Still, we should keep in mind that $\tilde{Q}_c^0$ and $\tilde{Q}_s^0$ are even and contribute to the reduced gluing conditions. In terms of $\tilde n_\mu$, the gluing conditions are reduced to
\begin{equation}
\begin{array}{cc}
{\mathbf a.}\;  \tilde{n}_c^0+ \tilde{n}_s^0=0  \qquad \qquad \;\;& \qquad  {\rm (mod 2)} \\
{\mathbf b.} \;  \tilde{n}_c^0+ \tilde{n}_c=0 \qquad\qquad \;\;  & \qquad {\rm (mod 2) } \\
{\mathbf c.}\;  \tilde{n}_c^0+ \tilde{n}_c+\tilde{n}_s^0+\tilde{n}_s=0 & \qquad {\rm (mod 4)} \\
\end{array}\; .
\end{equation}
The conditions {\bf a} and {\bf b} imply that $\tilde{n}_c^0$, $\tilde{n}_s^0$ and $\tilde{n}_c$ have the same parity. In addition, one can conclude that $\tilde{n}_s$ has the same parity as $\tilde{n}_c$ and $\tilde{n}_{c(s)}^0$ due to the condition {\bf c}. Hence
\begin{equation}
\label{eq:RGC-NN-n}
\tilde{n}_c+\tilde{n}_s=0\qquad  { \rm (mod 2)}
\end{equation}
becomes the only gluing condition for the variable $\tilde{n}_{c(s)}$. Naively, one may expect that condition {\bf c} should provide other constraints. However, it can be shown that the extra condition is redundant. We justify this statement below. 

First, because all the winding numbers, $n_\mu$, are of the same parity, we can set either $n_\mu=2m_\mu$ or $n_\mu=2 m_\mu+1$. For either case, the condition {\bf c} leads to
\begin{equation}
\tilde{m}_c^0+ \tilde{m}_c+\tilde{m}_s^0+\tilde{m}_s=0 \qquad {\rm (mod 2)}
\end{equation}
However, any combinations of integer $\tilde{m}_c$ and $\tilde{m}_s$ have relative sets of $\tilde{m}_c^0$ and $\tilde{m}_s^0$ which satisfy this gluing condition for $m_\mu$. Moreover, since the partition functions corresponding to center of mass modes provide an equal constant, set to be $1$, for any combinations of $\tilde{m}_{c(s)}$, this condition effectively provides no constraint on $\tilde{n}_{c(s)}$.

Now, the partition function, Eq.(\ref{eq:ReduNNparti1}), becomes
\begin{widetext}
\begin{equation}
Z_{NN,NN}= \frac{g_{NN}^2}{\eta(\tilde{q})^2} \sum^\prime \exp[-(\frac{l
\pi g_c}{2 \beta}(\tilde{n}_c)^2+\frac{l \pi g_s}{2
\beta}(\tilde{n}_s)^2)]\; ,
\end{equation}
in terms of variable $n_\mu$ with the reduced gluing condition, Eq.(\ref{eq:RGC-NN-n}). Further, the RGC leads to two separate sums
\begin{equation}
Z_{NN,NN}=\frac{g_{NN}^2}{\eta(\tilde{q})^2} \{\sum_{\tilde{m}_{c(s)} \in Z} \tilde{q}^{(\frac{g_c}{2}(\tilde{m}_c)^2+\frac{g_s}{2}(\tilde{m}_s)^2)} + \sum_{\tilde{m}_{c(s)} \in Z} \tilde{q}^{(\frac{g_c}{8}(2 \tilde{m}_c+1)^2+\frac{g_s}{8}(2\tilde{m}_s+1)^2)} \}\; ,
\end{equation}
without any constraint on $\tilde{m}_c$ and $\tilde{m}_s$. Upon the modular transformation, the partition function becomes
\begin{subequations}
\begin{eqnarray}
Z_{NN,NN}(q) &=& \frac{g_{NN}^2}{\sqrt{g_c g_s}}\frac{1}{\eta(q)^2} \sum_{m_c,m_s} q^{\frac{(m_c)^2}{2 g_c} +\frac{(m_s)^2}{2 g_s}} \times [1+e^{-i \pi (m_c+m_s)}]
\\
&=& \frac{2 g_{NN}^2 }{\sqrt{g_c g_s}}\frac{1}{\eta(q)^2} \sum_{m_c+m_s=0 \;\; {\rm(mod\;2)}} q^{\frac{(m_c)^2}{2 g_c} +\frac{(m_s)^2}{2 g_s}}\; .
\end{eqnarray}
\end{subequations}

Hence, the spectrum of boundary operators can be read off and the leading order perturbations have the scaling dimensions
\begin{equation}
\frac{1}{2 g_c}+\frac{1}{2 g_s}, \qquad \frac{2}{g_c}, \qquad \frac{2}{g_s}\; ,
\end{equation}
which agree with the previous results obtained using the DEBC method in Sec.~\ref{sec:DEBC}. Moreover, the ground state degeneracy, useful for determining the stability of the fixed point, can be found by Cardy's consistency condition as $g_{NN}=(g_cg_s/4)^{1/4}$.

\subsubsection{{\bf DD Boundary State}}

The reduced DD boundary state can be constructed as
\begin{equation}
|DD\rangle = g_{DD} \exp[-\sum_{n=1}^\infty a_{n,c}^{L\dag}a_{n,c}^{R\dag}-\sum_{n=1}^\infty a_{n,s}^{L\dag}a_{n,s}^{R\dag}] \sum^\prime |(0,Q_c)\rangle\otimes|(0,Q_s)\rangle,
\end{equation}
using Eq.(\ref{eq:B-S-D}) with ground state degeneracy $g_{DD}$. Again, the prime over the summation implies the reduced gluing conditions. The partition function corresponding to the DD boundary state is given by
\begin{equation}
\label{eq:ReduDDparti1}Z_{DD,DD} = \langle DD|e^{-lH_\beta^P}|DD\rangle=\frac{g_{DD}^2}{\eta(\tilde{q})^2} \sum^\prime \exp[-(\frac{l \pi}{8 \beta g_c}(Q_c)^2+\frac{l \pi}{8 \beta g_s}(Q_s)^2)]\; ,
\end{equation}
\end{widetext} 

Since $Q^0_{c(s)}=\tilde{Q}_{c(s)}=0$, the condition {\bf 1} of Eq.(\ref{eq:Gluing-Cond-2W}) leads to $\tilde{Q}^0_{c(s)}=2 \tilde{n}^0_{c(s)}$ and $Q_{c(s)}=2 n_{c(s)}$ for integer $n_\mu$. Hence the gluing conditions in terms of the quantum numbers $n_\mu$ become
\begin{equation}
\begin{array}{cc}
{\mathbf a.}\;  \tilde{n}_c^0+ \tilde{n}_s^0=0  \qquad \qquad \;\;& \qquad  {\rm (mod \; 2)} \\
{\mathbf b.} \;  \tilde{n}_c^0+ n_s=0 \qquad\qquad \;\;  & \qquad {\rm (mod \; 2) } \\
{\mathbf c.} \;  n_c+n_s=0 \qquad\qquad \;\;  & \qquad {\rm (mod \; 2) } \\
{\mathbf d.}\;  \tilde{n}_c^0+ n_c+\tilde{n}_s^0+n_s=0 & \qquad {\rm (mod \;4)} \\
\end{array}\; .
\end{equation}
Here conditions {\bf a}-{\bf c} imply that $n_{c(s)}$ and $\tilde{n}^0_{c(s)}$ have the same parity. Condition {\bf d} provides no constraint on the quantum number $n_{c(s)}$ by a similar reason as in the case of the NN boundary state.

Upon the modular transformation, the partition function corresponding to the DD boundary state can be evaluated as
\begin{equation}
Z_{DD,DD}(q)= g_{DD}^2 \frac{2 \sqrt{g_c g_s}}{\eta(q)^2} \sum^{\prime} q^{\frac{g_c}{2}(\tilde{m}_c)^2+\frac{g_s}{2}(\tilde{m}_s)^2}
\end{equation}
with the constraint $\tilde{m}_c+\tilde{m}_s=0$ (mod 2). Therefore, the scaling dimensions of the leading order perturbations can be read off as follows
\begin{equation}
\frac{g_c+g_s}{2}, \qquad 2g_c, \qquad 2g_s.
\end{equation}
This is in agreement with the previous results obtained using the DEBC method in Sec.~\ref{sec:DEBC}. Finally, the ground state degeneracy is given by $g_{DD}=(1/4 g_cg_s)^{1/4}$.

\begin{widetext}
\subsubsection{{\bf ND Boundary State}}

We first construct the reduced boundary state for the ND boundary condition
\begin{equation}
|ND\rangle = g_{ND}  \exp[\sum_{n=1}^\infty a_{n,c}^{L\dag}a_{n,c}^{R\dag}-\sum_{n=1}^\infty a_{n,s}^{L\dag}a_{n,s}^{R\dag}] \sum^\prime    |(\tilde{Q}_c,0)\rangle\otimes|(0,Q_s)\rangle,
\end{equation}
where the $g_{ND}$ is the ground state degeneracy. Now, the corresponding partition function can be computed as
\begin{eqnarray}
Z_{ND,ND} = \langle ND|e^{-lH_\beta^P}|ND\rangle
\label{eq:ReduNDparti1} =\frac{g_{ND}^2}{\eta(\tilde{q})^2} \sum^\prime \exp[-(\frac{l \pi g_c}{8 \beta}(\tilde{Q}_c)^2+\frac{l \pi}{8 \beta g_s}(Q_s)^2)]\; ,
\end{eqnarray}
with proper reduced gluing conditions indicated by the prime over the summation.

For obtaining the reduced gluing conditions on $\tilde{Q}_c$ and $Q_s$, we first observe that $\tilde{Q}^0_{c(s)}$, $\tilde{Q}_c$ and $Q_s$ are even integers. Hence, $\tilde{Q}^0_{c(s)}=2 \tilde{n}^0_{c(s)}$, $\tilde{Q}_c=2 \tilde{n}_c$ and $Q_s=2 n_s$. After some algebra,one finds the following reduced gluing condition,
\begin{equation}
\label{eq:RGC-ND-n}
\tilde{n}_c=n_s=0\qquad {\rm (mod \; 2) }
\end{equation}
Upon the modular transformation, the partition function becomes
\begin{equation}
Z_{ND,ND}(q)=\sqrt{\frac{g_s}{g_c}} \frac{g_{ND}^2}{\eta(q)^2} \sum q^{\frac{1}{2 g_c}(m_c)^2+\frac{g_s}{2}(\tilde{m}_s)^2} ,
\end{equation}
without any constraint on $m_c$ and $\tilde{m}_s$. Thus the scaling dimensions of boundary operators for the leading order perturbations are given by
\begin{equation}
\frac{g_s}{2}, \qquad \frac{1}{2g_c}\; .
\end{equation}
Again, this matches the results from the DEBC method in Sec.~\ref{sec:DEBC}. The ground state degeneracy becomes $(g_c/g_s)^{1/4}$.

\subsubsection{{\bf DN Boundary State}}

First, we construct the reduced boundary state for the DN boundary condition
\begin{equation}
|DN\rangle = g_{DN} \exp[-\sum_{n=1}^\infty a_{n,c}^{L\dag}a_{n,c}^{R\dag}+\sum_{n=1}^\infty a_{n,s}^{L\dag}a_{n,s}^{R\dag}] \sum^\prime |(0,Q_c)\rangle \otimes|(\tilde{Q}_s,0)\rangle
\end{equation}
with $g_{DN}$ defined as the ground state degeneracy. Then, the corresponding partition function can be written as
\begin{equation}
\label{eq:ReduDNparti1}Z_{DN,DN} = \langle DN|e^{-lH_\beta^P}|DN\rangle =\frac{g_{DN}^2}{\eta(\tilde{q})^2} \sum^\prime \exp[-(\frac{l \pi g_s}{8 \beta}(\tilde{Q}_s)^2+\frac{l \pi}{8 \beta g_c}(Q_c)^2)],
\end{equation}
\end{widetext}
with the proper gluing conditions. Again, all relevant momentum quantum number are even. Thus, $\tilde{Q}^0_{c(s)}=2 \tilde{n}^0_{n(s)}$, $Q_c=2 n_c$ and $\tilde{Q}_s=2 \tilde{n}_s$. One can show that the only reduced gluing condition for $n_c$ and $\tilde{n}_s$ is
\begin{equation}
n_c=\tilde{n}_s=0 \qquad {\rm (mod \; 2)}.
\end{equation}
Upon the modular transformation, the partition function becomes
\begin{equation}
Z_{DN,DN}(q)=  \sqrt{\frac{g_c}{g_s}} \frac{g_{DN}^2}{\eta(q)^2} \sum q^{\frac{g_c}{2}(\tilde{m}_c)^2+\frac{1}{2 g_s}(m_s)^2} ,
\end{equation}
with arbitrary integers, $\tilde{m}_c$ and $m_s$. Hence, the scaling dimensions of boundary operators for the lowest order perturbations can be read off as
\begin{equation}
\frac{g_c}{2}, \qquad \frac{1}{2g_s}.
\end{equation}
This is in agreement with the previous results from the DEBC method in Sec.~\ref{sec:DEBC}. The ground state degeneracy is given by $(g_s/g_c)^{1/4}$. 

In summary, we have computed explicitly in this subsection the spectrum of boundary operators and the ground state degeneracy for a junction of two quantum wires. The use from the onset of the $U_c(1)\times U_s(1)$ symmetry, corresponding to the total charge and spin conservation, leads to the reduced boundary states and reduced gluing conditions; these reduced relations largely simplify the computations. It is worthwhile to emphasize that this simplified scheme can be applied only to a system with the conservation of total charge and spin and without resonant tunneling.


\section{Junction of three quantum wires for spin-1/2 electrons-BCFT}
\label{sec:BCFT-three-junction}

In this section, we will apply the technique developed in the previous section to the case of a junction of three quantum wires for spin-1/2 electrons. We shall first derive the gluing conditions for a convenient basis and project out the center of mass modes of the charge and spin degrees of freedom. Then, the reduced partition functions for given boundary conditions and the scaling dimensions of boundary operators will be computed and used to determine the stability of the fixed points.

\subsection{Reduced gluing conditions}

Here, we start from the mode expansion of the bosons for a single quantum wire, Eq.(\ref{eq:Mode-Exp-varphi-int}) and Eq.(\ref{eq:Mode-Exp-theta-int}) with the gluing conditions in Eq.(\ref{eq:GluingCondition2}), and generalize the mode expansions and gluing conditions to another orthogonal basis:
\begin{eqnarray}
\nonumber \Phi_{c(s)}^0&=&\frac{1}{\sqrt{3}}(\varphi_{c(s)}^1+\varphi_{c(s)}^2+\varphi_{c(s)}^3)
\\
\label{eq:dynamical-field-3W}\Phi_{c(s)}^1&=&\frac{1}{\sqrt{2}}(\varphi_{c(s)}^1-\varphi_{c(s)}^2)
\\
\nonumber \Phi_{c(s)}^2&=&\frac{1}{\sqrt{6}}(\varphi_{c(s)}^1+\varphi_{c(s)}^2-2\varphi_{c(s)}^3),
\end{eqnarray}
and a corresponding set for $\theta$ fields.

The momentum quantum numbers of the total charge and spin modes follows the relations
\begin{subequations}
\begin{eqnarray}
Q_{c(s)}^0&=&Q_{c(s)}^1+Q_{c(s)}^2+Q_{c(s)}^3,
\\        
\tilde{Q}_{c(s)}^0&=&\tilde{Q}_{c(s)}^1+\tilde{Q}_{c(s)}^2+\tilde{Q}_{c(s)}^3.
\end{eqnarray}
\end{subequations}
Again, due to the conservation of total charge and spin, the N boundary condition should be imposed on the $\Phi_{c(s)}^0$ field. The corresponding boundary state always has the quantum number $Q_{c(s)}^0=0$. Hence, it is convenient to parameterize the vector of integers, $Q_{c(s)}^{i}$, as:
\begin{eqnarray}
&&(Q_{c(s)}^1,Q_{c(s)}^2,Q_{c(s)}^3) \nonumber
\\
&\equiv& m_{c(s)}^1(0,1,-1)+m_{c(s)}^2(-1,0,1), \label{eq:ParamQ}
\end{eqnarray}
where $m_{c(s)}^{1(2)} \in {\mathbf Z}$. Consequently, $m_{c(s)}^1=Q^2_{c(s)}$ and $m^2_{c(s)}=-Q_{c(s)}^1$. On the other hand, $\tilde{Q}_{c(s)}^0$ can be nonzero. So it is convenient to parameterize the $\tilde{Q}_{c(s)}^i$ as
\begin{eqnarray}
&&(\tilde{Q}_{c(s)}^1,\tilde{Q}_{c(s)}^2,\tilde{Q}_{c(s)}^3) \nonumber
\\
&\equiv& n_{c(s)}^0(1,1,1)-n_{c(s)}^1(0,1,1)-n_{c(s)}^2(1,0,1). \label{eq:ParamtildeQ}
\end{eqnarray}
Therefore, $n_{c(s)}^i$ can be expressed in terms of $\tilde{Q}_{c(s)}^i$ as
\begin{eqnarray}
&&(n_{c(s)}^0,n_{c(s)}^1,n_{c(s)}^2)
\\
\nonumber &=& (\tilde{Q}_{c(s)}^1+\tilde{Q}_{c(s)}^2-\tilde{Q}_{c(s)}^3, -\tilde{Q}_{c(s)}^3+\tilde{Q}_{c(s)}^1, -\tilde{Q}_{c(s)}^3+\tilde{Q}_{c(s)}^2).
\end{eqnarray}
Hence the $n_{c(s)}^i$ provide a representation of quantum numbers in an alternative basis.

Now, let us investigate the corresponding gluing conditions of the new variables $m_{c(s)}^i$ and $n_{c(s)}^i$. From the gluing conditions in Eq.(\ref{eq:GluingCondition2}) and the definition of $Q_{c(s)}^0$ and $\tilde{Q}_{c(s)}^0$, one concludes following gluing condition,
\begin{subequations}
\begin{eqnarray}
Q_c^0=Q_s^0=\tilde{Q}_c^0=\tilde{Q}_s^0  \qquad {\rm(mod \; 2)},
\\
\tilde{Q}_c^0+\tilde{Q}_s^0+Q_c^0+Q_s^0=0 \qquad {\rm (mod \;4)}.
\end{eqnarray}
Because $Q_{c(s)}^0=0$, the following gluing conditions hold
\begin{eqnarray}
\label{eq:GC-3W-zero-mode-1}\tilde{Q}_c^0=\tilde{Q}_s^0&=&0  \qquad {\rm(mod \; 2)}, 
\\
\label{eq:GC-3W-zero-mode-2}\tilde{Q}_c^0+\tilde{Q}_s^0&=&0 \qquad {\rm (mod \;4)}.
\end{eqnarray}
\end{subequations}
In terms of $n_{c(s)}^i$, the first condition Eq.(\ref{eq:GC-3W-zero-mode-1}) becomes
\begin{equation}
\tilde{Q}_{c(s)}^0 = 3 n_{c(s)}^0-2(n_{c(s)}^1+n_{c(s)}^2) = 0 \qquad {\rm(mod \;2)}.
\end{equation}
This implies that $n_{c(s)}^0=2 p_{c(s)}^0$ are even integers, where $p_{c(s)}^0$ are arbitrary integers. By using the gluing conditions for each of the quantum wires Eq.(\ref{eq:GluingCondition2}) and the fact that $n_{c(s)}^0$ are even, one can prove the following gluing condition
\begin{equation}
\label{eq:GC-m-n-1}n_c^i=n_s^i=m_c^i=m_s^i \qquad{\rm (mod\;  2)},
\end{equation}
from the relations between $m_{c(s)}^i(n_{c(s)}^i)$ and $Q_{c(s)}^j(\tilde{Q}_{c(s)}^j)$. Combining the gluing conditions Eq.(\ref{eq:GC-3W-zero-mode-2}) and Eq.(\ref{eq:GC-m-n-1}), a new gluing condition emerges
\begin{equation}
\label{eq:GC-pc-ps}
p_c^0+p_s^0=0\qquad {\rm(mod\; 2)}. 
\end{equation}
Finally, two nontrivial gluing conditions for $m_{c(s)}^i$ and $n_{c(s)}^i$ arise 
\begin{widetext}
\begin{subequations}
\begin{eqnarray}
\tilde{Q}_c^{1(2)}+\tilde{Q}_s^{1(2)}+Q_c^{1(2)}+Q_s^{1(2)}=0 \; {\rm (mod \;4)} 
&\Longrightarrow& -m_c^{2(1)} +n_c^0-n_c^{2(1)} -m_s^{2(1)} +n_s^0-n_s^{2(1)}=0 \; {\rm (mod \;4)},
\\
&\Longrightarrow& m_c^{2(1)} + n_c^{2(1)} + m_s^{2(1)} + n_s^{2(1)}=0 \; \qquad \qquad \;\;\;\;\;  {\rm (mod \;4)},
\end{eqnarray}
\end{subequations}
where the third equality holds because $n_c^0+n_s^0=2(p_c^0+p_s^0)= 0$ (mod 4). 

In summary, we found the following gluing conditions for the $n_{c(s)}^i$ and $m_{c(s)}^i$
\begin{equation}
\label{eq:GC-m-n}
\begin{array}{cc}
{\mathbf a.}\;  n_{c(s)}^0=2 p_{c(s)}^0,   \qquad p_c^0+p_s^0=0 \qquad  & \qquad  {\rm (mod \; 2)} \\
{\mathbf b.} \;  n_c^j=n_s^j=m_c^j=m_s^j  \qquad\qquad\qquad \;\;  & \qquad {\rm (mod \; 2) } \\
{\mathbf c.} \;  m_s^1+m_c^1+n_s^1+n_c^1=0 \qquad\qquad \;\;\;  & \qquad {\rm (mod \; 4) } \\
{\mathbf d.}\;  m_s^2+m_c^2+n_s^2+n_c^2=0  \qquad \qquad \;\;\; & \qquad {\rm (mod \;4)} \\
\end{array}\; .
\end{equation}
In particular, $n^0_{c(s)}$s disentangle from the rest of quantum numbers and do not have nontrivial gluing condition.

\subsection{Mode Expansions and center of mass mode projection}


We first define a vector field and a conjugate vector field, representing the dynamical boson fields,
\begin{equation}
\vec{\Phi}_{c(s)} = (\Phi_{c(s)}^1,\Phi_{c(s)}^2), \qquad \vec{\Theta}_{c(s)} = (\Theta_{c(s)}^1,\Theta_{c(s)}^2)
\end{equation}
where $\Phi_{c(s)}^{1(2)}$ are defined in Eq.(\ref{eq:dynamical-field-3W}). Using the definition of the dynamical field in Eq.(\ref{eq:dynamical-field-3W}), the periodicity along the spatial direction follows
\begin{subequations}
\begin{eqnarray}
\Delta \vec{\Phi}_{c(s)}&=& \vec{\Phi}_{c(s)}(\beta,t)- \vec{\Phi}_{c(s)}(0,t)= \sqrt{2} \pi (n_{c(s)}^1 \frac{\vec{R}_1}{2} +n_{c(s)}^2 \frac{\vec{R}_2}{2}),
\\
\Delta \vec{\Theta}_{c(s)}&=& \vec{\Theta}_{c(s)}(\beta,t)- \vec{\Theta}_{c(s)}(0,t)= \sqrt{2} \pi (m_{c(s)}^1 \vec{K}_1 +m_{c(s)}^2 \vec{K}_2),
\end{eqnarray}
\end{subequations}
where $n_{c(s)}^i$ and $m_{c(s)}^i$ are defined in the previous subsection and $K_{1(2)}$ and $R_{1(2)}$ are defined as
\begin{eqnarray}
\vec{K}_1&=& (-\frac{1}{2},+\frac{\sqrt{3}}{2}),\qquad  \vec{R}_1=\frac{2}{\sqrt{3}} (\vec{K}_1 \times \hat{z})=(+1 , +\frac{\sqrt{3}}{3}),
\\
\vec{K}_2&=&(-\frac{1}{2},-\frac{\sqrt{3}}{2}), \qquad \vec{R}_2=\frac{2}{\sqrt{3}} (\vec{K}_2 \times \hat z)=(-1 , +\frac{\sqrt{3}}{3}).
\end{eqnarray}
Hence the mode expansion of the two-component boson field becomes
\begin{eqnarray}
\nonumber \vec{\Phi}_{c(s)}&=&\vec{\hat{\Phi}}_{0,c(s)}+\frac{2 \pi}{ \beta} [ \frac{1}{\sqrt{2} g_{c(s)} } (\sum_{j=1,2} m^j_{c(s)} \vec{K}_j ) \, t+ \frac{1}{ \sqrt{2} } (\sum_{j=1,2} n^j_{c(s)} \frac{\vec{R}_j}{2} ) \, x]
\\
&+&\frac{1}{\sqrt{2 g_{c(s)} }} \sum_{n=1}^{\infty}\frac{1}{\sqrt{n}}[\vec{a}^{R}_{n,c(s)} e^{-i\frac{2 \pi n}{\beta} x_{-}} + \vec{a}^{L}_{n,c(s)}e^{-i\frac{2 \pi n}{\beta} x_{+}}+ h.c.]\; .
\end{eqnarray}
where the integers $n_{c(s)}^i$ and $m_{c(s)}^i$ are restricted by the gluing conditions. The corresponding energy is 
\begin{equation}
H^P_{\beta} = \frac{2\pi}{\beta} \left[ \frac{1}{ 4 g_c} | \sum_i m^i_c \vec{K}_i  |^2 +\frac{g_c}{4} |\sum_i n_c^i \frac{\vec{R}_i}{2}|^2 +  \frac{1}{ 4 g_s}|\sum_i m^i_s \vec{K}_i  |^2 +\frac{g_s}{4} |\sum_i n_s^i \frac{\vec{R}_i}{2}|^2  + \cdots  \right],
\end{equation}
where $\cdots$ represents the energy of the oscillator modes. 

Recalling the discussion of the last section for the case of the two quantum wires, the center of mass modes will only contribute a constant and can be projected out with a corresponding reduced gluing condition. Observe that $\tilde{Q}_{c(s)}^0=2(3 p^0_{c(s)}-n^1_{c(s)}-n^2_{c(s)} )$ implies that there are three classes of quantum numbers categorized by $\tilde{Q}_{c(s)}^0/2= -(n^1_{c(s)}+n^2_{c(s)} )$ (mod 3), we shall introduce auxiliary quantum numbers $k_{c(s)}= n^1_{c(s)}+n^2_{c(s)}$. Thus a general boundary state which is N with respect to $\Phi^0$, i.e $Q^0_{c(s)}=0$, takes the form
\begin{equation}
|B\rangle = \sum^{\prime}_{k_{c(s)}=-1,0,1} ( | N^0_{k_c} \rangle \otimes | N^0_{k_s}  \rangle )\otimes  \sum^{\prime}_{n^1_{c(s)}+n^2_{c(s)} =- k_{c(s)} \;({\rm mod}\; 3)}  ( |\psi(n^1_c,n^2_c,m^1_c,m^2_c) \rangle \otimes  |\psi(n^1_s,n^2_s,m^1_s,m^2_s) \rangle )
\end{equation}
where the prime over the summation indicates the gluing conditions. With the gluing condition, the boundary states corresponding to the $\Phi^0$ fields are given
\begin{equation}
| N^0_{k_c} \rangle \otimes | N^0_{k_s}  \rangle = (9 g_c g_s)^{1/4}\sum_{p^0_c+p^0_s=0 \;({\rm mod}\; 2)} | 2 (3 p^0_c +k_c), 0 \rangle \otimes | 2 (3 p^0_s +k_s), 0\rangle. 
\end{equation}
Hence the corresponding ``partial Neumann'' partition functions are readily calculated
\begin{eqnarray}
Z_{N^0_{k_c} N^0_{k_s}} &=& \frac{3\sqrt{g_c g_s} }{\eta(\tilde{q})^2} \sum_{p^0_c+p^0_s=0 \;({\rm mod}\; 2)} \tilde{q}^{\left[ g_c(3 p_c+k_c)^2/12 + g_s(3 p_s+k_s)^2/12  \right]}
\\
&=& \frac{1 }{\eta(q)^2} \sum_{Q^0_c(s)}^\prime \exp[- i \frac{\pi}{3} (k_cQ^0_c+k_sQ^0_s)] q^{ \frac{(Q^0_c)^2}{12 g_c}+ \frac{(Q^0_s)^2}{12 g_s}},
\end{eqnarray}
with the constraint $Q_c^0+Q_s^0=0$ (mod 2). As in the case of two quantum wires, this part of the partition function corresponds to boundary operators changing the total charge and spin of the system. Hence, $Z_{N^0_{k_c} N^0_{k_s}}$ can be projected out and replaced by unity. Now, the dimensions of all primary boundary operators involving only the dynamical fields can be obtained by using a reduced boundary state, which lives in the reduced Hilbert space of the 2-component boson fields, $\vec{\Phi}_{c(s)}$. Thus, the reduced boundary state becomes
\begin{equation}
|B\rangle \to \sum^{\prime}  ( |\psi(n^1_c,n^2_c,m^1_c,m^2_c) \rangle \otimes  |\psi(n^1_s,n^2_s,m^1_s,m^2_s) \rangle ),
\end{equation}
with the constraints given in Eq.(\ref{eq:GC-m-n}). Note that there is no gluing conditions between $p_{c(s)}^0$ and the rest of quantum numbers, hence we can simply ignore $p_{c(s)}^0$ and take the rest of gluing conditions as the constraint for the reduced boundary state.

\end{widetext}

\subsection{Reduced partition functions and the dimensions of boundary operators}

We first recall the reduced gluing conditions for the quantum numbers $n_{c(s)}^j$ and $m_{c(s)}^j$
\begin{equation}
\label{eq:RGC-m-n}
\begin{array}{cc}
{\mathbf a.} \;  n_c^j=n_s^j=m_c^j=m_s^j  \qquad \;\;  & \qquad {\rm (mod \; 2) } \\
{\mathbf b.} \;  m_s^j+m_c^j+n_s^j+n_c^j=0  \;\;\;\;  & \qquad {\rm (mod \; 4) } \\
\end{array},
\end{equation}
where $j=1,2$. In obtaining the reduced partition functions, these RGCs play crucial roles and lead to the nontrivial scaling dimensions of the operators. We shall investigate below the scaling dimensions of boundary operators for given boundary conditions. 

\subsubsection{{\bf NN boundary condition}}

As discussed in the last section, the corresponding boundary state of N BC has quantum numbers $Q_{c(s)}^{1(2)}=0$. This implies $m_{c(s)}^{1(2)}=0$ and the NN boundary state is given by
\begin{widetext}
\begin{equation}
|NN \rangle = g_{NN} \exp[\sum_{n=1}^\infty \vec{a}_{n,c}^{L\dag} \vec{a}_{n,c}^{R\dag}+\sum_{n=1}^\infty  \vec{a}_{n,s}^{L\dag} \vec{a}_{n,s}^{R\dag}] \sum^\prime |(n_c^1,n_c^2,0,0)\rangle \otimes|(n_s^1,n_s^2,0,0)\rangle,
\end{equation}
where the prime over the summation indicates the gluing conditions and $g_{NN}$ is the ground state degeneracy. The corresponding partition function is calculated as
\begin{equation}
Z_{NN} =\langle NN| e^{-l H_\beta^P}| NN\rangle =\frac{g_{NN}^2}{\eta(\tilde{q})^4}  \sum^{\prime} \exp( - \frac{\pi l}{8 \beta} \left[ g_c |\sum_j n_c^j \vec{R}_j|^2+g_s|\sum_j n_s^j \vec{R}_j|^2 \right] ).
\end{equation}

Because $m_{c(s)}^{1(2)}=0$, $n_{c(s)}^{1(2)}$s are even using the condition {\bf a} of Eq.(\ref{eq:RGC-m-n}), hence we conclude that $n_{c(s)}^{1(2)}=2 h_{c(s)}^{1(2)}$ with arbitrary integers for $h_{c(s)}^{1(2)}$. Rewriting the condition {\bf b} of RGC in terms of $h_{c(s)}^{1(2)}$, one obtains
\begin{equation}
 h_s^j+h_c^j=0  \qquad {\rm (mod \; 2), \; for \;  j=1,2 }.
\end{equation}
It is evident that $h_s^j$ and $h_c^j$ have the same parity. Hence, there are four plausible combinations of $(h_c^1,h_s^1,h_c^2,h_s^2)$ categorized by even or odd integers and listed below 
\begin{equation}
(h_c^1,h_s^1,h_c^2,h_s^2)=
\left\{
\begin{split}
& (e,e,e,e) \dots {\bf I} \\
& (o,o,o,o) \dots {\bf II}\\
& (e,e,o,o) \dots {\bf III}\\
& (o,o,e,e) \dots {\bf IV}
\end{split}\right.
\end{equation}
where $e$ stands for even while $o$ stands for odd. The total partition function is the sum of the partition functions corresponding to each combination above.

In case {\bf I}, we can introduce new integer quantum numbers $h_{c(s)}^{j}=2 \tilde{l}_{c(s)}^{j}$ and compute the corresponding partition function
\begin{eqnarray}
Z^{I}_{NN} = \frac{g_{NN}^2} {\eta(\tilde{q})^4}  \sum^{\prime} \exp[ - \frac{2 \pi l}{ \beta}  (g_c |\sum_j \tilde{l}_c^j \vec{R}_j|^2+g_s|\sum_j \tilde{l}_s^j \vec{R}_j|^2) ] = \frac{3}{4 g_c g_s}\frac{g_{NN}^2} {\eta(q)^4} \sum_{t_{c(s)}^j} q^{  (\frac{1}{2g_c} |\sum_j t_c^j \vec{K}_j|^2+\frac{1}{2 g_s} |\sum_j t_s^j \vec{K}_j|^2) },
\end{eqnarray}
where the second equality is obtained by multi-dimensional modular transformation, Appendix~\ref{sec:Modular-Trans}, and $t_{c(s)}^j$ are arbitrary integers. Similarly, one can execute the calculation for case {\bf II} by introducing $h_{c(s)}^{j}=2 \tilde{l}_{c(s)}^{j}+1$. Upon the modular transformation, the partition function is given by
\begin{equation}
Z^{II}_{NN} = \frac{3}{4 g_c g_s}\frac{g_{NN}^2} {\eta(q)^4} \sum_{t_{c(s)}^j} q^{  (\frac{1}{2g_c} |\sum_j t_c^j \vec{K}_j|^2+\frac{1}{2 g_s} |\sum_j t_s^j \vec{K}_j|^2) } \times e^{-i \pi ( t^1_c+t^1_s+t^2_c+t^2_s )},
\end{equation}
with arbitrary integers $t_{c(s)}^j$. For cases {\bf III} and {\bf IV}, the similar parameterization of $h_{c(s)}^{j}$ leads to  the following portions of partition function
\begin{eqnarray}
Z^{III}_{NN} &=& \frac{3}{4 g_c g_s}\frac{g_{NN}^2} {\eta(q)^4} \sum_{t_{c(s)}^j} q^{  (\frac{1}{2g_c} |\sum_j t_c^j \vec{K}_j|^2+\frac{1}{2 g_s} |\sum_j t_s^j \vec{K}_j|^2) } \times e^{-i \pi ( t^1_c+t^1_s )},
\\
Z^{IV}_{NN} &=& \frac{3}{4 g_c g_s}\frac{g_{NN}^2} {\eta(q)^4} \sum_{t_{c(s)}^j} q^{  (\frac{1}{2g_c} |\sum_j t_c^j \vec{K}_j |^2+\frac{1}{2 g_s} |\sum_j t_s^j \vec{K}_j|^2) } \times e^{-i \pi ( t^2_c+t^2_s)}.
\end{eqnarray}
\end{widetext}

By adding contributions from each part, the full partition function is given by
\begin{equation}
Z_{NN} =\frac{3 g_{NN}^2} { g_c g_s\eta(q)^4} \sum^{\prime} q^{  (\frac{|\sum_j t_c^j \vec{K}_j|^2}{2g_c} +\frac{|\sum_j t_s^j \vec{K}_j|^2}{2 g_s} ) }.
\end{equation} 
Observe that the unit vectors $\vec{K}_j$ form a triangular lattice. In general, the scaling dimensions can be calculated by finding the length square of linear combinations of $\vec{K}_j$ with the constraints $t_c^j+t_s^j=0$ (mod 2) for both $j=1,2$. The dimensions of the boundary operators corresponding to the leading order perturbation is given
\begin{equation}
\Delta_{NN}: \; \frac{1}{2 g_c}+\frac{1}{2 g_s}, \qquad \frac{2}{g_c}, \qquad \frac{2}{g_s}\; ,
\end{equation}
which are the same as that for junction of two quantum wires with NN BC and agree with the results from the DEBC method. Moreover, the ground state degeneracy is $g_{NN} =\sqrt{g_c g_s /3}$.

\begin{widetext}
\subsubsection{{\bf DD boundary condition}}

The corresponding boundary state with DD BC takes the form with $n^{1(2)}_{c(s)}=0$
\begin{equation}
|DD \rangle = g_{DD} \exp[- ( \sum_{n=1}^\infty \vec{a}_{n,c}^{L\dag} \vec{a}_{n,c}^{R\dag}+\sum_{n=1}^\infty  \vec{a}_{n,s}^{L\dag} \vec{a}_{n,s}^{R\dag} ) ] \sum^\prime |(0,0,m_c^1,m_c^2)\rangle \otimes|(0,0,m_s^1,m_s^2)\rangle,
\end{equation}
with the proper gluing conditions indicated by the prime over the summation. The diagonal partition function is
\begin{equation}
Z_{DD} =\langle DD | e^{-l H_\beta^P}| DD \rangle = \frac{g_{DD}^2}{\eta(\tilde{q})^4} \sum^{\prime} \exp[ -\frac{2 \pi l}{4 \beta} ( \frac{1}{g_c}|\sum_i m_c^i \vec{K}_i |^2 +\frac{1}{g_s} |\sum_i m_s^i \vec{K}_i |^2 ) ]. 
\end{equation}

Let us investigate the reduced gluing conditions here. Again, the condition {\bf a} of Eq.(\ref{eq:RGC-m-n}) and $n_{c(s)}^{1(2)}=0$ lead to even $m_{c(s)}^{j}$, hence $m_{c(s)}^{j}=2 \omega_{c(s)}^j$ for arbitrary integers $\omega_{c(s)}^j$. Then condition {\bf b} of RGC can be written in terms of $\omega_{c(s)}^{j}$ as
\begin{equation}
\omega_{c}^{j}+\omega_{s}^{j}=0 \qquad  {\rm (mod \; 2), for \; j=1,2 }.
\end{equation}
Similarly to the case of the NN BC, there are four possible combinations of $(\omega_c^1,\omega_s^1,\omega_c^2,\omega_s^2)$ categorized by even or odd integers and listed below
\begin{equation}
(\omega_c^1,\omega_s^1,\omega_c^2,\omega_s^2)=
\left\{
\begin{split}
& (e,e,e,e) \dots {\bf I} \\
& (o,o,o,o) \dots {\bf II}\\
& (e,e,o,o) \dots {\bf III}\\
& (o,o,e,e) \dots {\bf IV}
\end{split}\right. .
\end{equation}

Again, we will investigate the contributions to the partition functions of each combination. In case {\bf I}, we can introduce new integer quantum numbers $\omega^j_{c(s)}= 2 l_{c(s)}^j$ and compute the corresponding part of the partition function
\begin{equation}
Z^{I}_{DD} = \frac{g_{DD}^2} {\eta(\tilde{q})^4}  \sum^{\prime} \exp[ - \frac{8 \pi l}{ \beta}  (\frac{1}{g_c} |\sum_j l_c^j \vec{K}_j|^2+\frac{1}{g_s}|\sum_j l_s^j \vec{K}_j|^2) ] = \frac{g_c g_s }{12 }\frac{g_{DD}^2} {\eta(q)^4} \sum_{\tilde{t}_{c(s)}^j} q^{  (\frac{g_c}{8} |\sum_j \tilde{t}_c^j \vec{R}_j|^2+\frac{g_s}{8} |\sum_j \tilde{t}_s^j \vec{R}_j|^2) },
\end{equation}
for arbitrary integers $\tilde{t}_{c(s)}^j$. Since $\vec{R}_i$ forms a triangular lattice with the lattice spacing $|\vec{R}_i |= \frac{2}{\sqrt{3}}$, it is convenient to introduce scaled vectors $\vec{R'}_i= \frac{\sqrt{3}}{2} \vec{R}_i $ with unit length, $|\vec{R'}_i|=1$. Hence the partition functions in terms of these unit vectors are given by
\begin{equation}
Z^{I}_{DD} = \frac{g_c g_s }{12 }\frac{g_{DD}^2} {\eta(q)^4} \sum_{\tilde{t}_{c(s)}^j} q^{  (\frac{g_c}{6} |\sum_j \tilde{t}_c^j \vec{R'}_j|^2+\frac{g_s}{6} |\sum_j \tilde{t}_s^j \vec{R'}_j|^2) }.
\end{equation}
For other cases, we can introduce new quantum numbers $l_{c(s)}^j$ such that $\omega^j_{c(s)}= 2 l_{c(s)}^j+1$ for odd integers and $\omega^j_{c(s)}= 2 l_{c(s)}^j$ for even integers. Hence the partition functions for each case are given by
\begin{subequations}
\begin{eqnarray}
Z^{II}_{DD} &=& \frac{g_c g_s }{12 }\frac{g_{DD}^2} {\eta(q)^4} \sum_{\tilde{t}_{c(s)}^j} q^{  (\frac{g_c}{6} |\sum_j \tilde{t}_c^j \vec{R'}_j|^2+\frac{g_s}{6} |\sum_j \tilde{t}_s^j \vec{R'}_j|^2) } e^{-i \pi ( \tilde{t}^1_c+ \tilde{t}^1_s+ \tilde{t}^2_c+ \tilde{t}^2_s )},
\\
Z^{III}_{DD} &=& \frac{g_c g_s }{12 }\frac{g_{DD}^2} {\eta(q)^4} \sum_{\tilde{t}_{c(s)}^j} q^{  (\frac{g_c}{6} |\sum_j \tilde{t}_c^j \vec{R'}_j|^2+\frac{g_s}{6} |\sum_j \tilde{t}_s^j \vec{R'}_j|^2) } e^{-i \pi (  \tilde{t}^1_c+ \tilde{t}^1_s )},
\\
Z^{IV}_{DD} &=& \frac{g_c g_s }{12 }\frac{g_{DD}^2} {\eta(q)^4} \sum_{\tilde{t}_{c(s)}^j} q^{  (\frac{g_c}{6} |\sum_j \tilde{t}_c^j \vec{R'}_j|^2+\frac{g_s}{6} |\sum_j \tilde{t}_s^j \vec{R'}_j|^2) } e^{-i \pi ( \tilde{t}^2_c+ \tilde{t}^2_s )}. 
\end{eqnarray}
\end{subequations}
\end{widetext}

Now, let us add all contributions and obtain the full partition function
\begin{equation}
Z_{DD} =\frac{g_c g_s g_{DD}^2} {3 \eta(q)^4} \sum^{\prime} q^{  (\frac{g_c |\sum_j \tilde{t}_c^j \vec{R'}_j|^2}{6} +\frac{g_s |\sum_j \tilde{t}_s^j \vec{R'}_j|^2 }{6} ) },
\end{equation}
with the constraints $\tilde{t}_c^j+\tilde{t}_s^j=0$ (mod 2) for $j=1,2$. The scaling dimensions of boundary operators for leading order perturbations can be read off directly from the partition function
\begin{equation}
\Delta_{DD}: \; \frac{g_c}{6}+\frac{g_s}{6}, \qquad \frac{2 g_c}{3}, \qquad \frac{2 g_s}{3}. 
\end{equation}
Again, the results agree with that obtained by the DEBC method. Finally, the ground state degeneracy of the DD boundary state reads $g_{DD}= \sqrt{3/ g_c g_s}$.

\subsubsection{{\bf ND and DN boundary condition}}

We will focus on the ND BC first. (The scaling dimensions and partition function corresponding to the DN BC can be calculated by exchanging the charge and spin sectors of the ND case.) The boundary state corresponding to the ND BC can be constructed with $m_c^{1(2)}=0$ and $n_s^{1(2)}=0$
\begin{widetext}
\begin{equation}
|ND \rangle = g_{ND} \exp[  +\sum_{n=1}^\infty \vec{a}_{n,c}^{L\dag} \vec{a}_{n,c}^{R\dag}-\sum_{n=1}^\infty  \vec{a}_{n,s}^{L\dag} \vec{a}_{n,s}^{R\dag}  ] \sum^\prime |(n_c^1,n_c^2,0,0)\rangle \otimes|(0,0,m_s^1,m_s^2)\rangle,
\end{equation}
with proper constraints and ground state degeneracy $g_{ND}$. Since the gluing condition { \bf a} in Eq.(\ref{eq:RGC-m-n}) implies that $n_c^{1(2)}$ and $m_s^{1(2)}$ are even when $m_c^{1(2)}= n_s^{1(2)}=0$, we shall parameterize $n_c^{1(2)}= 2 h_c^{1(2)}$ and $m_s^{1(2)}=2 \omega_s^{1(2)}$. The partition function thus can be computed, 
\begin{equation}
Z_{ND} =\langle ND | e^{-l H_\beta^P}| ND \rangle = \frac{g_{ND}^2}{\eta(\tilde{q})^4} \sum^{\prime} \exp[ -\frac{2 \pi l}{\beta} ( \frac{g_c}{4}|\sum_i h_c^i \vec{R}_i |^2 +\frac{1}{g_s} |\sum_i \omega_s^i \vec{K}_i |^2 ) ],
\end{equation}
\end{widetext}
with the gluing conditions $h_c^j+\omega_s^j=0$ (mod 2) for $j=1,2$. 

After some algebra, similar to the case of the DD and NN boundary conditions, one obtains the full partition function
\begin{equation}
\label{eq:partition-func-ND}Z_{ND} =\frac{g_s g_{ND}^2} { g_c \eta(q)^4} \sum^{\prime} q^{  (\frac{|\sum_j t_c^j \vec{K}_j|^2}{2 g_c} +\frac{g_s |\sum_j \tilde{t}_s^j \vec{R'}_j|^2 }{6} ) },
\end{equation}
with the constraints $t_c^j+\tilde{t}_s^j =0$ (mod 2) for $j=1,2$. Then the leading order boundary operators have the scaling dimensions
\begin{equation}
\Delta_{ND}: \; \frac{1}{2 g_c}+\frac{g_s}{6},\qquad \frac{2}{g_c}, \qquad \frac{2g_s}{3},
\end{equation}
which match the results from the DEBC scheme. Moreover, the ground state degeneracy reads $g_{ND}= \sqrt{g_c/g_s}$.

Due to the symmetric structure between the charge and spin part of the boundary states, one can exchange $c \leftrightarrow s$ in Eq.(\ref{eq:partition-func-ND}) and the corresponding constraints to obtain the partition function given the DN BC. Then, the dimensions of boundary operators of leading order perturbations with the DN BC are given by
\begin{equation}
\Delta_{DN}: \; \frac{1}{2 g_s}+\frac{g_c}{6},\qquad \frac{2}{g_s}, \qquad \frac{2g_c}{3}.
\end{equation}
The results agree with that of the DEBC scheme. Further, the ground state degeneracy reads $g_{DN}= \sqrt{g_s/g_c}$.

\subsubsection{{\bf $\chi_{\pm}\chi_{\pm}$ and $\chi_{\pm}\chi_{\mp}$ boundary conditions}}

We will first study the case of $\chi_+\chi_+$ boundary condition. The $\chi_-\chi_-$ boundary condition can be calculated in the same manner. Indeed, one can show that the relevant scaling dimensions have exactly the same structure for both $\chi_\pm \chi_\pm$ boundary conditions. Then, we will comment on the case of $\chi_{\pm}\chi_{\mp}$.

Generically, a comformally invariant boundary condition can be expressed as
\begin{equation}
\vec{\phi}^R= R_{\xi} \vec{\phi}^L, \;  {\rm with} \; R_\xi =\left(\begin{array}{cc}\cos \xi & -\sin \xi \\ \sin \xi & \cos \xi \end{array}\right)\; . 
\end{equation}
Since the winding along the boundary can be written as 
\begin{subequations}
\begin{eqnarray}
\Delta \vec{\phi}_{c(s)}^L &=& \frac{1}{2}(\sqrt{g_{c(s)}} \Delta \vec{\Phi}_{c(s)}+ \frac{\Delta \vec{\Theta}_{c(s)}}{ \sqrt{g_{c(s)} } } ),
\\
\Delta \vec{\phi}_{c(s)}^R &=& \frac{1}{2}(\sqrt{g_{c(s)}} \Delta \vec{\Phi}_{c(s)} - \frac{\Delta \vec{\Theta}_{c(s)} }{ \sqrt{g_{c(s)} } } ),
\end{eqnarray}
\end{subequations}
arbitrary conformal boundary conditions satisfy
\begin{eqnarray}
\nonumber
\sqrt{g_{c(s)}} \Delta \vec{\Phi}_{c(s)}- \frac{\Delta \vec{\Theta}_{c(s)}}{ \sqrt{g_{c(s)} } } 
= R_{\xi} (\sqrt{g_{c(s)}} \Delta \vec{\Phi}_{c(s)}+ \frac{\Delta \vec{\Theta}_{c(s)}}{ \sqrt{g_{c(s)} } } ).
\end{eqnarray}
Observe that $\Delta \vec{\Theta}_{c(s)}=0$ for the NN BC leads to $\xi=0$ and $\Delta \vec{\Phi}_{c(s)}=0$ for the DD BC leads to $\xi=\pi$. For $\xi \ne 0,\;  \pi$, it is clear that we need both $\Delta \vec{\Phi}_{c(s)}$ and $\Delta \vec{\Theta}_{c(s)}$ nonvanishing. Moreover, $R_\xi$ is a rotation matrix which preserves the length of the vector, hence $\Delta \vec{\Phi}_{c(s)}$ and $\Delta \vec{\Theta}_{c(s)}$ have to be mutually orthogonal. For satisfying this constraint, the general oscillator vacua can be constructed from 
\begin{equation}
|(\alpha_c \eta_c^1, \alpha_c \eta_c^2, \beta_c \eta_c^1, \beta_c \eta_c^2 ) \rangle \otimes |(\alpha_s \eta_s^1, \alpha_s \eta_s^2, \beta_s \eta_s^1, \beta_s \eta_s^2 ) \rangle \nonumber ,
\end{equation}
for arbitrary integers $\eta_{c(s)}^{1(2)}$ with proper gluing conditions on $\alpha$ and $\beta$. For instance, the condition {\bf a} in Eq.(\ref{eq:RGC-m-n}) leads to $\alpha_{c(s)}=\beta_{c(s)}$ (mod 2). Moreover, the chiral rotation angle is fixed by $(\alpha_{c(s)},\beta_{c(s)})$ as
\begin{equation}
\tan\frac{\xi_{c(s)}}{2} =\frac{\beta_{c(s)}}{\alpha_{c(s)}} \frac{\sqrt{3}}{g_c(s)}.
\end{equation}
Notice that the chiral rotation angle is ``quantized''.

As shown in Ref.~\cite{COA}, the $\chi_+$ fixed point corresponds to the choice of $\alpha_{c(s)}= \beta_{c(s)} =1$ while the $\chi_-$ fixed point corresponds to the choice of $\alpha_{c(s)}= -\beta_{c(s)} =1$. Hence, the boundary state corresponding to the $\chi_+\chi_+$ boundary condition can be constructed as
\begin{widetext}
\begin{equation}
|\chi_+\chi_+ \rangle = g_{\chi_+ \chi_+} \exp[  \sum_{n=1}^\infty \vec{a}_{n,c}^{R\dag} \cdot R_{\xi_c} \vec{a}_{n,c}^{L\dag} + \sum_{n=1}^\infty  \vec{a}_{n,s}^{R\dag} \cdot R_{\xi_s}\vec{a}_{n,s}^{L\dag}  ] \sum^\prime |(\eta_c^1,\eta_c^2,\eta_c^1,\eta_c^2)\rangle \otimes|(\eta_s^1,\eta_s^2,\eta_s^1,\eta_s^2)\rangle ,
\end{equation}
where the prime over the summation indicates the gluing conditions. The diagonal partition function is given by
\begin{eqnarray}
\nonumber Z_{\chi_+ \chi_+}&=& \frac{g_{\chi_+ \chi_+}^2}{\eta(\tilde{q})^4} \sum^\prime \exp[ -\frac{2 \pi l}{\beta} ( \frac{g_c}{4}|\sum_i \eta_c^i \frac{\vec{R}_i}{2} |^2 +\frac{1}{4g_c} |\sum_i \eta_c^i \vec{K}_i |^2 ) ] \exp[ -\frac{2 \pi l}{\beta} ( \frac{g_s}{4}|\sum_i \eta_s^i \frac{\vec{R}_i}{2} |^2 +\frac{1}{4g_s} |\sum_i \eta_s^i \vec{K}_i |^2 ) ]
\\
\label{eq:partition-chi+chi+} &=& \frac{g_{\chi_+ \chi_+}^2}{\eta(\tilde{q})^4} \sum^\prime \exp[ -\frac{\pi l}{\beta} |\eta_c^1\vec{A}^1_{c}+ \eta_c^2 \vec{A}^2_{c}  |^2 ]
\exp[ -\frac{\pi l}{\beta} |\eta_s^1\vec{A}^1_{s}+ \eta_s^2 \vec{A}^2_{s}  |^2 ]\end{eqnarray}.
\end{widetext}
where the second equality hold because $\vec{K}_i \cdot \vec{R}_j=\epsilon_{ij}$ and the new vectors are defined as
\begin{equation}
\vec{A}^j_{c(s)}= \sqrt{g_{c(s)} } \frac{\vec{R}_j}{2 \sqrt{2}} + \frac{\vec{K}_j}{\sqrt{2 g_{c(s)}} }.
\end{equation}
These new vectors form a triangular lattice with lattice spacing 
\begin{equation}
|\vec{A}_{c(s)}^j|=\sqrt{ \frac{g_{c(s)} }{6} + \frac{1}{2 g_{c(s)}}  }.
\end{equation}

Using Eq.(\ref{eq:RGC-m-n}), there is only one constraint
\begin{equation}
\eta_c^j=\eta_s^j \;{\rm (mod\; 2), \qquad for\; } j=1,2 .
\end{equation}
Therefore, there are four possible combinations of the winding numbers classified by even or odd integers which are listed below
\begin{equation}
(\eta_c^1,\eta_s^1,\eta_c^2, \eta_s^2)=
\left\{
\begin{split}
& (e,e,e,e) \dots {\bf I} \\
& (o,o,o,o) \dots {\bf II}\\
& (e,e,o,o) \dots {\bf III}\\
& (o,o,e,e) \dots {\bf IV}
\end{split}\right. .
\end{equation}

Again, parameterizing the even and odd variables as $\eta=2 l$ and $\eta=2 l+1$, respectively, we are able to compute the diagonal partition function corresponding to each set. Upon a modular transformation, the partial partition function for each case is given by
\begin{widetext}
\begin{subequations}
\begin{eqnarray}
Z_{\chi_+\chi_+}^I&=& \frac{12 g_c g_s}{ ( 3+ g_c^2) ( 3+ g_s^2)}\frac{g_{\chi_+\chi_+}^2 }{\eta(q)^4} \sum_{t_{c(s)}^j \in Z } q^{| \sum_j t_c^j \vec{W}_c^j|^2 + | \sum_j t_s^j \vec{W}_s^j|^2},
\\
Z_{\chi_+\chi_+}^{II}&=& \frac{12 g_c g_s}{ ( 3+ g_c^2) ( 3+ g_s^2)}\frac{g_{\chi_+\chi_+}^2 }{\eta(q)^4} \sum_{t_{c(s)}^j \in Z } q^{| \sum_j t_c^j \vec{W}_c^j|^2 + | \sum_j t_s^j \vec{W}_s^j|^2} e^{- i \pi (t_c^1+t_s^1+t_c^2+t_s^2)},
\\
Z_{\chi_+\chi_+}^{III}&=& \frac{12 g_c g_s}{ ( 3+ g_c^2) ( 3+ g_s^2)}\frac{g_{\chi_+\chi_+}^2 }{\eta(q)^4} \sum_{t_{c(s)}^j \in Z } q^{| \sum_j t_c^j \vec{W}_c^j|^2 + | \sum_j t_s^j \vec{W}_s^j|^2} e^{- i \pi (t_c^2+t_s^2)},
\\
Z_{\chi_+\chi_+}^{IV}&=& \frac{12 g_c g_s}{ ( 3+ g_c^2) ( 3+ g_s^2)}\frac{g_{\chi_+\chi_+}^2 }{\eta(q)^4} \sum_{t_{c(s)}^j \in Z } q^{| \sum_j t_c^j \vec{W}_c^j|^2 + | \sum_j t_s^j \vec{W}_s^j|^2} e^{- i \pi (t_c^1+t_s^1)},
\end{eqnarray}
\end{subequations}
\end{widetext}
where $\vec{W}_{c}^j$ and $\vec{W}_{s}^j$ are the dual vectors of $\vec{A}_{c}^j$ and $\vec{A}_{s}^j$ respectively, and form two sets of dual triangular lattices with different lattice spacing, $\sqrt{\frac{2 g_c}{g_c^2+ 3}}$ and $\sqrt{\frac{2 g_s}{g_s^2+ 3}}$, respectively. With the identification of the ground state degeneracy as $ g_{\chi_+\chi_+}= \frac{ ( 3+ g_c^2) ( 3+ g_s^2)}{48 g_c g_s}$, the full partition function is
\begin{equation}
Z_{\chi_+\chi_+}= \frac{1 }{\eta(q)^4} \sum^\prime q^{| \sum_j t_c^j \vec{W}_c^j|^2 + | \sum_j t_s^j \vec{W}_s^j|^2},
\end{equation}
with the constraints $t_c^j+t_s^j=0$ (mod 2) for $j=1,2$. Hence the dimensions of the boundary operators can be read off directly and the leading order perturbations have the dimension
\begin{equation}
\Delta_{\chi_+ \chi_+} : \; \frac{2 g_c}{g_c^2+3} +\frac{2 g_s}{g_s^2+3}. 
\end{equation}

The only difference of the boundary state of the $\chi_{-}\chi_{-}$ boundary condition comes from the parameterization of $\alpha_{c(s)}=- \beta_{c(s)}=0$. Consequently, the partition function is similar to the case of the $\chi_+\chi_+$ BC with a minor variation,
\begin{equation}
\vec{A}_{c(s)}^j \to \vec{A'}_{c(s)}^j=  \sqrt{g_{c(s)} } \frac{\vec{R}_j}{2 \sqrt{2}} - \frac{\vec{K}_j}{ \sqrt{ 2 g_{c(s)}}  }.
\end{equation}
in Eq.(\ref{eq:partition-chi+chi+}). Also, the $\vec{A'}_{c(s)}^j$s form a triangular lattice with lattice spacing equal to $|\vec{A}_{c(s)}^j|$. Moreover, the constraint, $\eta_c^j+\eta_s^j=0$ (mod 2), still holds. Hence, we can conclude that the partition function upon the modular transformation has a similar structure to the $\chi_+\chi_+$ case. Hence the leading order perturbations have the same scaling dimension.

Since the scaling dimensions of the boundary operators are the same for the cases of $\chi_{\pm} \chi_{\pm}$ BCs, one may wonder if the $\chi_\pm \chi_\mp$ boundary conditions will have similar behaviors. However, one can see from the argument below that the scaling dimensions of the boundary operators are rather different. We have learned that the non-trivial scaling dimensions come from the gluing conditions. Even in the case of the same dual lattice structure and spacing, the gluing conditions may provide nontrivial constraints. In the case of the $\chi_\pm \chi_\mp$ BCs, the only gluing condition $\eta_{c(s)}^j=0$ (mod 4) will not lead to any constraint on the integer $t_{c(s)}^j$. As a result, the leading order perturbations have scaling dimensions
\begin{equation}
\Delta_{\chi_\pm \chi_\mp} : \; \frac{2 g_c}{g_c^2+3},\qquad \frac{2 g_s}{g_s^2+3}, 
\end{equation}
which is always smaller than one for any $g_{c(s)}$ and leads to an instability of the fixed point. So, we conclude that the $\chi_\pm \chi_\mp$ fixed points are not stable.

\subsubsection{{\bf $D_AD_A$ boundary condition}}

From the results of the DEBC approach, the asymmetric boundary conditions could be stable in some regions of the interaction parameter space. Hence, it becomes important to construct the corresponding boundary state for checking the instability of the $D_A D_A$ boundary condition. Without loss of generality, we choose to impose Dirichlet boundary condition on the dynamical field, $\Phi_{c(s)}^1=(\varphi_{c(s)}^1-\varphi_{c(s)}^2 )/\sqrt{2}$, between first and second wire, and Neumann boundary condition at the third wire. Indeed, this set of boundary conditions is equivalent to having D BC on $\Phi_{c(s)}^1$ and N BC on $\Phi_{c(s)}^2$. Using the parameters defined in this section, the D BC of the dynamical field leads to $n_{c(s)}^1=n_{c(s)}^2\equiv n_{c(s)}$ while the N BC of the $\varphi_{c(s)}^3$ leads to $m_{c(s)}^1=m_{c(s)}^2 \equiv m_{c(s)}$. Hence, the reduced boundary state can be constructed as
\begin{widetext}
\begin{equation}
|D_A D_A \rangle = g_{D_A D_A} \exp[  \sum_{n=1}^\infty \vec{a}_{n,c}^{R\dag} \cdot R_{A} \vec{a}_{n,c}^{L\dag} + \sum_{n=1}^\infty  \vec{a}_{n,s}^{R\dag} \cdot R_{A} \vec{a}_{n,s}^{R\dag}  ] \sum^\prime |(n_c,n_c,m_c,m_c)\rangle \otimes|(n_s,n_s,m_s,m_s)\rangle ,
\end{equation}
with the rotation matrix and the appropriate gluing conditions written as
\begin{equation}
R_A=\left(\begin{array}{cc}-1 & 0 \\0 & 1\end{array}\right)\;, \qquad
\begin{array}{cc}
{\mathbf a.}\;  n_c= n_s=m_c=m_s  \;\;\;\;\;\; & \qquad  {\rm (mod \; 2)} \\
{\mathbf b.}\;  n_c+ n_s+m_c+m_s=0 & \qquad {\rm (mod \;4)} \\
\end{array}\; .
\end{equation}
The partition function is given by
\begin{eqnarray}
Z_{D_A,D_A}&=&\langle D_AD_A| e^{-lH_\beta^P}| D_A D_A \rangle  
\\
&=&\frac{ g_{D_AD_A}^2 }{\eta(\tilde{q})^4} \sum^{\prime}  \exp[ -\frac{2 \pi l}{\beta} ( \frac{1}{4 g_c} |m_c \sum_i \vec{K}_i |^2 +\frac{g_c}{4} | \frac{n_c}{2} \sum_i \vec{R}_i |^2 ) ] \exp[ -\frac{2 \pi l}{\beta} ( \frac{1}{4 g_s} |m_s \sum_i \vec{K}_i |^2 +\frac{g_s}{4} | \frac{n_s}{2} \sum_i \vec{R}_i |^2 ) ] \nonumber
\\
&=&\frac{ g_{D_AD_A}^2 }{\eta(\tilde{q})^4} \sum^{\prime}  \exp[ -\frac{2 \pi l}{\beta} ( \frac{m_c^2}{4 g_c} +\frac{g_c n_c^2}{12} ) ] \exp[ -\frac{2 \pi l}{\beta} ( \frac{m_s^2}{4 g_s}  +\frac{g_s n_s^2 }{12}  ) ],  \nonumber
\end{eqnarray}
\end{widetext}
where we use $|\sum_i \vec{K}_i|^2=1$ and $| \sum_i \vec{R}_i/2|^2=1/3$ for the second equality and the prime over the summation indicates the constraints of the integers.

Now, we can separate $n_{c(s)}$ and $m_{c(s)}$ into two independent sets, all even or all odd,
\begin{equation}
\left\{
\begin{split}
& (a) \; n_{c(s)}=2 h_{c(s)},\qquad m_{c(s)}=2 \omega_{c(s)}
\\
& (b) \; n_{c(s)}=2 h_{c(s)}+1,\;\; m_{c(s)}=2 \omega_{c(s)}+1
\end{split}\right. ,
\end{equation}
with a constraint
\begin{equation}
h_c+\omega_c+h_s+\omega_s=0\qquad {\rm (mod \; 2)}.
\end{equation}
Hence the partition function can be decomposed in terms of new variables as
\begin{eqnarray}
Z_{D_A,D_A}&=&\frac{ g_{D_AD_A}^2 }{\eta(\tilde{q})^4}\{  \sum^{\prime}  \tilde{q}^{  \frac{\omega_c^2}{2 g_c} +\frac{g_c h_c^2}{6} + \frac{\omega_s^2}{2 g_s}  +\frac{g_s h_s^2 }{6} }
\\
&+& \tilde{q}^{  \frac{(2 \omega_c+1)^2 }{8 g_c} +\frac{g_c (2 h_c+1)^2}{24} + \frac{(2\omega_s+1)^2}{8 g_s}  +\frac{g_s (2h_s+1)^2 }{24} } \}.  \nonumber
\end{eqnarray}
Upon the modular transformation, one obtains
\begin{equation}
Z_{D_A,D_A}=\frac{3g^2_{D_AD_A}}{\eta(q)^4}\sum^{\prime}  q^{(\frac{g_c (\tilde{t}_c)^2}{8}+ \frac{3 (t_c)^2}{8 g_c}  +\frac{g_s (\tilde{t}_s)^2}{8}+ \frac{3 (t_s)^2}{8 g_s}  )   },
\end{equation}
with the constraint 
\begin{equation}
\begin{array}{cc}
{\mathbf a.}\;  t_c= t_s=\tilde{t}_c=\tilde{t}_s  \;\;\;\;\;\; & \qquad  {\rm (mod \; 2)} \\
{\mathbf b.}\;  t_c+ t_s+\tilde{t}_c+\tilde{t}_s=0 & \qquad {\rm (mod \;4)} \\
\end{array}\; .
\end{equation}
Then the scaling dimensions of the boundary operators corresponding to the leading order perturbations can be read off,

\begin{eqnarray}
\Delta_{D_AD_A}: && \frac{g_c^2+3}{8 g_c} +\frac{g_s^2+3}{8 g_s},
\nonumber\\
&& \frac{g_s}{2}+\frac{g_s}{2}, \qquad \frac{3}{2}(g_c^{-1} + g_s^{-1}), \nonumber
\\
&& \frac{g_c}{2}+\frac{3}{2g_s}, \qquad   \frac{g_s}{2}+\frac{3}{2g_c}, 
\nonumber\\
&& 2g_c,\qquad 2g_s.
\end{eqnarray}
These results agree with the conclusions from the DEBC method. Moreover, the ground state degeneracy is $g_{D_AD_A}= 1/\sqrt{3}$.

\begin{figure}
\includegraphics[width=0.80\linewidth]{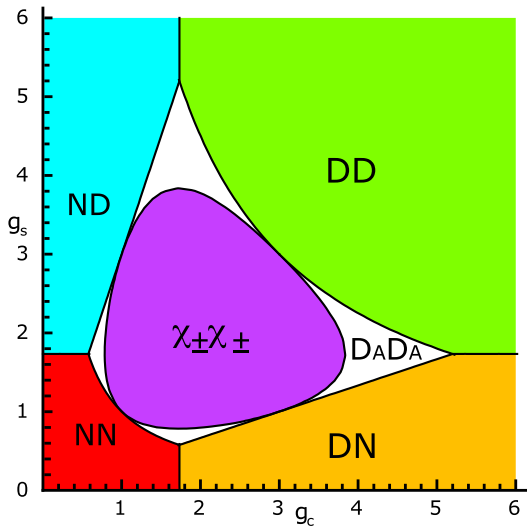}
\caption{[Color online] The areas with red, green, blue, orange, purple and white indicate the regions in parameter space where the ground state degeneracy corresponding to the NN, DD, ND, DN, $\chi_\pm\chi_\pm$ and $D_AD_A$ boundary conditions, respectively, is the minimum. The borders with black lines are determined by the condition that the ground state degeneracies for adjacent regions corresponding to two different boundary conditions are equal.}
\label{fig:phase-GSD}
\end{figure}

We conclude this section with the discussion about how to determine the instability of the phases by using the ground state degeneracy of the boundary states. Recall the ground state degeneracy of the different boundary states:
\begin{equation}
\label{eq:GSD}
\begin{split}
&g_{NN}=\sqrt{\frac{g_c g_s}{3}}, \qquad g_{DD}=\sqrt{\frac{3}{g_c g_s}},
\\
&g_{ND}=\sqrt{\frac{g_c}{g_s}},  \qquad g_{DN}=\sqrt{\frac{g_s}{g_c}},
\\
&g_{\chi_+\chi_+}= \frac{ ( 3+ g_c^2) ( 3+ g_s^2)}{48 g_c g_s}, \qquad g_{D_AD_A}= 1/\sqrt{3}.
\end{split}
\end{equation}
The universal non-integer ground state degeneracy $g_{B_cB_s}$ always
decreases under the renormalization from a less stable to a more
stable fixed point in the same bulk universality class~\cite{gtheorem}
(g-theorem). If there are two stable phases for a given value of
interaction parametres $(g_c,g_s)$, then an unstable fixed point must
lie in between these two stable points (with a value of $g_{B_cB_s}$
that is larger than those at the two stable fixed points). While one
cannot resolve to which fixed point one flows in these overlapping
regions of stability (it dependes on the strengths of the bare
couplings), it is instructive to look at the phase boundaries as
determined from the condition of minimal boundary entropy. The phases
with minimum ground state degeneracy $g_{B_cB_s}$ [computed from
Eq.~(\ref{eq:GSD})] for given $(g_c,g_s)$ are shown in
Fig.~\ref{fig:phase-GSD}. Comparing with the phase diagram proposed
based on the scaling dimensions of the leading order perturbations in
Fig.~\ref{fig:phase-All}, the borderlines in Fig.~\ref{fig:phase-GSD}
are located where there is overlap between two or more phases.

\section{Conclusion}
\label{sec:Conclusion}

In this paper we studied the problem of a junction of three quantum
wires for spin-1/2 electrons connected by a ring, through which a
magnetic flux can be applied. The bulk of the wires was formulated as
Tomonaga-Luttinger liquids with interaction parameters $g_c$ and $g_s$
(for charge and spin sectors, respectively). The problem was studied
using two different methods: delayed evaluation of boundary conditions
and boundary conformal field theory. These methods bypass the
difficulty that normally occurs with the inclusion of Klein factors to
ensure the proper fermionic statistics for different species of
fermions. We reached consistent results for the stability of the
phases obtained from the two different methods.

We computed the low energy and low temperature charge and spin
conductance tensors corresponding to the fixed points for the junction
as a function of the interaction parameters $g_{c(s)}$. These
conductance tensors, $G(g_c,g_s)$, presented in
Sec.~\ref{sec:DEBC-three-junction} (and summarized in
Sec.\ref{sec:results}), characterize the response of the junction to
the externally applied voltages. We have presented a simple one-to-one relation between the conductance tensor and the rotation matrix
$\mathcal{R}$ that encodes the types of boundary conditions in the
DEBC method.

When the Y-junctions are attached to Fermi liquid leads, the fixed
points are still controlled by the $g_c$ and $g_s$ in the
wires. However, the conductance tensor is altered due to the contact
resistances at the lead/wire interfaces. Similarly to what was found
in Ref~\cite{COA}, the conductance tensor in the presence of the leads
is the one determined by the appropriate BC (which is controlled by
the $g_c$ and $g_s$), but evaluated at $g_c=g_s=1$ instead. For
instance, in the case when the chiral fixed point is the stable one,
one plugs $g_c=g_s=1$ into Eq.~(\ref{eq:chi-Cond}). Interestingly, for
the chiral fixed point, the switching of the current in the presence
of the leads is then perfect, circulating the current from one lead
completely into one of the other two leads. Indeed, this
renormalization of the conductance tensor is the 3-wire analog of what
was found in Tarucha {\it et al.}'s experiments on Luttinger liquids
coupled to reservoirs~\cite{Tarucha} and explained theoretically in
Refs.~\cite{Maslov-Stone,Safi-Schulz}.

The phase diagram, as a function of the interaction parameters
$g_{c(s)}$, is contained in Fig.~\ref{fig:phase-All}. Among the
possible phases, we find one corresponding to a chiral fixed point
similar to the case of spinless electrons. In this phase, the flow of
current is sensitive to the flux through the ring, and we find that
the charge and spin degrees of freedom must circulate with the same
chirality.

We have also found that the inclusion of the spin degree of freedom
allows for the existence of a stable fixed point where current flows
only between two wires, while the third remains uncoupled. Such fixed
point was always unstable in the case of spinless electrons. Thus, in
more realistic models that include the electron spin into account, it
may be possible that controlling small anisotropies in Y-junctions of
quantum wires may lead to sensitive current switches.


\section*{Acknowledgements}

We would like to thank Ian Affleck and Masaki Oshikawa for several
illuminating discussions on this problem, and Armin Rahmanisisan for
his careful reading of the manuscript and many useful
suggestions. This work is supported by the DOE Grant DE-FG02-06ER46316.


\appendix

\renewcommand{\theequation}{A\arabic{equation}}
\setcounter{equation}{0}  
\section{Boundary conditions and conductance}  
\label{sec:conductance}

We will show in this section how the conductance tensor is extracted from the boundary conditions. Within the linear response theory, we obtain the Kubo formula for the conductance tensor of multiple wires introduced in Eq.(\ref{eq:Def-of-conduc-tensor}) as in~\cite{COA}
\begin{widetext}
\begin{equation}
\label{eq:App-Kubo-1} G_{jk,c(s)}=\lim_{\omega\rightarrow0^+}\frac{e^2}{h \pi \omega L}\int^L_0 dx \int_{-\infty}^\infty  d\tau  e^{i \omega \tau}\langle T_\tau J_{j,c(s)}(y,\tau)J_{k,c(s)}(x,0)\rangle ,
\end{equation}
where the currents $J_{c(s)}(x,\tau)=-i  \sqrt{2} \partial_\tau \theta_{c(s)}(x,\tau)$. The currents can be separated into two chiral currents, $J_{j,c(s)}^R=\sqrt{2}\partial \theta_{j,c(s)} \equiv(\partial_x-i\partial_\tau)\theta_{j,c(s)}/\sqrt{2}$ and $J_{j,c(s)}^L=\sqrt{2} \bar{\partial}
\theta_{j,c(s)} \equiv (\partial_x+i\partial_\tau) \theta_{j,c(s)}/\sqrt{2}$. In terms of these chiral currents, $J_{j,c(s)}=J_{j,c(s)}^R-J_{j,c(s)}^L$ and the Kubo formula Eq.(\ref{eq:App-Kubo-1}) becomes
\begin{eqnarray}
\label{eq:App-Kubo-2} G_{jk,c(s)}=\lim_{\omega\rightarrow0^+} \frac{e^2}{h
\pi \omega L}\int_{-\infty}^\infty d\tau  e^{i \omega
\tau} \int^{L}_{0} dx &[&\langle T_\tau J_{j,c(s)}^R(y,\tau)J_{k,c(s)}^R(x,0)\rangle+ \langle T_\tau J_{j,c(s)}^L(y,\tau)J_{k,c(s)}^L(x,0)\rangle
\\
\nonumber&-& \langle T_\tau J_{j,c(s)}^R(y,\tau)J_{k,c(s)}^L(x,0) \rangle-\langle T_\tau J_{j,c(s)}^L(y,\tau)J_{k,c(s)}^R(x,0)\rangle ].
\end{eqnarray}
\end{widetext}
We will use the rotation matrix $\mathcal{R}$ corresponding to the boundary conditions in DEBC method to evaluate the correlation functions. 

\subsection{Boundary conditions and correlation functions}

We first consider the correlation functions of the chiral currents in an infinite quantum wire. The off-diagonal components $\langle J_{j,c(s)}^R J_{j,c(s)}^L \rangle$ vanish and the diagonal components are given by
\begin{subequations}
\begin{eqnarray}
\langle J_{c(s)}^R(y,\tau)J_{c(s)}^R(x,0) \rangle = - 2 \partial^2 \langle \theta_{c(s)}(z,\bar{z}) \theta_{c(s)}(0) \rangle
\\
\langle J_{c(s)}^L(y,\tau)J_{c(s)}^L(x,0) \rangle = - 2 \bar{\partial}^2 \langle \theta_{c(s)}(z,\bar{z})\theta_{c(s)}(0) \rangle
\end{eqnarray}
\end{subequations}
where $z=i\tau+(y-x)$. Since the $\theta$-correlation function is
\begin{equation}
\langle \theta_{c(s)}(z,\bar{z})\theta_{c(s)}(0)\rangle = - \frac{g}{2} \ln{|z|^2} \; ,
\end{equation}
we obtain
\begin{subequations}
\begin{eqnarray}
\label{eq:CCCR} \langle J_{c(s)}^R(y,\tau)J_{c(s)}^R(x,0) \rangle&=&\frac{g_{c(s)}}{z^2}
\\
\label{eq:CCCL} \langle J_{c(s)}^L(y,\tau)J_{c(s)}^L(x,0) \rangle&=&\frac{g_{c(s)}}{\bar{z}^2}
\end{eqnarray}
\end{subequations}


Let us recall how the boundary conditions can be written in terms of the rotation matrix $\mathcal{R}$
\begin{equation}
\vec{\phi}^R(x)=\mathcal{R}^T \vec{\phi}^L(x)|_{x=0},
\end{equation}
where $\vec{\phi}^R$ and$\vec{\phi}^L$ are defined as
\begin{eqnarray}
\vec{\phi}_{c(s)}^R
=\left(%
\begin{array}{c}
  \phi^R_{1,c(s)} \\ \vdots \\ \phi^R_{N,c(s)}
\end{array}%
\right), {\rm \; and} \qquad  \vec{\phi}_{c(s)}^L
=\left(%
\begin{array}{c}
  \phi_{1,c(s)}^L \\ \vdots \\ \phi^L_{N,c(s)}
\end{array}%
\right)
\end{eqnarray}
for N quantum wires. Because $\theta_{c(s)}=\sqrt{g^{}_{c(s)}} (\phi_{c(s)}^L-\phi_{c(s)}^R)$, the boundary conditions can be translated to
\begin{equation}
\vec{J}_{c(s)}^L(0)=\mathcal{R} \vec{J}_{c(s)}^R(0).
\end{equation}
A convenient trick to respect boundary conditions is to analytically continue the right mover currents to $x<0$ and identify
\begin{equation}
\label{eq:CurBC} J_{i,c(s)}^L(x,\tau)=\mathcal{R}_{ij} J_{j,c(s)}^R(-x,\tau),
\end{equation}
for $x>0$. With this identification, the chiral current correlation functions between different wires can be evaluated in terms of the matrix elements
\begin{subequations}
\label{eq:CCCF-BC}
\begin{eqnarray}
\langle J_{i,c(s)}^R(y,\tau)J_{j,c(s)}^R(x,0)\rangle&=&\frac{g_{c(s)} }{z^2} \delta_{ij},
\\
\langle J_{i,c(s)}^L(y,\tau)J_{j,c(s)}^L(x,0)\rangle&=&\frac{g_{c(s)}}{\bar{z}^2}\delta_{ij},
\end{eqnarray}
for the diagonal correlation functions, $\langle RR \rangle $ and $\langle LL \rangle$, while
\begin{eqnarray}
\langle  J_{i,c(s)}^R(y,\tau)J_{j,c(s)}^L(x,0) \rangle =\frac{ \mathcal{R}_{ji}g_{c(s)} }{(i\tau+(x+y))^2}
\\
\langle J_{i,c(s)}^L(y,\tau)J_{j,c(s)}^R(x,0) \rangle = \frac{ \mathcal{R}_{ij} g_{c(s)} }{(i\tau-(x+y))^2}
\end{eqnarray}
\end{subequations}
for the off-diagonal correlation functions, $\langle RL \rangle $ and $\langle LR \rangle$.

\subsection{Conductance tensor}

Now, we can insert the correlation functions Eq.(\ref{eq:CCCF-BC}) into the Kubo formula Eq.(\ref{eq:App-Kubo-2}) to evaluate the conductance tensor. With the aid of the integral formula,
\begin{equation}
\label{eq:tauintegral}\int_{-\infty}^{\infty} d\tau e^{i\omega\tau} \frac{1}{(i\tau+u)^2}=2\pi\omega \Theta(u)  e^{-\omega u},
\end{equation}
where $\Theta(u)$ is the Heaviside step function, the Kubo formula reads
\begin{widetext}
 \begin{equation}
G_{ij,c(s)}=\frac{2 g_{c(s)} e^2}{hL} \int_0^L
dx\;[\delta_{ij}\,(\Theta(x-y)+\Theta(y-x))-\mathcal{R}_{ji}\Theta(x+y)-\mathcal{R}_{ij}\Theta(-x-y)].
\end{equation}
\end{widetext}
The integration of the combined first and second $\Theta$-functions gives a constant $L$. In addition, since both $x,y>0$, the integration over the third $\Theta$-function gives a constant $L$, while that over the fourth one vanishes. The conductance tensor evaluated from the Kubo formula is therefore given by
\begin{equation}
\label{eq:conductance} G_{ij,c(s)}=2 g_{c(s)}\frac{e^2}{h} (\delta_{ij}-\mathcal{R}_{ji}).
\end{equation}

As an example of this generic formula for junctions of multiple quantum wires, let us compute the conductance for a junction of two wires. First, the N BC is governed by the rotation matrix $\mathcal{R}_{N,ij}=\delta_{ij}$. Hence, all elements of  the conductance $G_{N}$ vanish, indicating a total decoupled junction. Second, the D BC is governed by the rotation matrix 
\begin{equation}
\mathcal{R}_D= \left(
\begin{array}{cc}
  0 & 1 \\
  1 & 0 \\
\end{array}%
\right).
\end{equation}
Inserting this into Eq.(\ref{eq:conductance}), one obtains 
\begin{equation}
G_D= 2 g_{c(s)} \frac{e^2}{h}
\left(%
\begin{array}{cc}
  1 & -1 \\
  -1 & 1 \\
\end{array}%
\right), 
\end{equation}
and 
\begin{equation}
I_1=-I_2=2 g\frac{e^2}{h}(V_1-V_2), 
\end{equation}
from the definition $I_{i}=G_{ij}V_j$. This is, as one would expect, the conductance for perfect transmission in a 1D quantum wire.

\renewcommand{\theequation}{B\arabic{equation}}
\setcounter{equation}{0}  
\section{Multi-Dimensional Modular Transformation}  
\label{sec:Modular-Trans}

Since modular transformations are useful in calculations using BCFT, we will in this appendix define and provide the general formulation of the the multi-dimensional modular transformation.

Generically, a d-dimensional partition function $Z_{\tilde{q}}$ is proportional to
\begin{equation}
\label{eq:d-dim-partition}
Z=\frac{1}{\eta(\tilde{q})^d} \sum_{\vec{u} \in \Lambda} \tilde{q}^{\frac{1}{4}\vec{u}^2}=\frac{1}{\eta(\tilde{q})^d} \sum_{\vec{u} \in \Lambda} e^{-\frac{l \pi}{\beta} |\vec{u}|^2}
\end{equation}
where $\tilde{q} \equiv e^{-\frac{4 \pi l}{\beta}}$ and $\Lambda$ indicates the lattice points. In order to check the finite size spectrum with a given boundary condition, we have to rewrite the partition function in terms of $q\equiv e^{-\frac{\pi \beta}{l}}$, i.e. perform a modular transformation. First, the modular transformation of the Dedekind $\eta$-function reads
\begin{equation}
\label{eq:modular-trans-Dedekind}\eta(\tilde{q})=\sqrt{\frac{\beta}{2 l}}\eta(q).
\end{equation}
Then the modular transformation of the summation in Eq.(\ref{eq:d-dim-partition}) can be achieved by using the Poisson summation formula, which replaces the summation by a integration with the periodic $\delta$-function, $\delta_{\vec{u} \in \Lambda}(\vec{x})$
\begin{equation}
Z=\frac{1}{\eta(\tilde{q})^d} \int d^d{x} \;\delta_{\vec{u}\in \Lambda}(\vec{x}) \;e^{-\frac{l \pi}{\beta} |\vec{x}|^2}\; .
\end{equation}
The $\delta$-function can be further written as the sum of $\Lambda'$, reciprocal lattice of $\Lambda$
\begin{equation}
\delta_{\vec{u}}(\vec{x})=\frac{1}{V_0(\Lambda)}\sum_{\vec{u}' \in \Lambda'} \exp[i 2 \pi (\vec{u}' \cdot \vec{x})]\; ,
 \end{equation}
where $V_0(\Lambda)$ is the volume of the unit cell. The partition function becomes
\begin{equation}
\label{eq:d-dim-partition-1} Z=\frac{1}{V_0(\Lambda) \eta(\tilde{q})^d} \sum_{\vec{u}' \in\Lambda'} \int d^d{x} \;e^{i 2 \pi (\vec{u}' \cdot \vec{x})}\; e^{-\frac{l \pi}{\beta}|\vec{x}|^2},
\end{equation}
cast using a standard Gaussian integral.

To proceed, we shall use the following identity
\begin{eqnarray}
&&\int d^d x \; \exp[-\frac{1}{2} \sum_{i,j=1}^d x_i A_{ij} x_j+ \vec{b} \cdot \vec{x}] \nonumber
\\
&=& \frac{(2 \pi)^{d/2} }{\sqrt{\det{A}} } \; \exp[W(\vec{b})],
\end{eqnarray}
where $A_{ij}$ is a $d\times d$ matrix and $W(\vec{b})=\frac{1}{2} \sum_{i,j=1}^d b_i (A^{-1})_{ij} b_j$. In our case Eq.(\ref{eq:d-dim-partition-1}), one can identify that $A_{ij}=\frac{2 \pi l}{\beta}\delta_{ij}$ and $\vec{b}=i 2 \pi \vec{u}'$ and obtain
\begin{eqnarray}
(\det{A})^{-\frac{1}{2}}&=&(\frac{2 \pi l }{\beta})^{-\frac{d}{2}}=(\frac{\beta}{2 \pi l})^{\frac{d}{2}}
\\
W(\vec{b})&=&-\frac{(2 \pi)^2}{2} \cdot \frac{\beta}{2 \pi l}|\vec{u}'|^2= -\frac{\pi \beta}{l}   |\vec{u}'|^2.
\end{eqnarray}   
Inserting these results into Eq.(\ref{eq:d-dim-partition-1}), the partition function becomes
\begin{eqnarray}
\nonumber Z&=&\frac{(2 \pi)^{\frac{d}{2}}}{V_0(\Lambda) \eta(\tilde{q})^d}\times (\frac{\beta}{2 \pi l})^{\frac{d}{2}} \sum_{\vec{u}' \in \Lambda'} e^{-\frac{\pi \beta}{l}|\vec{u}'|^2}
\\
&=&\frac{(2)^{\frac{d}{2}}}{V_0(\Lambda) \eta(q)^d} \sum_{\vec{u}' \in \Lambda'} q^{|\vec{u}'|^2},
\end{eqnarray}
where the identity in Eq.(\ref{eq:modular-trans-Dedekind}) is used for second equality.


\end{document}